\documentclass[manuscript]{aastex}

\usepackage{natbib}
\shorttitle{Evolution of Gas Clouds}
\shortauthors{Kwak, Shelton and Raley}

\begin{document}


\title{The Evolution of Gas Clouds Falling \\ in the Magnetized Galactic Halo: \\ 
	High Velocity Clouds (HVCs) \\ Originated in the Galactic Fountain}	


\author{Kyujin Kwak, Robin L. Shelton and Elizabeth A. Raley} 
\affil{Department of Physics and Astronomy, University of Georgia, 
Athens, GA 30602}







\begin{abstract}

In the Galactic fountain scenario, supernovae and/or stellar winds propel material into 
the Galactic halo. As the material cools, it condenses into clouds. 
By using FLASH three-dimensional magnetohydrodynamic simulations, we model and study 
the dynamical evolution of these gas clouds after they form and begin to 
fall toward the Galactic plane. In our simulations, we assume that the gas clouds 
form at a height of $z=5~\mbox{kpc}$ above the Galactic midplane, 
then begin to fall from rest. 
We investigate how the cloud's evolution, dynamics, and interaction with 
the interstellar medium (ISM) are affected 
by the initial mass of the cloud. 
We find that clouds with sufficiently large initial densities 
($n \ge 0.1 ~ \mbox{H atoms} ~ \mbox{cm}^{-3}$) 
accelerate sufficiently and maintain sufficiently large column densities 
as to be observed and identified as high-velocity clouds (HVCs)
even if the ISM is weakly magnetized ($1.3~\mu$G). 
However, the ISM can provide noticeable 
resistance to the motion of a low-density cloud 
($n \le 0.01 ~ \mbox{H atoms} ~ \mbox{cm}^{-3}$) 
thus making it more probable 
that a low-density cloud will attain the speed of 
an intermediate-velocity cloud rather than the speed of an HVC. 
We also investigate the effects of various possible magnetic field configurations. 
As expected, the ISM's resistance 
is greatest when the magnetic field is strong and 
perpendicular to the motion of the cloud. 
The trajectory of the cloud is guided by the magnetic field lines 
in cases where the magnetic field is oriented diagonal to the 
Galactic plane. 
The model cloud simulations show that the interactions between 
the cloud and the ISM can be understood via analogy to the shock 
tube problem which involves shock and rarefaction waves. 
We also discuss accelerated ambient gas, streamers of material 
ablated from the clouds, and the cloud's evolution from a 
sphere-shaped to a disk- or cigar-shaped object. 

\end{abstract}





\section{Introduction}

Neutral hydrogen (\ion{H}{1}) clouds that have large local standard of rest velocities 
($\vert V_{\mbox{LSR}} \vert \geq 90 ~\mbox{km} ~ \mbox{s}^{-1}$) 
were first discovered by \citet{muller1963} and
identified as high-velocity clouds (HVCs). 
For HVCs at high latitude, these velocities are incompatible 
with a simple model of the differential rotation of the 
Galaxy. 
Thus, a 40 yr long search for their origins began. 
The three leading models propose that HVCs originate from: 
the Galactic fountain; the infall of matter stripped from 
smaller galaxies that orbit the Milky Way; and the accretion of primordial 
gas left over from the epoch of galaxy formation. 
In the fountain scenario \citep{shapiro1976}, 
material from the Galactic disk is pushed into the lower Galactic halo 
by superbubbles (made by a series of co-located supernovae in an OB cluster), 
radiatively cools, and then falls back to the Galactic disk due to the force of gravity. 
This model is capable of explaining both upward-moving and downward-moving 
HVCs, and intermediate-velocity clouds (IVCs) 
which are \ion{H}{1} clouds with intermediate velocities 
\citep[$40 ~\mbox{km}~\mbox{s}^{-1} \leq \vert V_{\mbox{LSR}} 
\vert \leq 90 ~\mbox{km} ~ \mbox{s}^{-1}$;][]{KuntzDanly1996}. 
In the second model, gas is stripped from nearby galaxies. 
The Magellanic Stream (MS), which is associated with 
the Large Magellanic Cloud (LMC) and 
the Small Magellanic Cloud (SMC), is an example. 
In addition, there are more than a dozen dwarf galaxies
within $100~\mbox{kpc}$ of the Milky Way 
(see \citet{belokurov2007}, and references therein). 
These dwarf galaxies may have been 
stripped of their gas during previous passages through the Milky Way 
as the Sagittarius Dwarf Galaxy is thought to 
have been \citep{putman2004}. 
In the accretion scenario, some of the gas left over from the formation of 
the Galaxy is currently being accreted \citep{oort1966}. 
\citet{Blitz1999} developed this idea further, 
claiming that HVCs are material moving under the gravitational potential of the
Local Group of galaxies. According to their model, nearby HVCs are falling onto 
the Galactic disk due to tidal instabilities, although fewer HVCs probably 
have intergalactic origins than \citet{Blitz1999} originally expected. 

For any given HVC, the applicability of different models can be 
distinguished by the distance to the HVC and the HVC's metallicity, 
but metallicity is widely believed to be the better discriminant. 
Clouds made by the Galactic fountain should be relatively 
near to the Galactic plane ($\leq 5 ~\mbox{kpc}$) and 
have substantial metallicities (comparable to the solar metallicity). 
Generally, the clouds that originated from extragalactic sources may 
be further away than clouds produced by the Galactic fountain, 
although it is possible that externally produced HVCs 
could eventually approach the Galactic plane after falling for a long time. 
For such HVCs, a measurement of their low metallicities 
would be needed in order to assign the primordial gas accretion model. 
For HVCs that are composed of gaseous material that was  
stripped off of the satellite galaxies orbiting the Milky Way, their metallicities 
should be strongly correlated with those of the satellite galaxies. 
Generally they should be intermediate between those of the Milky Way's disk 
and those of intergalactic clouds. 

According to the current measurements of distances and metallicities, 
not all HVCs have the same origin. For example, 
the metallicity of the MS is reported to be $\sim 0.25$ solar, 
which is consistent with the metallicity 
of the SMC and LMC \citep{Lu1998,Gibson2000}. 
This consistency together with the geometrical association of the MS 
with the SMC and LMC provides strong evidence that the MS 
material came from the SMC and LMC. 
The distance upper limit 
\citep[$z< 4$ kpc;][]{danly1993,keenan1995} 
and solar-comparable metallicity 
\citep[$\sim 0.8$ solar;][]{Tufte1998,Wakker2001} 
of Complex M put it in the list of those made by Galactic fountains. 
Complex C, which is one of the most extensively studied HVCs, 
is more likely to 
have an extragalactic origin because of its relatively low metallicity. 
The most recently measured metallicity and distance bracket are 
$\sim 0.13$ solar \citep{collins2007} and approximately 
between 4 and 11 kpc 
\citep{wakker2007_complexC,Thometal2008ApJ}, respectively. 
Primordial inter-galactic medium (IGM) gas is thought to have metallicities 
of $\le 0.1$ solar \citep{Davisetal1996ApJ}, thus clouds formed 
from the IGM may have similarly low metallicities. 

In this paper, we further examine clouds produced in the Galactic fountain. 
By using three-dimensional magnetohydrodynamic (MHD) simulations made 
with FLASH version 2.5 \citep{fryxell2000},
we model and  study the dynamical evolution of gas clouds 
that form in the Galactic fountain process and start to fall from rest. 
The goal of our simulations is to investigate whether these clouds could develop 
the characteristics of HVCs when they fall back toward the Galactic plane, 
specifically whether they could accelerate to the velocities of HVCs 
while retaining sufficiently large column densities to be observable. 
The initial clouds are assumed to form at the height $z=5~\mbox{kpc}$ 
above the Galactic midplane 
when the radiatively cooled gas condenses at the peak of the fountain process. 
The peak location at which the dense clouds form is not directly constrained by observations, 
but the hydrodynamic (HD) simulations of \citet{avillez2000HVCs} suggest 
that $5$ kpc is within the reasonable height range. 
Besides, in order for a cloud to accelerate from rest to 
$\vert v \vert > 90~\mbox{km}~\mbox{s}^{-1}$ 
under the Galaxy's gravity, it must fall from a height of at least a few kiloparsec
above the Galactic midplane. 
Note that the clouds in our simulations fall down directly toward 
an observer so that the observed velocities are the same as the
vertical velocities. However, in the cases in which the cloud's 
velocity vector is not parallel to the sight line to the observer, 
the observed velocity is smaller than the cloud's real velocity. 

Since the initial cloud falls through the Galactic halo, 
which is filled with a low density interstellar 
medium (ISM) as well as a magnetic field, its interaction with the ISM also affects 
the dynamical evolution of the cloud. Because the conditions of 
the Galactic halo and the initial cloud
are not precisely constrained by the observations, 
we survey a large range of possible conditions. 
We pay special attention to the effects of 
both the strength and the orientation of the interstellar magnetic field. 
In order to determine the effects of magnetic field geometries, 
we consider three different magnetic field configurations in the Galactic halo: 
parallel, perpendicular, and $45\degr$ with respect to the Galactic plane. 
Cases with no magnetic field are also performed and two initial cloud 
densities are examined. 

Previous work demonstrated that HD and 
MHD simulations are well developed 
and useful tools for studying HVCs 
\citep{tenorio1986,tenorio1987,santillan1999,kudoh2004}. 
The two-dimensional HD simulations by \citet{tenorio1986,tenorio1987} 
implied that very massive HVCs shocked and perturbed 
the Galactic disk. The two-dimensional HD simulations of \citet{kudoh2004} 
implied that HVC-galaxy interactions could explain unusual 
observed structures in our galaxy, such as the mushroom-shaped 
GW 123.4-1.5. \citet{santillan1999} used the two-dimensional MHD simulations 
to examine the important role of the magnetic field. They found 
that magnetic field lines that are oriented perpendicular to the cloud's 
motion became sufficiently stretched and compressed as to prevent 
HVCs from falling closer to the Galactic disk. 
In their simulations, the ISM is assumed to be isothermal 
and in hydrostatic equilibrium (HSE) in which the gravity is 
balanced by the gradients both in the thermal and 
magnetic pressures. Note that at the beginning of their simulations, 
the HVCs were already close to the Galactic plane and usually 
moving rapidly. This initial configuration is based upon the assumption that 
either HVCs had originated at large $z$-heights but did not 
interact significantly with the extended halo ISM or HVCs originated 
close to the Galactic plane. Our simulations are different from 
these cloud--disk interaction simulations because we simulate all 
three dimensions and because the primary goals 
of our simulations are to determine whether gas clouds that form 
in the Galactic fountain process could accelerate to HVC velocities 
and to determine how the interstellar magnetic field and initial 
cloud density affect the outcome. 

This paper is organized as follows. 
The model parameters used in our simulations 
are summarized in Section \ref{model_para_S}. 
We will briefly describe the 
numerical methods used for our simulations in Section \ref{methods_S}. 
The results of the simulations are presented in Section \ref{results_S}. 
We will discuss our results further and their connection to 
observations in Section \ref{discuss_S}. 
In Section \ref{discuss_S}, we will also show how the model geometries 
that we choose for the magnetic field 
in the Galactic halo can be related to realistic configurations 
in the context of the Galactic fountain model. 
Our conclusions are explained in Section \ref{conclusion_S}. 

\section{Modeled Physical Processes and Parameters} \label{model_para_S}

\begin{figure}
\plottwo{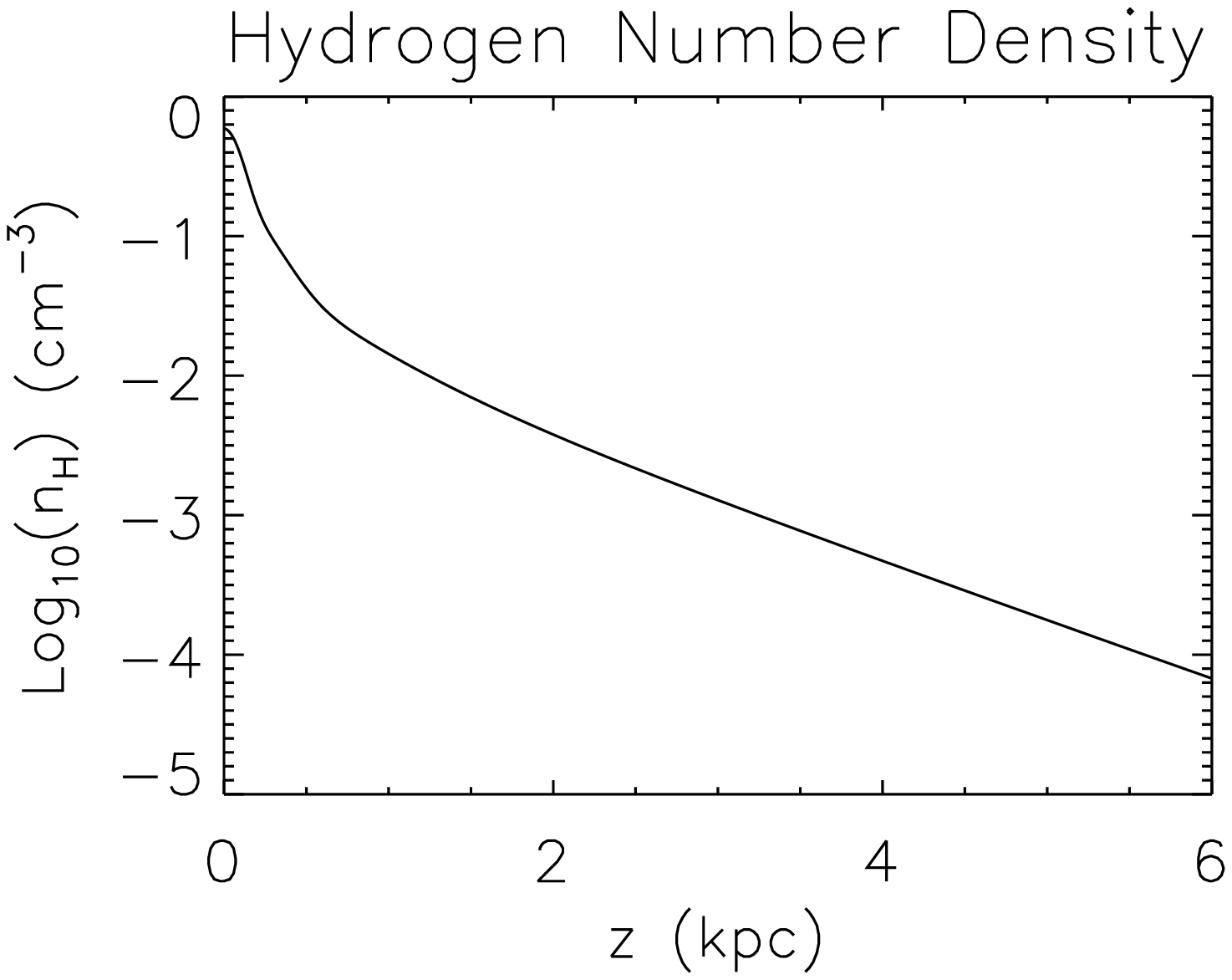}{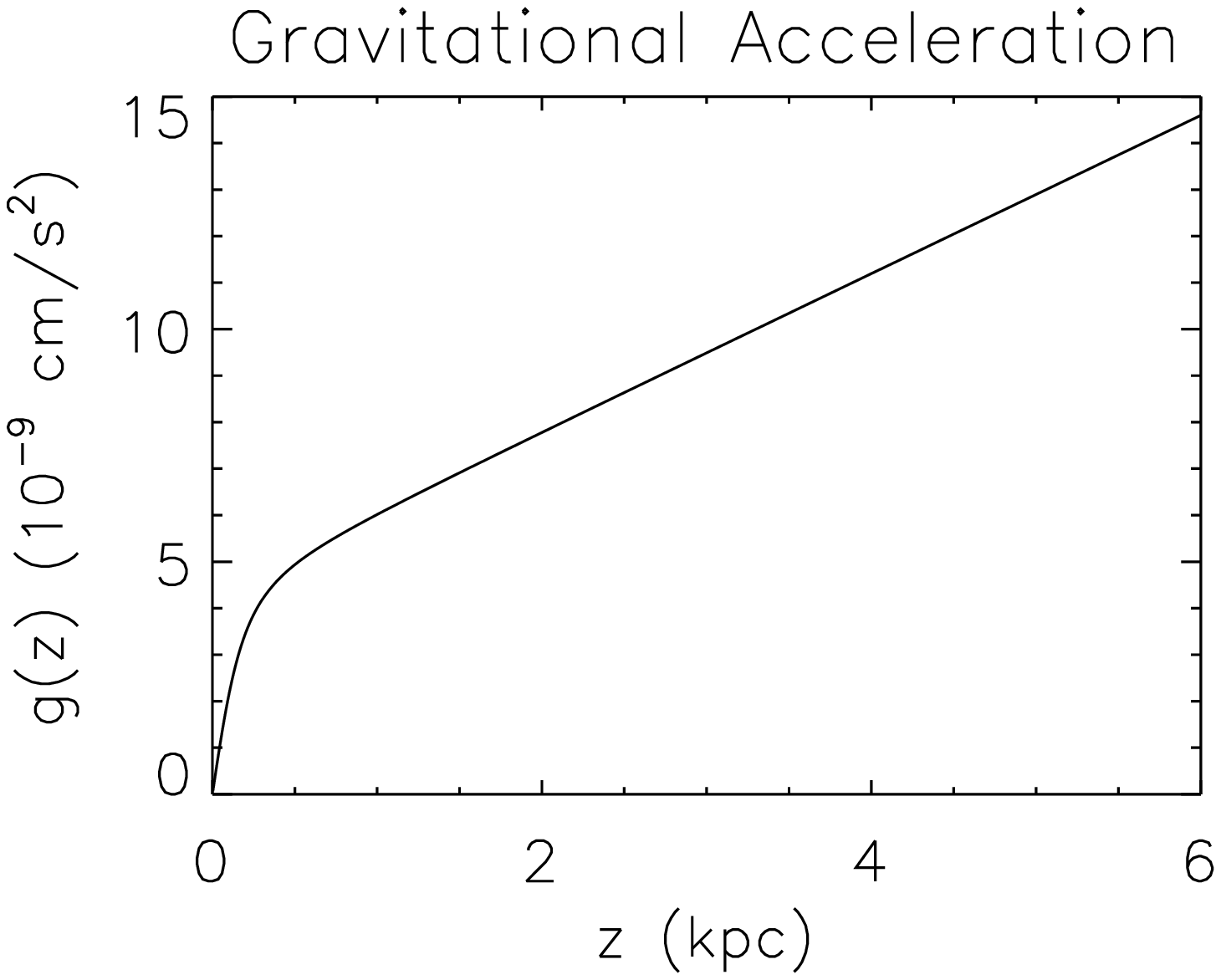}\\
\plottwo{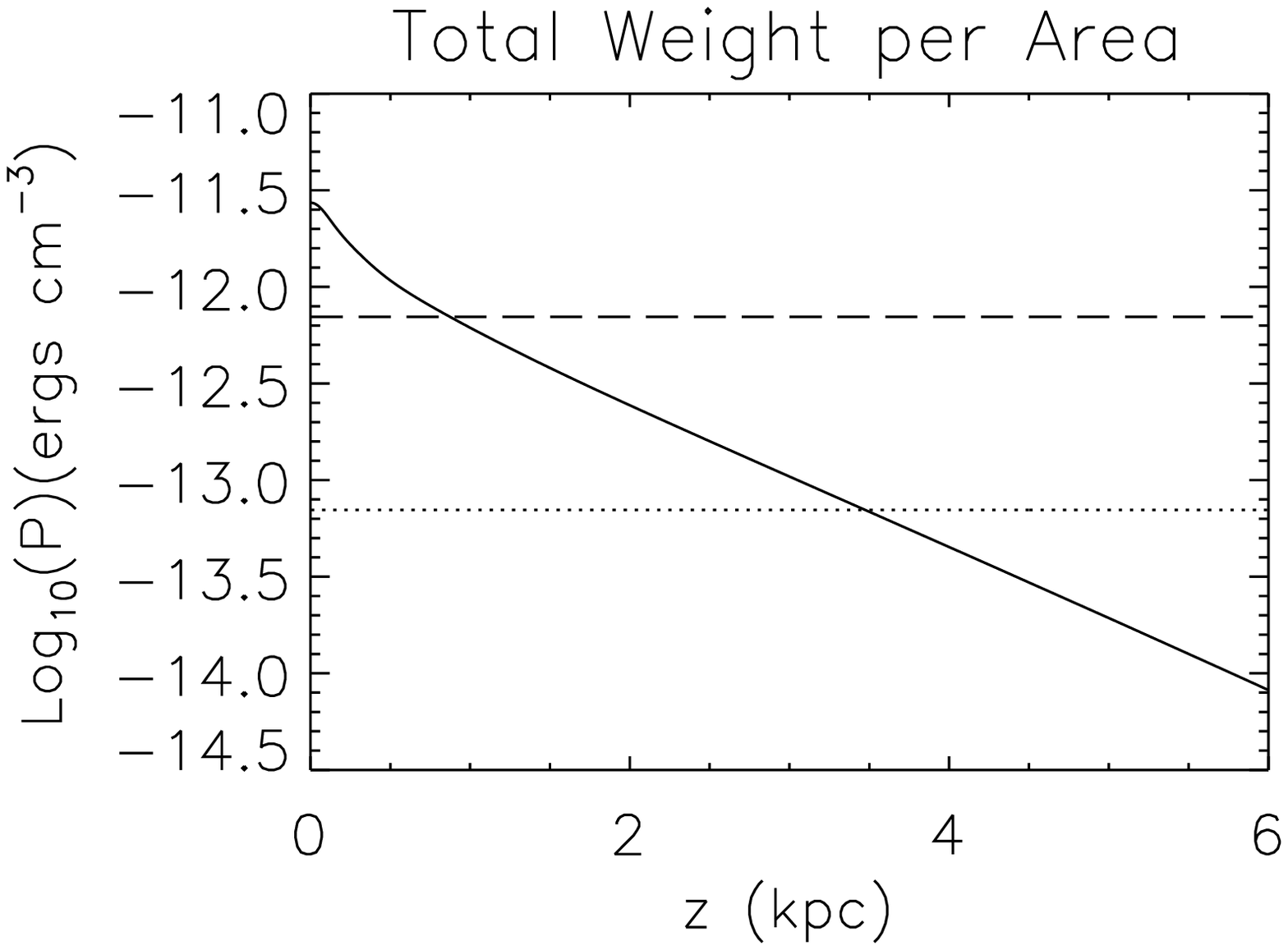}{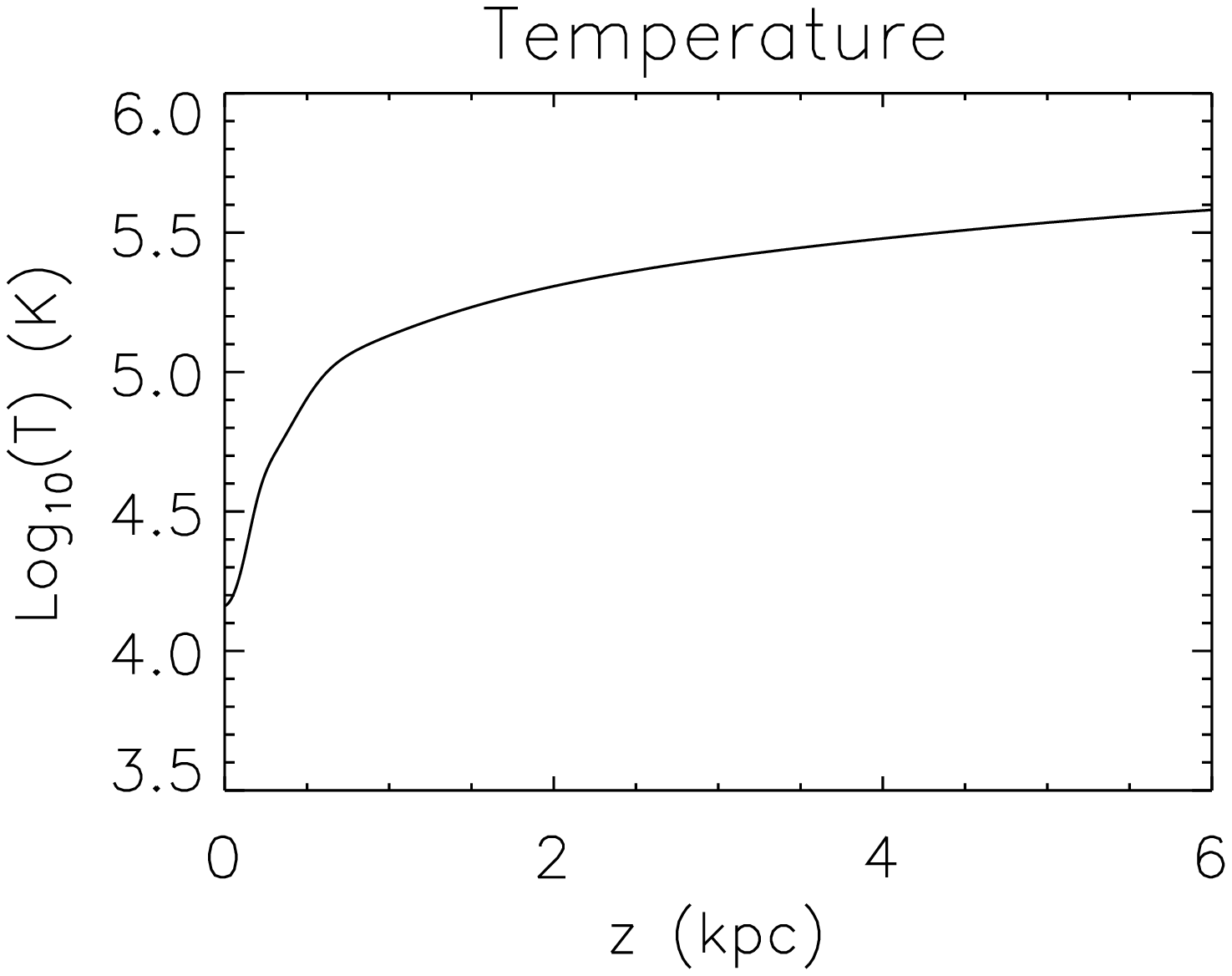}
\caption{The physical state of the Galactic halo gas assumed in the simulations 
is shown. 
Top panels from left to right: hydrogen number density and gravitational acceleration 
as a function of height for regions at the Sun's galactocentric radius, 
from \citet{ferriere1998}. 
Molecular hydrogen is excluded from the density profile. 
We assume that the helium abundance in the halo gas is 10\% of the 
hydrogen abundance. 
Bottom left panel: the total weight per unit area ($\mbox{erg}~\mbox{cm}^{-3}$) 
and two magnetic pressure values used in some simulations. 
The total weight per unit area is marked by the solid line, a magnetic pressure of 
$7.0\times10^{-14}~\mbox{erg}~\mbox{cm}^{-3}$ (corresponding to a magnetic 
field strength of $1.3~\mu\mbox{G}$) is marked by the 
dotted line, and a magnetic pressure of 
$7.0\times10^{-13}~\mbox{erg}~\mbox{cm}^{-3}$ (corresponding to a magnetic 
field strength of $4.2~\mu\mbox{G}$) is marked by the dashed line. 
Bottom right panel: assumed ambient temperature as a function of height. 
\label{ferriere_profile}}
\end{figure}

The physical state of the ISM 
at the height of the Galactic halo is less well known than
that of the ISM in the Galactic disk. 
Nonetheless, estimates have been made from a wide variety of 
observational evidence. 
In our simulations, we use the density and gravitational 
acceleration equations of \citet{ferriere1998} and assume that 
their functional forms can be extrapolated to large heights 
above the plane, i.e., $z \approx 5$ kpc. 
We ignore the molecular hydrogen density because
most molecular clouds reside 
close to the Galactic plane. 
Figure \ref{ferriere_profile} shows 
the hydrogen number density and gravitational acceleration profiles 
within the computational domain of our simulations. 

The total weight per unit area of the gaseous material residing above 
the height, $z$, can be calculated from the density and 
acceleration profiles and is shown in Figure \ref{ferriere_profile}. 
In our calculation, we assume that the total weight per unit area 
of the ISM above $z=20~\mbox{kpc}$ is zero. 
We also assume that the ISM in the Galactic halo 
is in HSE, which allows us to 
calculate the gradient in the total pressure by setting 
it equal to the gradient in the weight per unit area 
(i.e., the gravitational force per volume). 
Generally, the total pressure is composed of three components; 
thermal, magnetic, and cosmic-ray pressure. 
In our simulations, we include only thermal and magnetic 
components because our primary interest is the dynamical 
effect of the magnetic field and because it is difficult to model 
the cosmic-ray pressure in the numerical simulations. 
However, in our simulations, the interstellar magnetic field 
strength does not vary with height.  
Thus, the gradient in the total pressure is 
entirely due to the gradient in the thermal pressure. 
Therefore, the ISM in our simulations is in HSE 
if the gradient in the weight per unit area is 
balanced by the gradient in the thermal pressure alone. 
Additionally, we assume that there is no constant offset 
between these two variables. Thus, we set the thermal 
pressure at height $z$, equal to the weight per unit area 
of the gaseous material above height $z$. 

The thermal pressure, $p_{th} (z)$, at the height $z$ is given by 
$p_{th}(z)=n(z) ~ k_B ~ T(z)$, 
where $k_B$ is the Boltzmann's constant 
and $n(z)$ and $T(z)$ are the total number of particles per unit volume 
and the temperature at the height $z$, respectively. 
The density of atomic nuclei is taken to be $110\%$ of the density 
of hydrogen because the helium abundance in the ISM is $10\%$ of 
the hydrogen abundance. 
In order to apply the MHD simulations, we
also assume that both hydrogen and helium are fully ionized. 
As a result, the density of particles is $n(z)=2.3~n_H (z)$, 
where $n_H (z)$ is hydrogen number density from \citet{ferriere1998} 
that is shown in Figure \ref{ferriere_profile}. 
The temperature, $T(z)$, in Figure \ref{ferriere_profile} 
is calculated from $p_{th}(z)$ and $n(z)$. 
Because $n(z)$ incorporates contributions from cold dense gas, 
warm moderately dense gas, and hot rarefied gas, $n(z)$ and 
$T(z)$ are phase-averaged quantities. 
Therefore, the temperature of the ISM 
used in our simulations is less than a million degrees Kelvin. 
We will discuss the effect of the ISM temperature 
after discussing our simulation results (see 
Section \ref{hot_ISM_effect_S}). 

Given the current uncertainty about the magnetic field 
in the Galactic halo, we consider only simple cases, in which 
the strength of the magnetic field, $B$, is initially constant 
throughout the simulation domain. 
This allows us to choose different values of the 
constant magnetic field strength for different simulation models 
without violating the HSE condition. 
In order to analyze the effect of the magnetic field strength, 
we choose two different values for the magnetic pressure, $7.0 \times 
10^{-14}$ and $7.0 \times 10^{-13}$ $\mbox{erg}~\mbox{cm}^{-3}$
which correspond to $B=$ $1.3$ and $4.2$ $\mu \mbox{G}$, respectively. 
These two values are consistent with the strength of the magnetic field 
in the lower halo. 
The two magnetic pressure values are plotted together with the 
total weight per unit area in Figure \ref{ferriere_profile}. 

We consider three different configurations for the initial magnetic field
orientation: parallel, perpendicular, and tilted 
at $45 \degr$ with respect to the Galactic plane. In the three-dimensional Cartesian 
coordinates of our simulations, 
where $x$ and $y$ axes are in the Galactic plane, 
we set up the ``parallel magnetic field" to be parallel to 
the $z$-axis and the direction of the cloud's motion 
($B_z = \mbox{constant}$ and $B_x = B_y = 0$). 
The ``perpendicular magnetic field" is perpendicular to 
the $z$-axis and the direction of the cloud's motion 
($B_y = \mbox{constant}$ and $B_x = B_z = 0$). 
In the case where the magnetic field 
is titled at $45 \degr$ we set it up along 
the $y$-$z$ plane ($B_y=B_z=\mbox{constant}$ and $B_x=0$). 
We also simulate cases with no magnetic field. 
In all, we present nine cases ( A1 and A2, which have no magnetic 
field, B1-B4, which have a magnetic field that is 
perpendicular to $\hat{z}$, C1 and C2, which have a magnetic 
field that is parallel to $\hat{z}$, and D, in which the magnetic 
field orientation is tilted). 
The model parameters for the ambient ISM, magnetic 
pressure, and orientation are summarized in Table \ref{models_T}. 

In our simulations, a spherical gas cloud with a radius 
of $0.25~\mbox{kpc}$ falls down from rest through the ambient ISM. 
The center of the gas cloud is initially located 
at a height $z=5~\mbox{kpc}$. The cloud is homogeneous 
within the sphere and its hydrogen number density is either 
$0.01~\mbox{cm}^{-3}$ for a low-density cloud 
or $0.1~\mbox{cm}^{-3}$ for a high-density cloud. 
The corresponding cloud's mass is approximately 
$2.26 \times 10^4$ and $2.26 \times 10^5~\mbox{M}_{\sun}$ 
for the low- and high-density cloud, respectively. 
For a sight line passing through the center of the cloud, the hydrogen 
column density is approximately $1.5 \times 10^{19}$
and $1.5 \times 10^{20} ~ \mbox{cm}^{-2}$, respectively, if the cloud's radius 
is 0.25 kpc and the whole cloud is 
made of gas with hydrogen volume densities of 
$0.01$ and $0.1~\mbox{cm}^{-3}$, respectively. 
These column densities are common of observed HVCs. 
The $1.5 \times 10^{19}$ case is near the center of the observed range 
in column densities, while $1.5 \times 10^{20} ~ \mbox{cm}^{-2}$ is 
close to the largest observed HVC column density \citep{Wakker2001}. 
The modeled cloud is initially in pressure equilibrium 
with the ambient ISM, that is, 
the thermal pressure of the cloud at the height $z$ is the same as that of
the ambient ISM. Therefore, the initial temperature 
of the gas in the cloud varies 
with height, from about $5000$ to $7400$ K for the low-density cloud 
and from about $500$ to $740$ K for the high-density cloud. 
The magnetic field lines of the ambient ISM 
pass through the initial cloud without change in strength or orientation. 
The self-gravity of our model cloud can be ignored because it is 
much less than the Galactic gravity. 

Because the primary goal of the our numerical study is 
to determine whether or not clouds made by the fountain process can 
gravitationally accelerate to HVC velocities, we must model gravity. 
Gravity also acts on the ambient gas, so to prevent it from collapsing, 
we assume that it is in HSE, as is frequently 
done in numerical studies of cloud--ISM interactions 
\citep{tenorio1987,santillan1999,kudoh2004}. 
However, in order to maintain HSE, the background gas cannot be subject to 
net heating or cooling. For this reason, we disallow radiative cooling, 
radiative heating, and thermal conduction.

Although we do not include radiative cooling, external heat sources (aside 
from the cloud impact), or heat conduction, other authors have considered 
their effects. 
\citet{CowieMcKee1977ApJ} and \citet{McKeeBegelman1990ApJ} 
analytically calculated 
the evaporation and condensation rates experienced by a cool, static cloud 
embedded in a hot (a few million degree) ambient gas 
due to thermal conduction and radiative cooling. 
Their analytic calculations showed that the cool static cloud evaporates 
even when the rate of heat conduction is saturated. 
\citet{VieserHensler2007static} 
tested these analytic results by numerical 
simulations of static clouds in a hot ambient medium, 
and their numerical simulations showed a new result; 
the clouds experience delayed evaporation or the ambient 
gas condenses onto the clouds when saturated heat conduction or 
radiative heating and cooling are considered. 
In the dynamical case, in which the cloud interacts with 
hot ($5 \times 10^6$ K) streaming ambient gas \citep{VieserHensler2007dyn}, 
the evaporation of the cloud due to thermal 
conduction is even further delayed. Given their results, 
the clouds in our simulations should not be expected to 
evaporate easily if thermal conduction were allowed. 
In addition, the cloud--ISM temperature 
gradient in our simulations is $\sim~1/10$ of that 
in \citet{VieserHensler2007dyn}, making 
the clouds that we examine even less likely to evaporate due to 
thermal conduction. 

In dynamical situations, thermal conduction plays the role of 
a diffusion process. For this reason, it is sometimes ignored in 
cases in which there are more dominant diffusion processes. For example, 
\citet{deAvillezBreitschwerdt2007ApJL} found turbulent diffusion 
to be more significant than thermal conduction in their simulations. 
\citet{Slavinetal1993ApJ} also argued that turbulent mixing together 
with radiative cooling cools the gas more efficiently 
than thermal conduction. 
In our simulations, modeling the diffusion of heat 
in the region between the cloud 
and the shock front would have the effect of decreasing the thermal 
pressure gradient and therefore the resistance to the cloud's 
downward motion. 
Its effect would be very small when the magnetic field is 
oriented perpendicular to the cloud's motion 
because in this case the resisting force due to the magnetic pressure 
gradient is much larger than that due to the thermal pressure gradient. 
Note that in this case thermal conduction is reduced significantly 
because electrons that are responsible for thermal conduction 
are prevented from carrying the thermal energy across the magnetic 
field lines. 
In the other cases, those in which there is no magnetic field or 
the field lines are parallel to the cloud's motion, 
we anticipate that the clouds would fall slightly faster 
if thermal conduction were modeled. 

In our MHD simulations, the effect of turbulent mixing 
due to shear instabilities is reduced because the magnetic field 
suppresses the growth of shear instabilities 
\citep{MacLowetal1994ApJ,Jonesetal1996ApJ}. 
We expect that the inclusion 
of thermal conduction would further reduce turbulent mixing because 
\citet{VieserHensler2007dyn} found that thermal conduction suppresses 
the growth of shear instabilities when the cool cloud moves through 
hot gas.

\section{Numerical Methods} \label{methods_S}

We use the MHD module of FLASH version 2.5 
\citep{fryxell2000} for our simulations. 
FLASH is an Eulerian grid-based code which was found to model 
cloud instabilities and fragmentation better than smoothed 
particle hydrodynamics (SPHs) codes \citep{Agertzetal2007MNRAS}. 
The external gravity module in FLASH is used 
to deal with the Galaxy's gravitational force on the Galactic halo. 
The non-magnetic simulations of Model A are performed with 
the MHD module by setting the initial magnetic field to zero. 

FLASH uses a block-structured adaptive mesh refinement (AMR) 
technique implemented through PARAMESH \citep{mac00}. 
When AMR is used, only the ``interesting" region 
within the computational domain 
is resolved better. It is resolved with smaller meshes 
than the other regions, 
thus saving computing time. 
In our simulations, we choose the density gradient as the refinement 
criterion. Because the density gradient is large in the vicinity of 
the cloud, this region is always modeled with high resolution 
during the simulation. The other regions are modeled with low resolution. 
For this reason, the AMR in FLASH is well suited for our simulations. 
 
Our simulations are performed in three-dimensional Cartesian coordinates 
where $x$ and $y$ are parallel to the Galactic plane 
and $z$ is the height above the plane. 
The computational domain is composed of $x,y\in[-0.75,~0.75]~\mbox{kpc}$ and 
$z\in[1.5,~6.0]~\mbox{kpc}$. 
Initially, the computational domain is partitioned into three identical 
cubes, called blocks, the sizes of which are 
$1.5 ~\mbox{kpc} \times 1.5 ~\mbox{kpc} \times 1.5 ~\mbox{kpc}$. 
This partitioning is considered to be the first refinement in 
AMR terminology. Before the simulations begin, each block is further 
refined two to four additional times, bringing the total level of refinement 
to three to five (i.e., \verb,lrefine_min=3, and \verb,lrefine_max=5, in FLASH). 
Each of these refinements subdivides each affected block into 
eight equal-sized smaller blocks. Thus, when we use five levels of refinement, 
the maximum possible number of blocks in our computational domain 
is $3 \times 8^{(5-1)}=12,288$ blocks. Each block is automatically subdivided 
into $8 \times 8 \times 8$ zones. The length, width, and height of each of 
these zones is 1/8 of those of the block in which it resides. 
Thus, when we use five levels of refinement, the maximum possible number 
of zones in our computational domain is 
$3 \times 8^{(5-1)} \times 8^3 \approx 6.3 \times 10^6$. In a fully refined domain, 
the zone sizes would be 
$\sim 12 ~\mbox{pc} \times 12 ~\mbox{pc} \times 12 ~\mbox{pc}$. 
In order to determine if our simulations are sufficiently resolved, 
we perform higher resolution simulations 
by increasing the maximum level of refinement from five to seven 
for some of our models. 
The maximum number of zones increases to
$512 \times 512 \times 1536$ with $7$ levels of refinement. 
Although these simulations are a factor of 64 more finely resolved than
the simulations made with five levels of refinement, 
we do not find any qualitative differences. 
So we run the remainder of our simulations 
with three to five levels of refinement in order to save computing time. 

Outflow boundary conditions are used for the boundaries 
along the $x$ and $y$ directions. 
For the boundaries along the $z$-direction, 
we create boundary conditions that would maintain the HSE condition.
This is done by assigning suitable values of density, pressure, 
magnetic field, and gravity to the zones outside the boundaries. 
 
\section{Results} \label{results_S}

In preparation for the discussion of the MHD simulations, we first 
discuss analytic calculations of ballistic cloud motion, which provide 
useful estimates of cloud kinematics (see Section \ref{freefall_S}) 
and we discuss one-dimensional shock tube problems, which provide 
useful analogies for understanding the HDs of infalling clouds 
(see Section \ref{shocktube_S}). Following these subsections, we present 
the results of our MHD simulations of falling clouds 
(see Section \ref{model_sim_S}). 

\subsection{Analytic Estimates for a Freely Falling Cloud}\label{freefall_S}

\begin{figure}
\epsscale{1.0}
\plotone{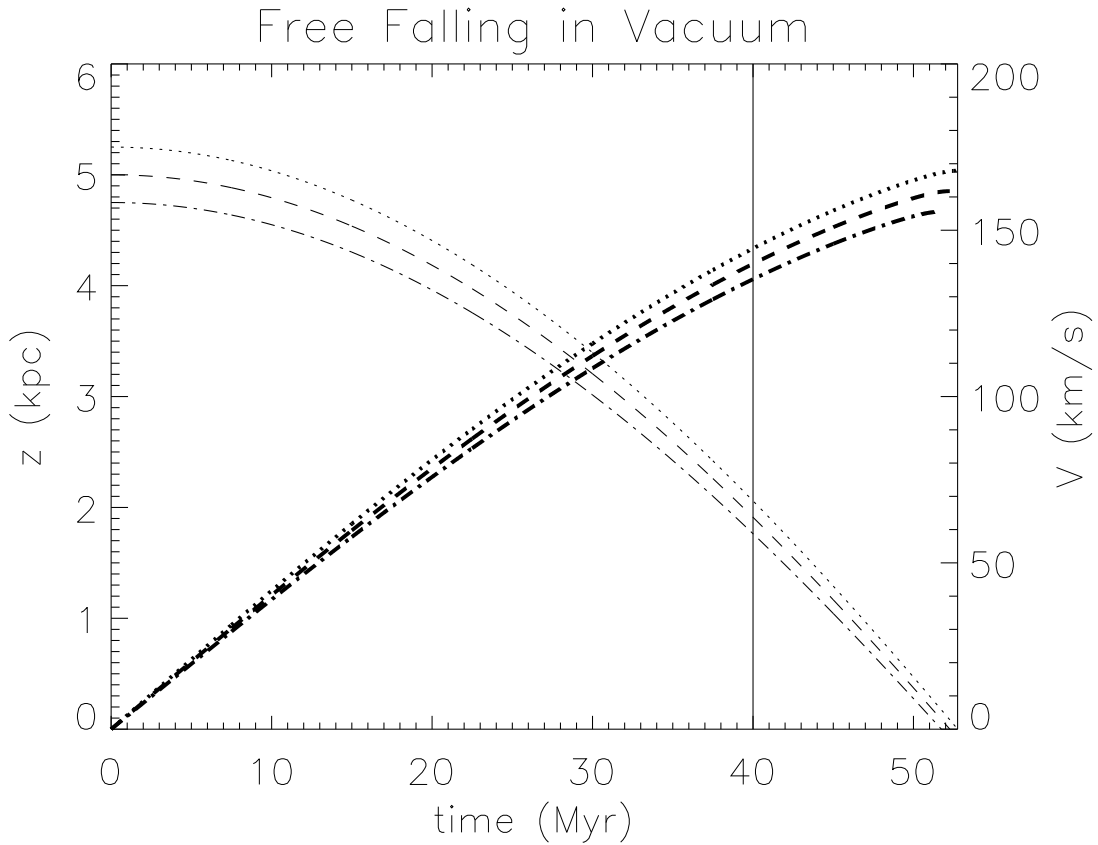}
\caption{Analytically calculated height and velocity of a cloud of the same 
size as the clouds used in our simulations as it falls in a vacuum 
under the gravitational acceleration shown in Figure \ref{ferriere_profile}. 
The heights (marked by thin curves that decrease from left to right) 
and downward speeds (marked by thick curves that increase from left to right) 
are shown as a function of time after the cloud begins to fall 
from rest. 
The dotted, dashed, and dot-dashed lines correspond to the top, center, 
and bottom of the cloud, respectively. At the beginning of the time period, 
the top, center, and bottom of the cloud are located at 5.25, 5.0, and 4.75 
kpc, respectively. The solid vertical line marks 40 Myr. 
\label{freefall_high_fig}}
\end{figure}

Figure \ref{freefall_high_fig} shows the height and velocity profiles 
of a cloud that falls freely through a vacuum, i.e., this is a theoretical 
calculation where the acceleration in the $z$-direction is set equal 
to the gravitational acceleration shown in Figure \ref{ferriere_profile} 
and there are no drag, buoyancy, or magnetic forces. 
The cloud imagined here has the same characteristics as the cloud 
in our simulations, i.e., it has a radius of $0.25~\mbox{kpc}$ 
and falls from rest starting at a height of $z=5~\mbox{kpc}$. 
The initial heights of the top, center, 
and bottom of the cloud are set to 
$5.25$, $5.0$, and $4.75$ kpc, respectively. 
By $40$ Myr, the top, center, and bottom of the cloud have fallen to 
$z=$ $2.06$, $1.91$, and $1.76$ kpc, respectively, 
with downward speeds of
$144.5$, $139.9$, and $135.3$ $\mbox{km}~\mbox{s}^{-1}$, respectively. 
If the cloud is composed of free particles which do not interact with each other
during the fall, then all of the particles should be confined between the top 
and bottom of the cloud and the cloud should look 
like an oblate spheroid at $40$ Myr. 

For the sake of roughly estimating the characteristics of the cloud, 
here, we assume that the cloud does not spread laterally 
during its fall. Thus, it maintains its original column density 
and radius. If this were to be the case, then, by the time that 
the cloud's speed reached $\sim140~\mbox{km}~\mbox{s}^{-1}$, 
the cloud would be 2 kpc from the Galactic midplane and subtend 
an angle of $14\degr$. This size is similar to that of Complex M I 
($10\degr$-$20\degr$, \citet{Wakker2001}). Note that 
Complex M I is known to reside at a distance of $\le4$ kpc 
\citep{Schwarzetal1995AA}. Given its characteristics, our cloud 
would be classified as an HVC. 

\subsection{Comparing with the Shock Tube Problems}\label{shocktube_S}

In the beginning of the simulations, 
the gravitational force exerted on the gaseous material inside the cloud 
exceeds the upward force due to the gradient in the thermal 
and magnetic pressures, so the cloud accelerates downward. 
As the cloud falls, its lower side pushes into 
the ambient ISM while its upper side pulls away from the ambient gas. 
These interactions instigate disturbances that propagate into 
the cloud and into the ISM as time passes. 

\begin{figure}

\plottwo{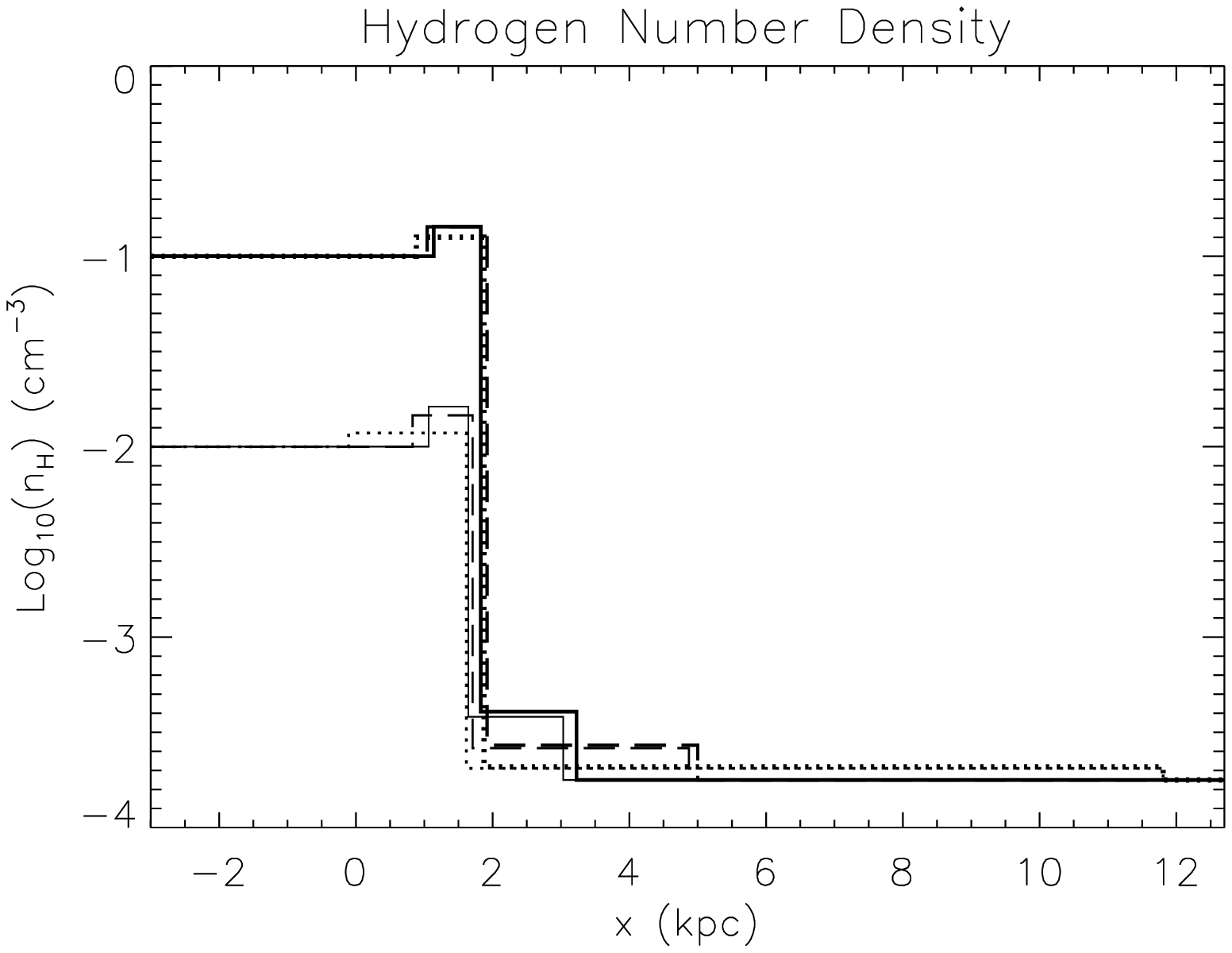}{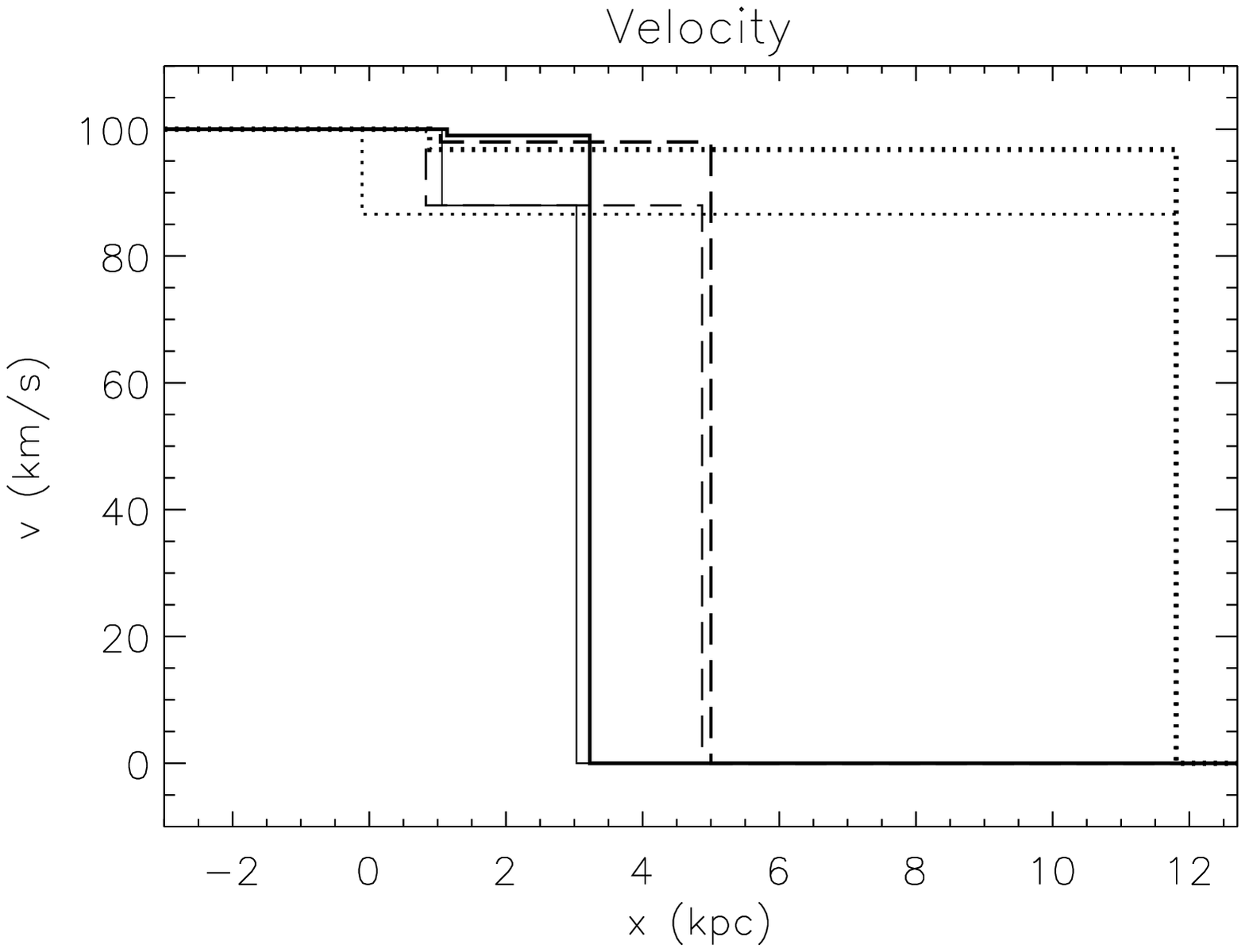} \\
\plottwo{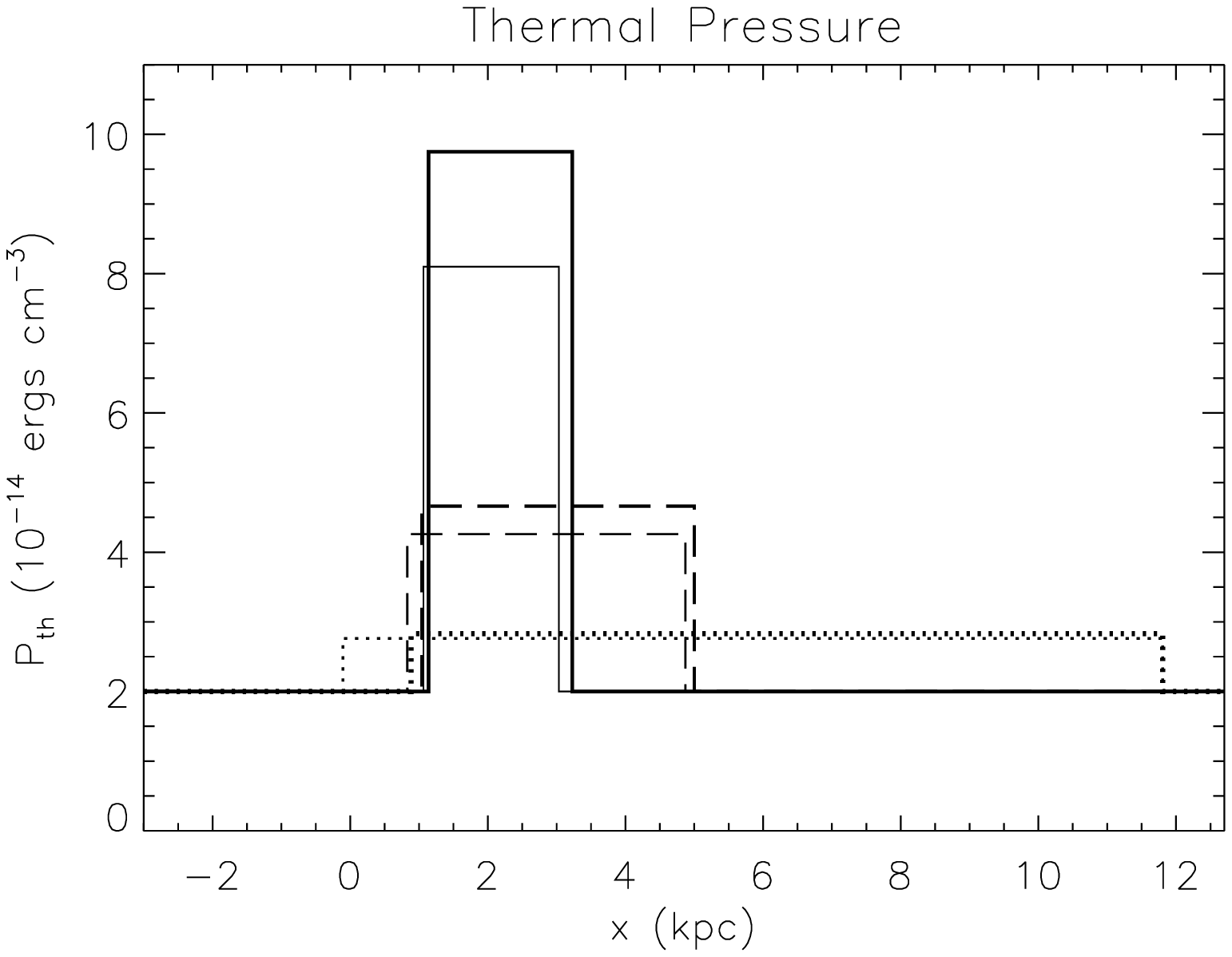}{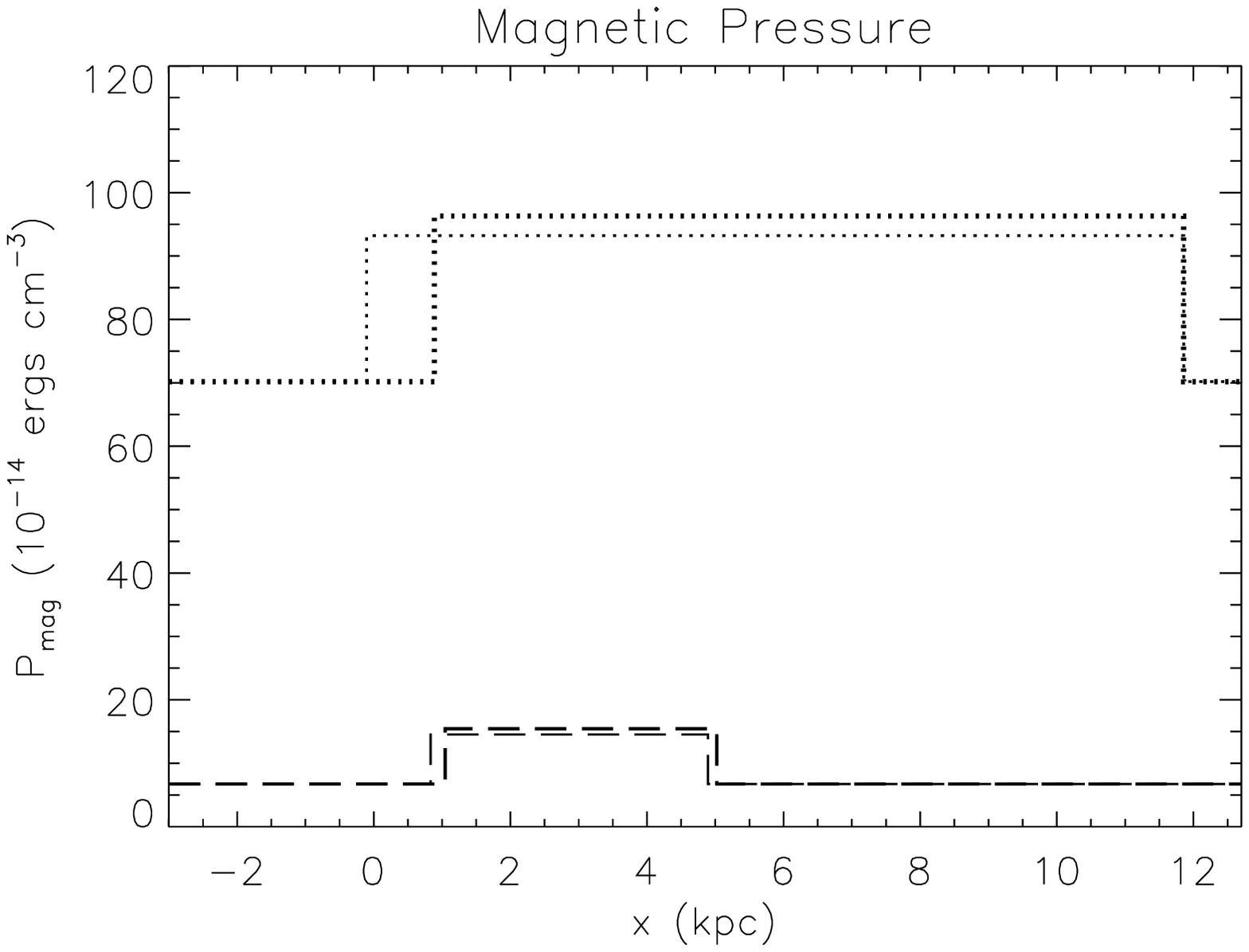}
\caption{One-dimensional shock tube profiles corresponding to the lower interface 
between the cloud and the ambient ISM in our simulations. The initial 
contact discontinuity is located at $x=0$ kpc and the plots are made 
at $t=18$ Myr. Different lines correspond to shock tube 
models that have the same densities and magnetic field strengths as 
we use in our HVC simulations; 
thick solid line (Model A2), thick dashed line (Model B2), 
thick dotted line (Model B4), thin solid line (Model A1), 
thin dashed line (Model B1), and thin dotted line (Model B3). 
In the hydrogen number density profile (top left panel), 
the reverse shock, contact discontinuity, and forward shock appear as 
each sudden change from left, respectively. Note that the contact discontinuity 
is the cloud-ISM interface and its locations in different models 
do not vary much. The velocity, thermal pressure and magnetic pressure 
profiles are shown in the top right, bottom left, and bottom right panels, 
respectively. Note that these physical variables do not change 
across the contact discontinuity. 
\label{shock_tube} }
\end{figure}

In the lower cloud--ISM interface, 
a shock wave propagates into the cloud and 
into the ISM, respectively. This situation is similar to one case 
of the one-dimensional shock tube problem \citep{leveque2002}, 
in which a high-density region moves toward a low-density region 
of the same pressure. 
Because the cloud's dynamics are affected by the physical conditions 
in the lower cloud--ISM interface, it is helpful to examine 
the shock tube problem relevant to this case. 
Figure \ref{shock_tube} shows the density, 
velocity, thermal pressure, and magnetic field profiles of 
the one-dimensional shock tube problem obtained numerically for 
the same physical parameters as in our cloud simulations. 
The cloud's density increases and the thermal and 
magnetic pressure jumps across the cloud-ISM interface. 
An important characteristic of 
our shock tube profiles is that when the magnetic field lines 
are perpendicular to the fluid's motion the jump in the magnetic 
pressure is much larger than 
that in the thermal pressure and it increases 
with the initial magnetic field's strength. 
For example, when the initial magnetic strength is $1.3~\mu\mbox{G}$, 
the thermal pressure increases by about 
$2.4\times10^{-14}~\mbox{erg}~\mbox{cm}^{-3}$ while the magnetic pressure 
increases by about $8.3\times10^{-14}~\mbox{erg}~\mbox{cm}^{-3}$. 
When the initial magnetic field is stronger ($4.2~\mu\mbox{G}$), 
the jump in the thermal pressure is even smaller 
($\sim 0.8\times10^{-14}~\mbox{erg}~\mbox{cm}^{-3}$) while the 
jump in the magnetic pressure is even larger 
($\sim 2.3\times10^{-13}~\mbox{erg}~\mbox{cm}^{-3}$). 

\begin{figure}
\centering
\includegraphics[scale=0.25]{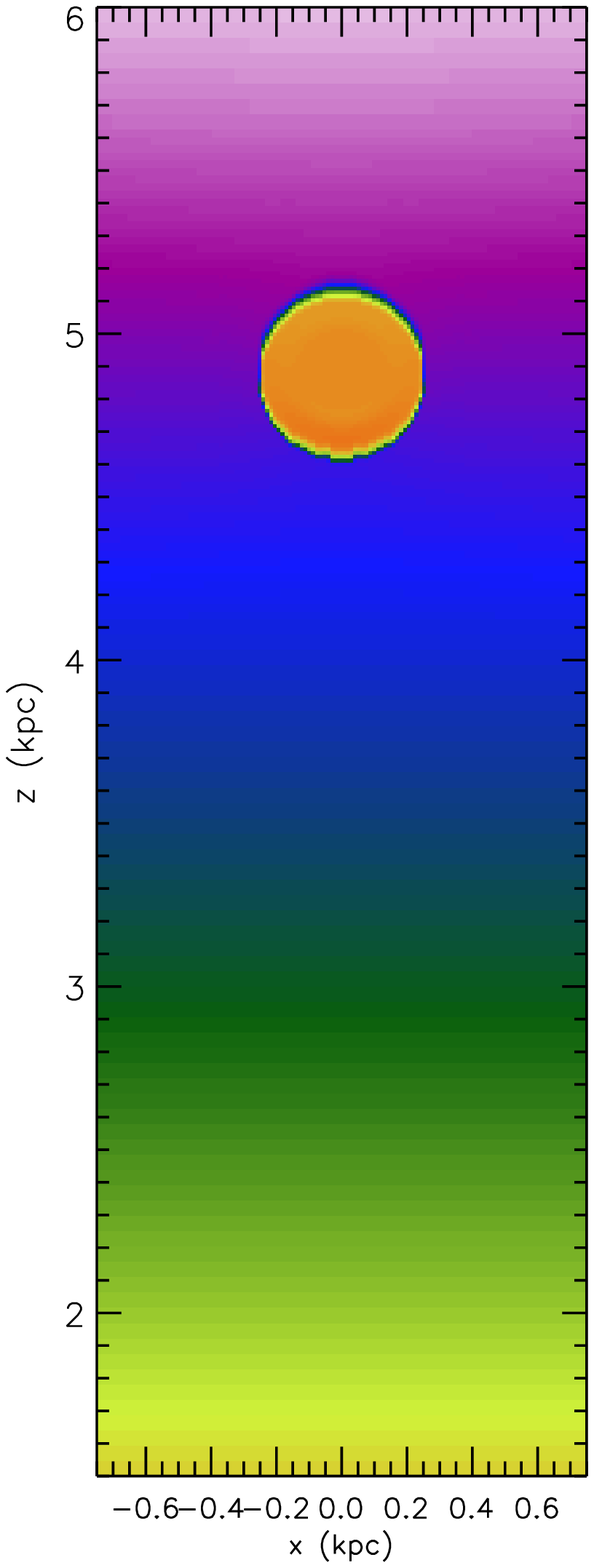}
\includegraphics[scale=0.25]{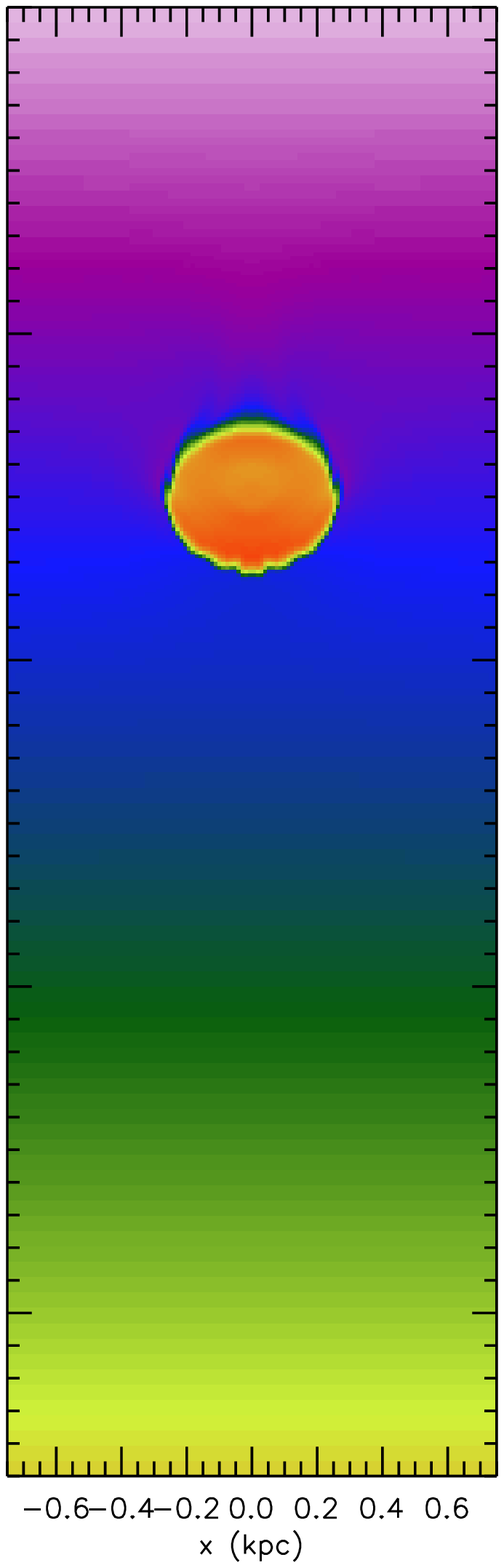}
\includegraphics[scale=0.25]{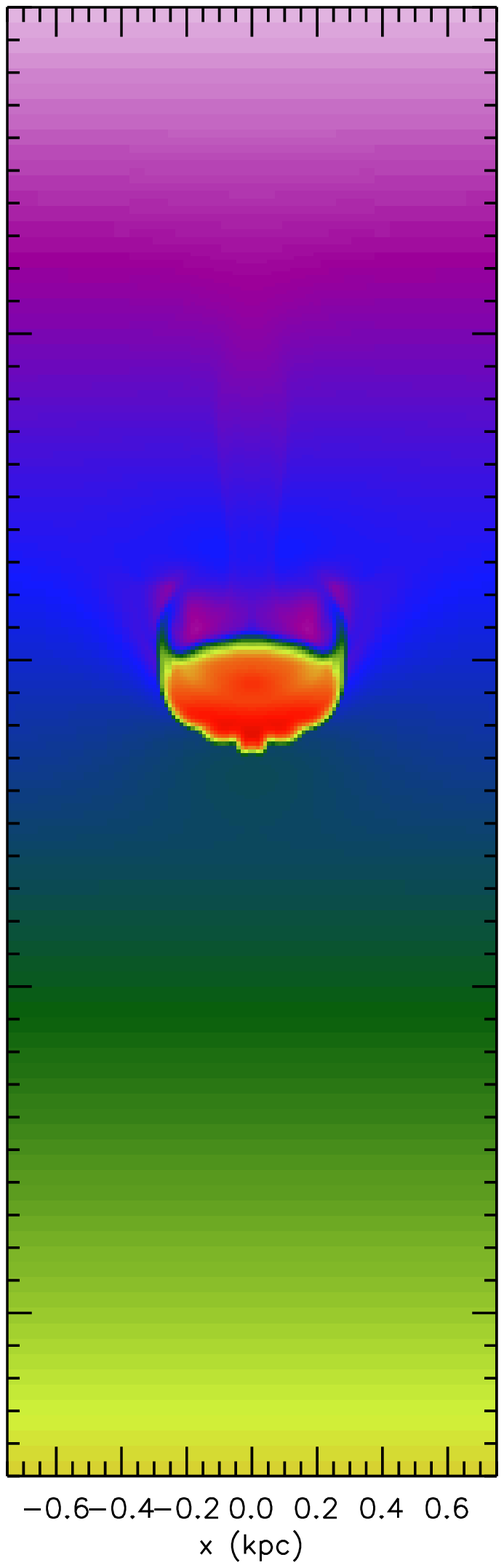}
\includegraphics[scale=0.25]{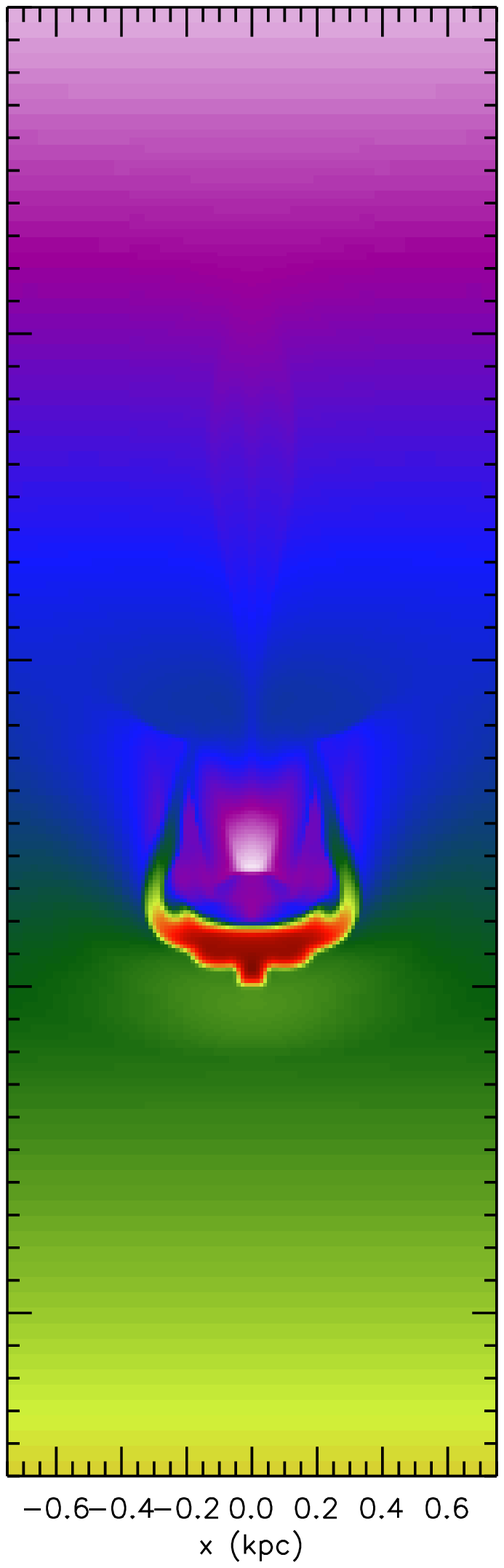}
\includegraphics[scale=0.25]{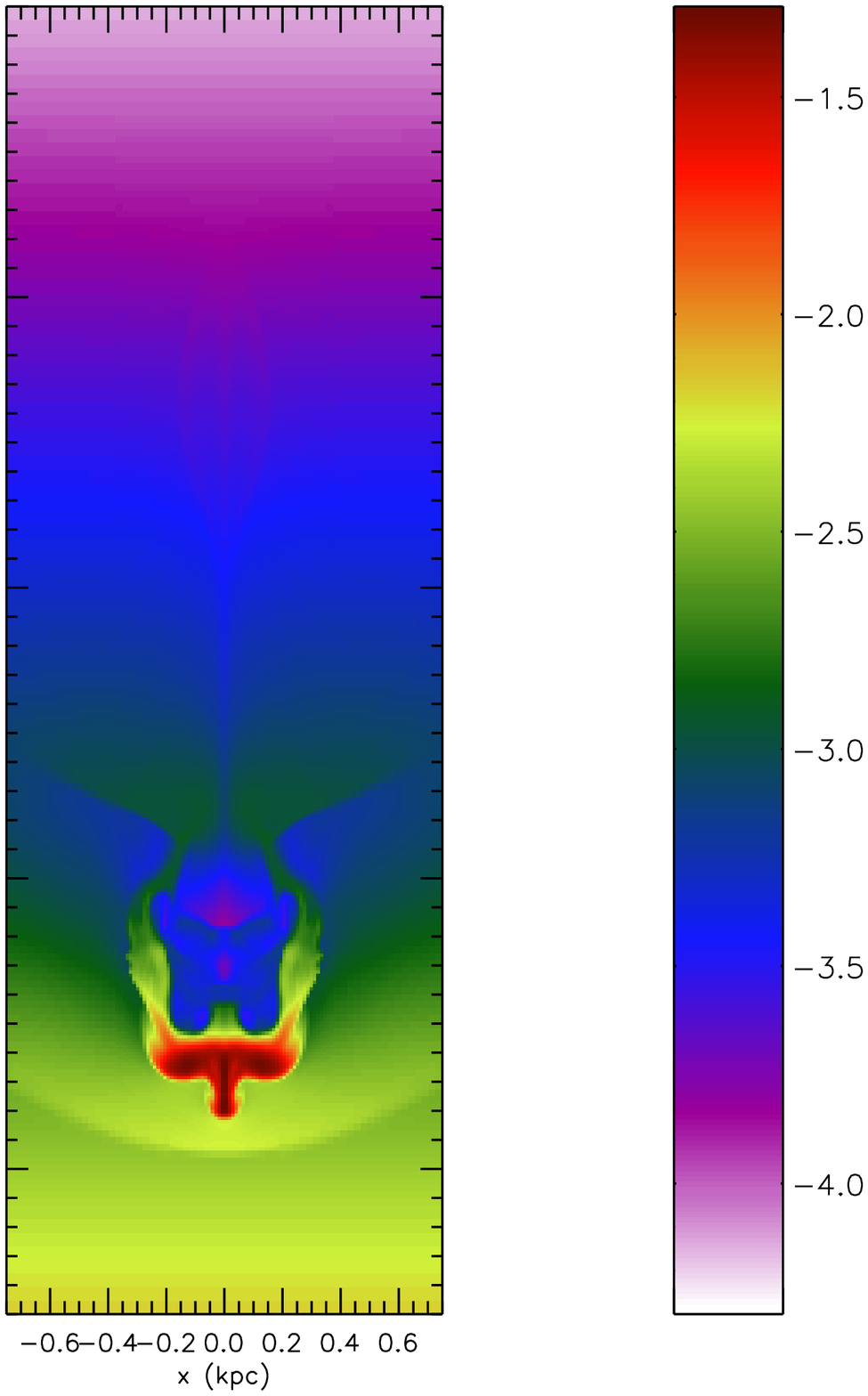}
\includegraphics[scale=0.25]{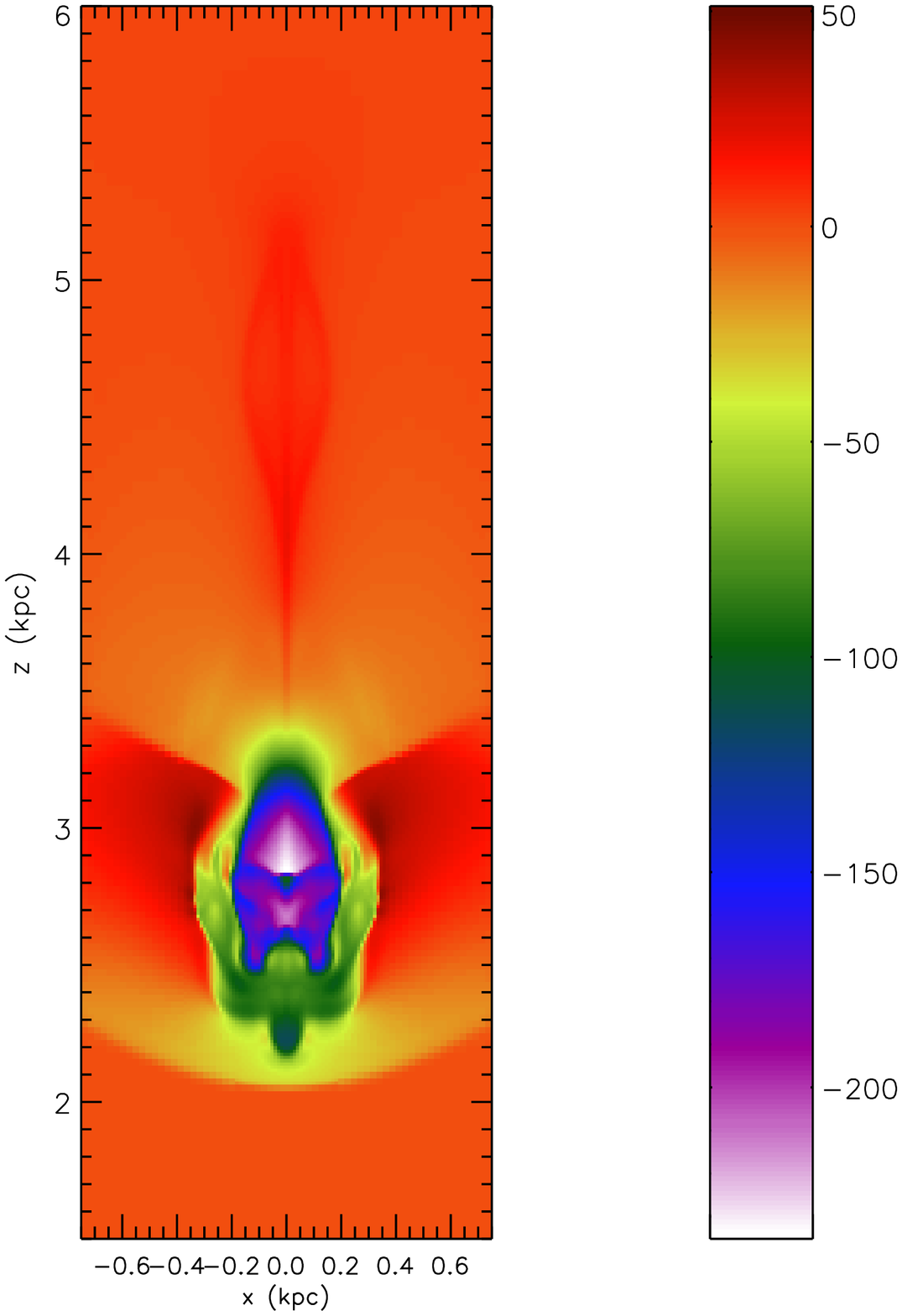}
\caption{Model A1 ($n_{cloud}=0.01 ~\mbox{H atoms} ~ \mbox{cm}^{-3}$, 
$\vert \mbox{{\boldmath$B$}} \vert = 0.0 ~ \mu \mbox{G}$). 
The leftmost five plots show the $\mbox{log}_{10}$ of 
the hydrogen number density in units of atoms $\mbox{cm}^{-3}$ for 
a cut along the $y=0$ plane at $t=8,~16,~24,~32$, and $40~\mbox{Myr}$. The right-most 
plot shows $v_z$ in units of $\mbox{km}~\mbox{s}^{-1}$ 
for the material on the $y=0$ plane at 
$t=40~\mbox{Myr}$. 
\label{modelA1_P}}
\end{figure}

\begin{figure}
\centering
\includegraphics[scale=0.25]{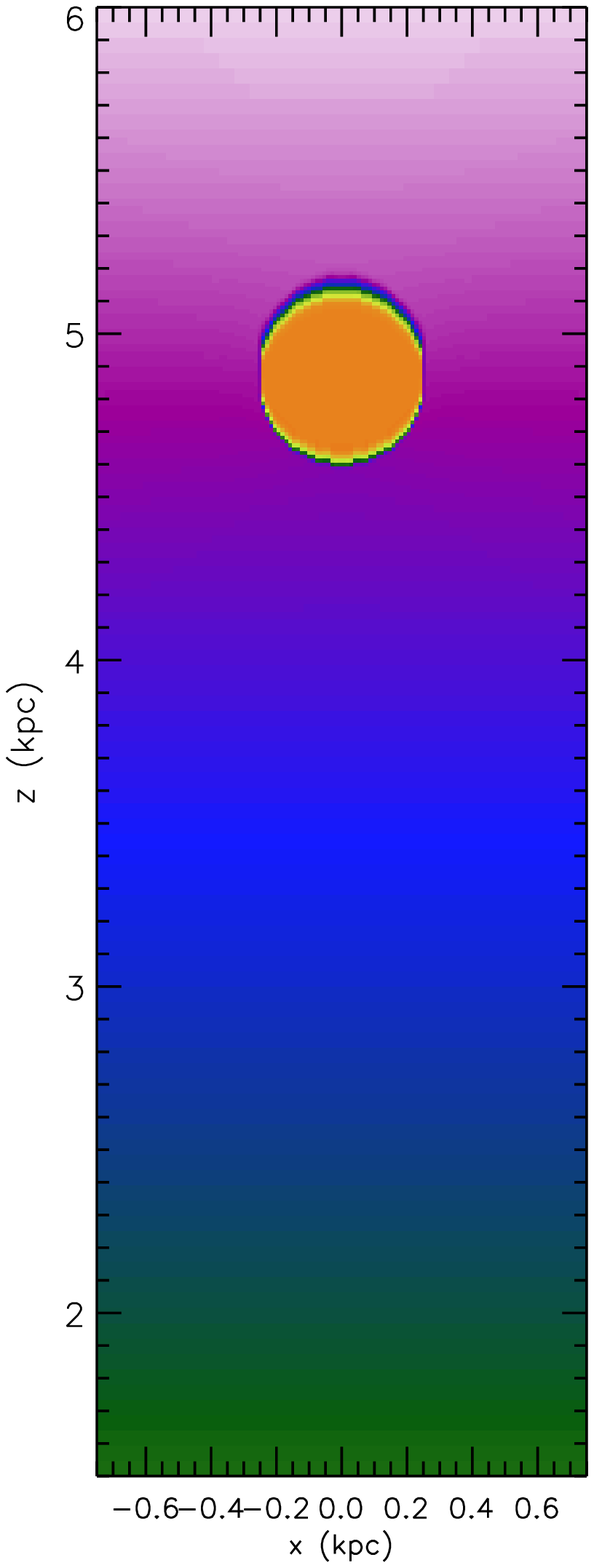}
\includegraphics[scale=0.25]{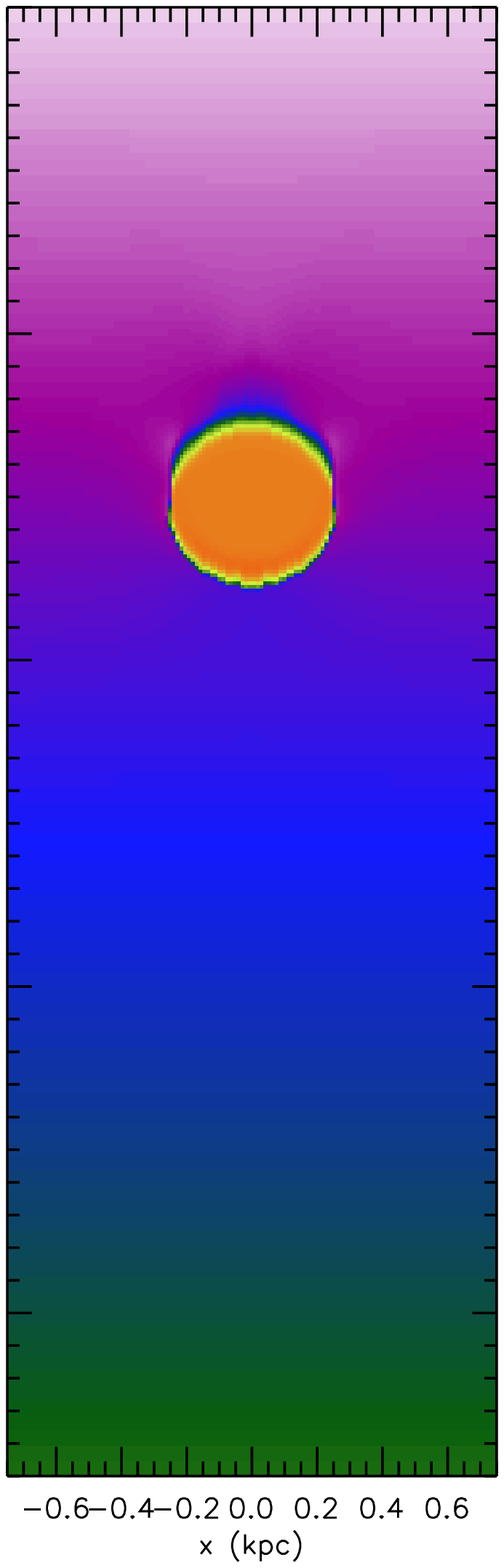}
\includegraphics[scale=0.25]{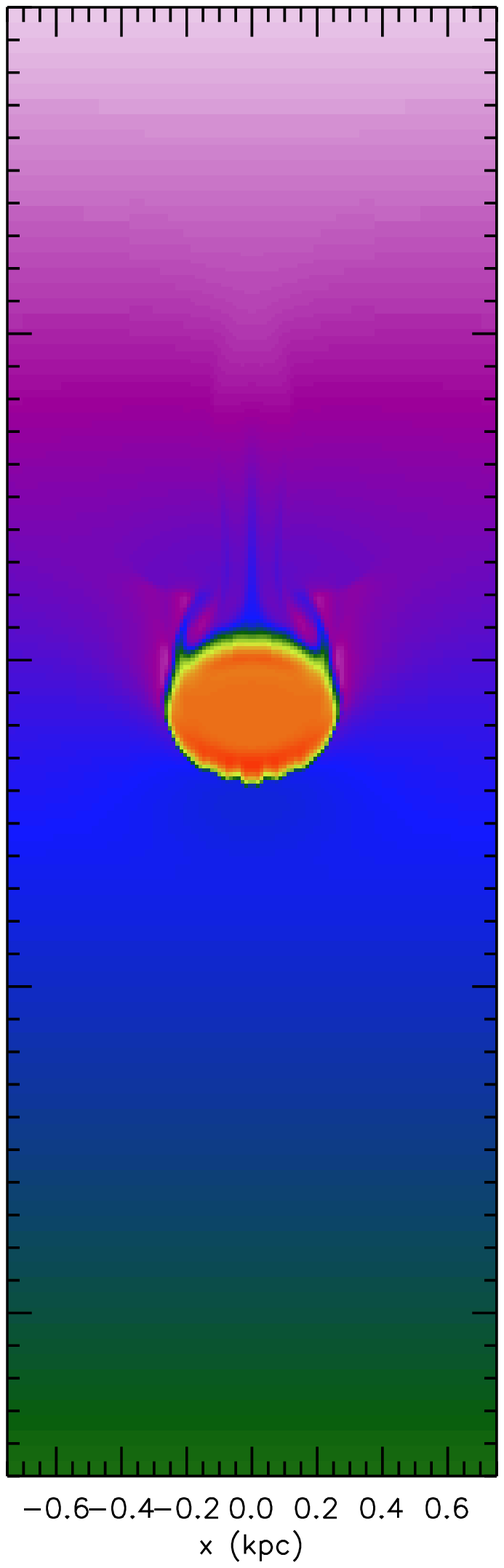}
\includegraphics[scale=0.25]{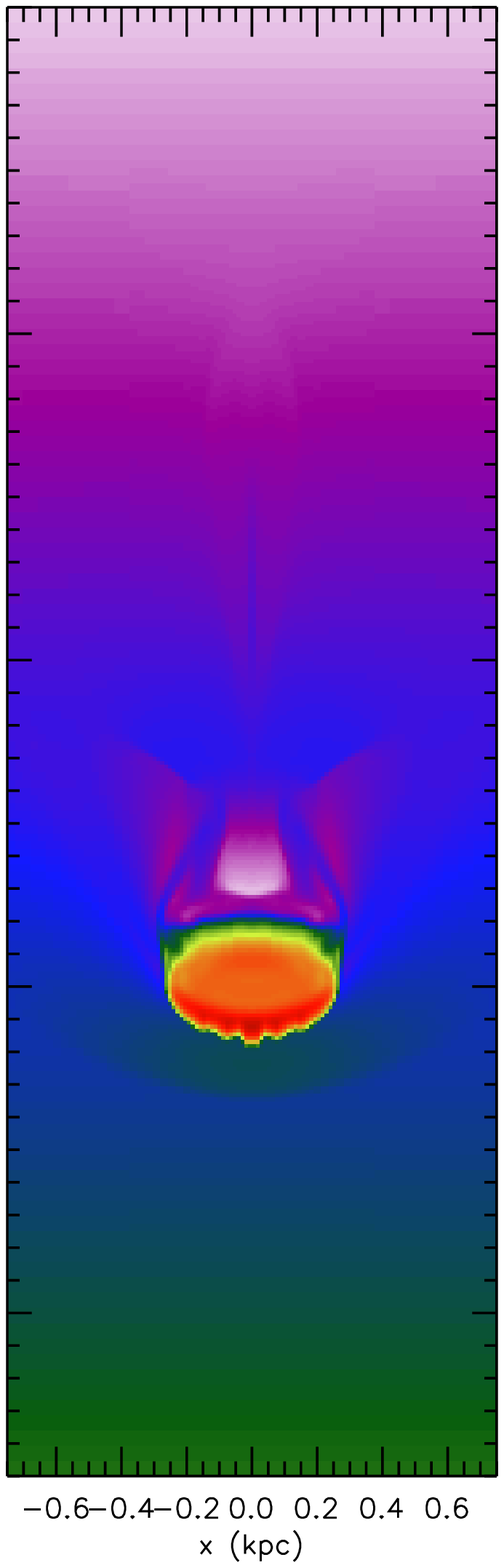}
\includegraphics[scale=0.25]{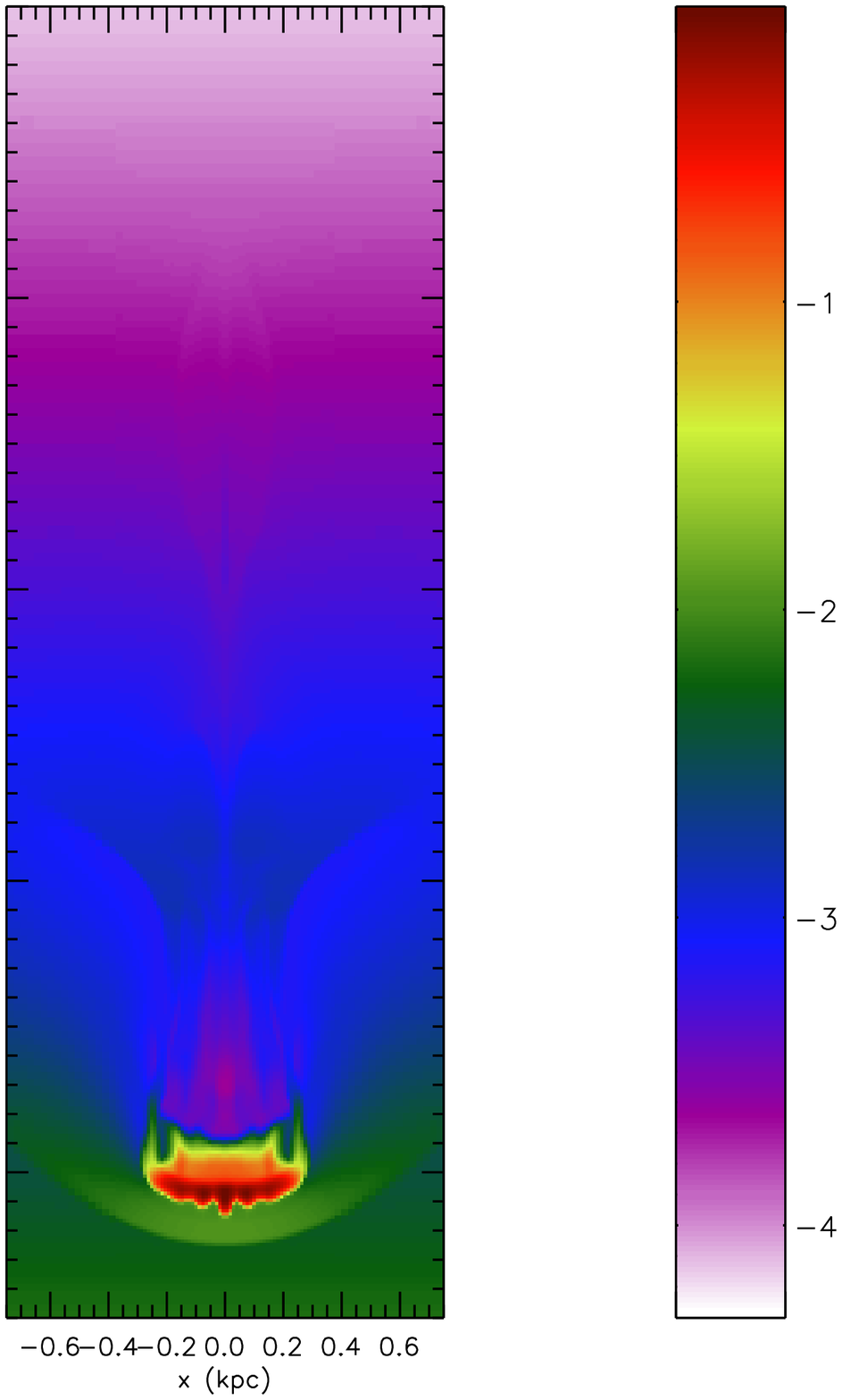}
\includegraphics[scale=0.25]{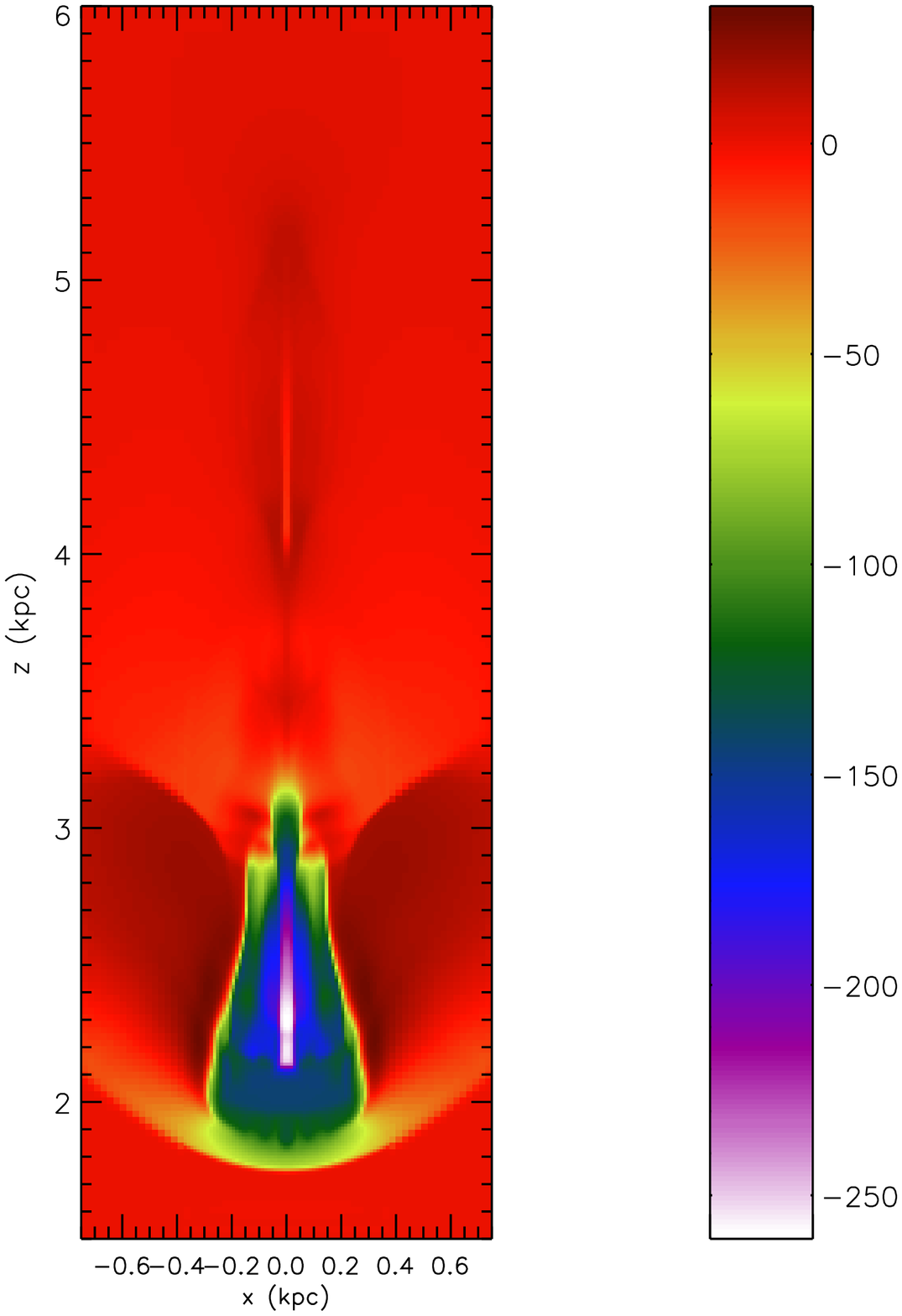}
\caption{Same as Figure \ref{modelA1_P}, but for Model A2 
($n_{cloud}=0.1 ~\mbox{H atoms} ~ \mbox{cm}^{-3}$, 
$\vert \mbox{{\boldmath$B$}} \vert = 0.0 ~ \mu \mbox{G}$). 
\label{modelA2_P}}
\end{figure}

\begin{figure}
\centering
\includegraphics[scale=0.25]{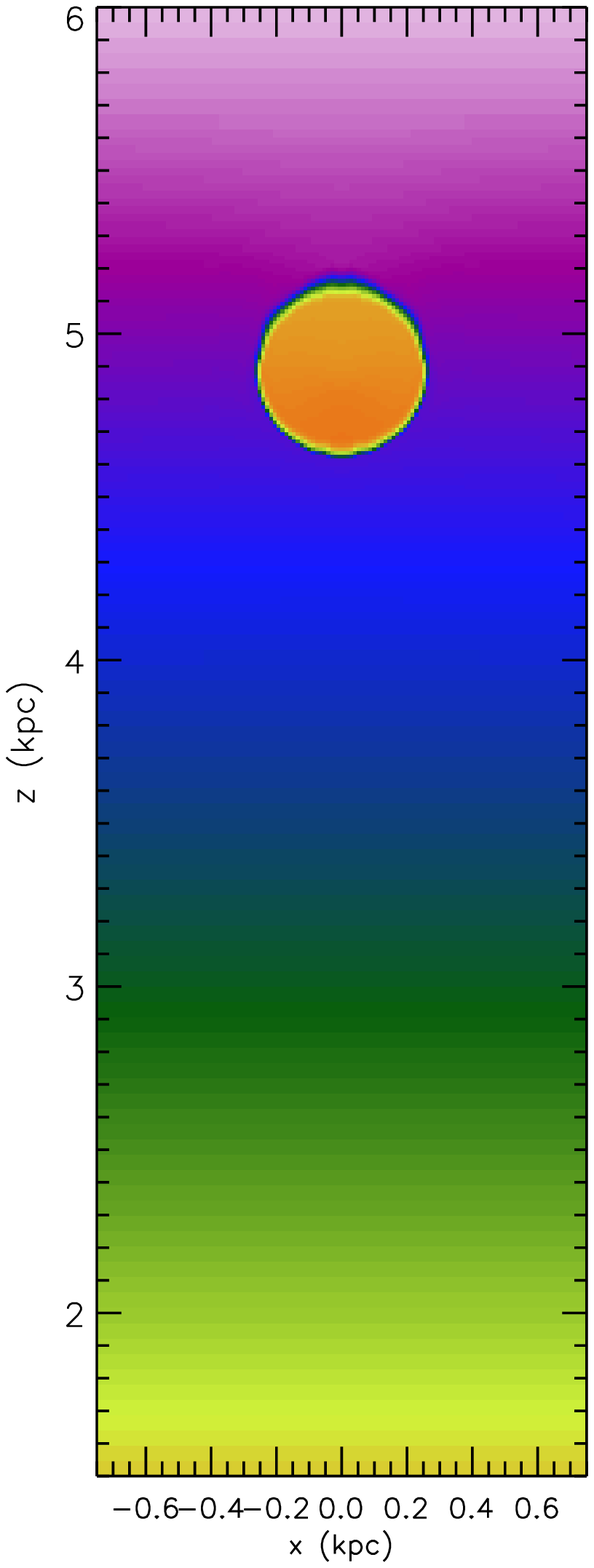}
\includegraphics[scale=0.25]{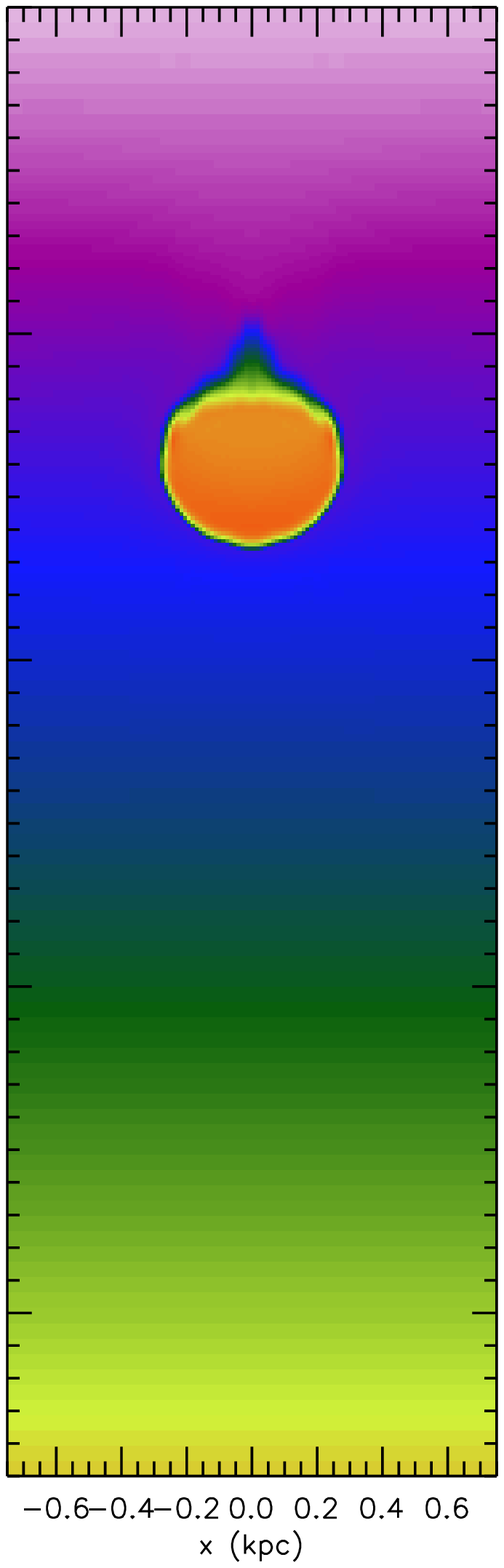}
\includegraphics[scale=0.25]{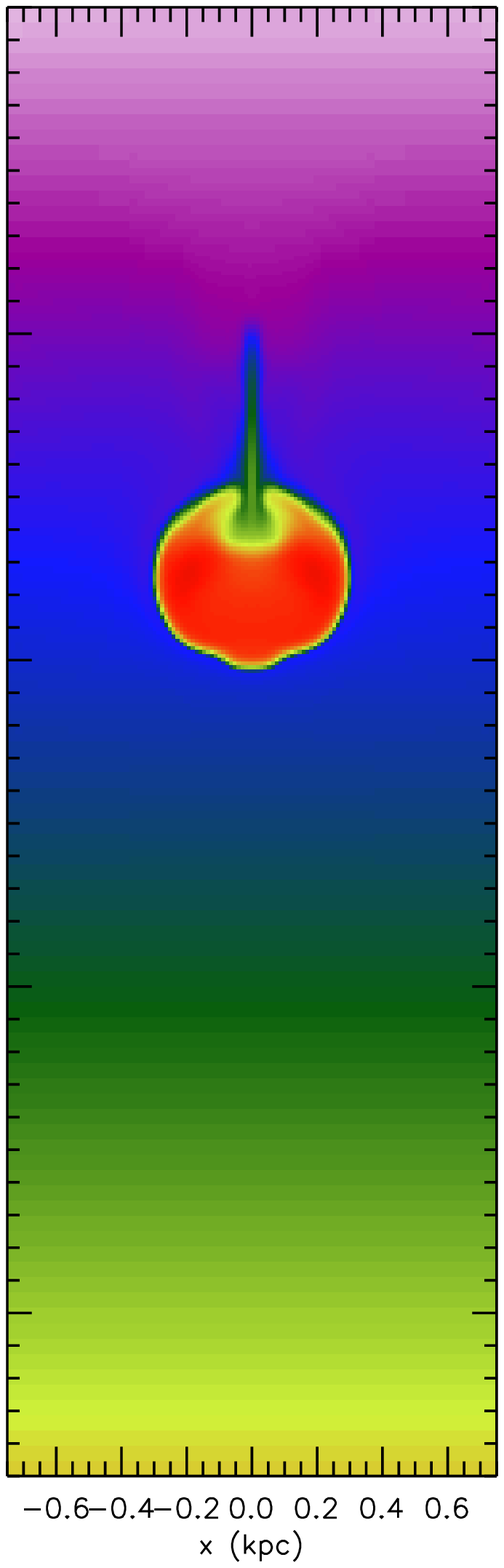}
\includegraphics[scale=0.25]{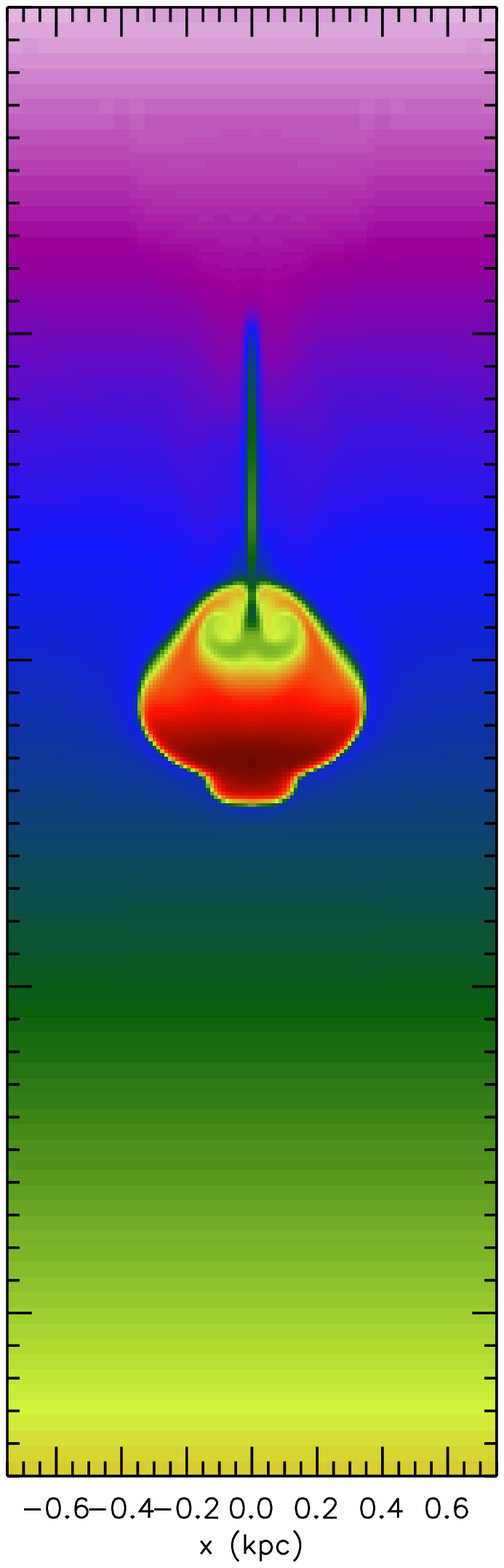}
\includegraphics[scale=0.25]{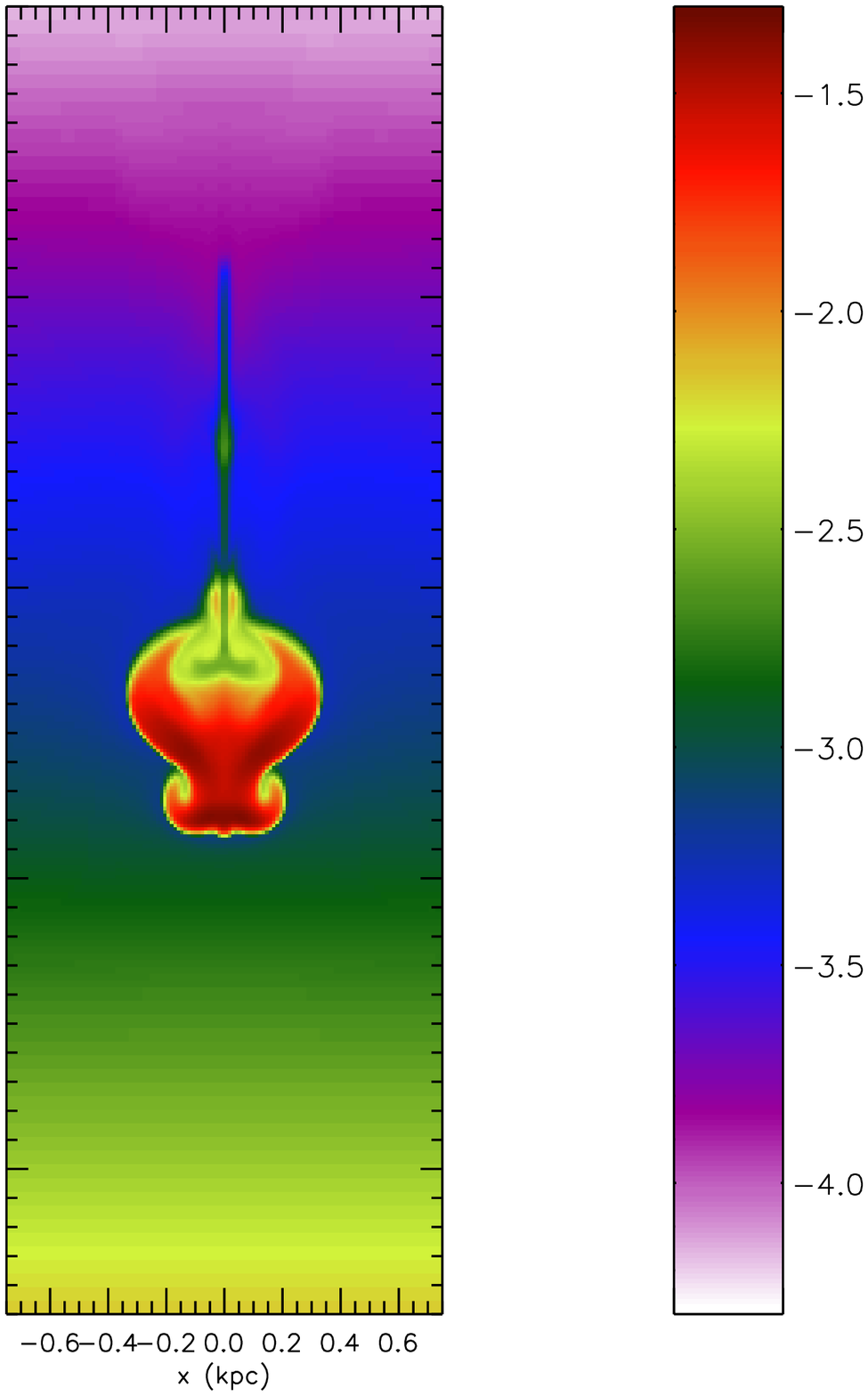}
\includegraphics[scale=0.25]{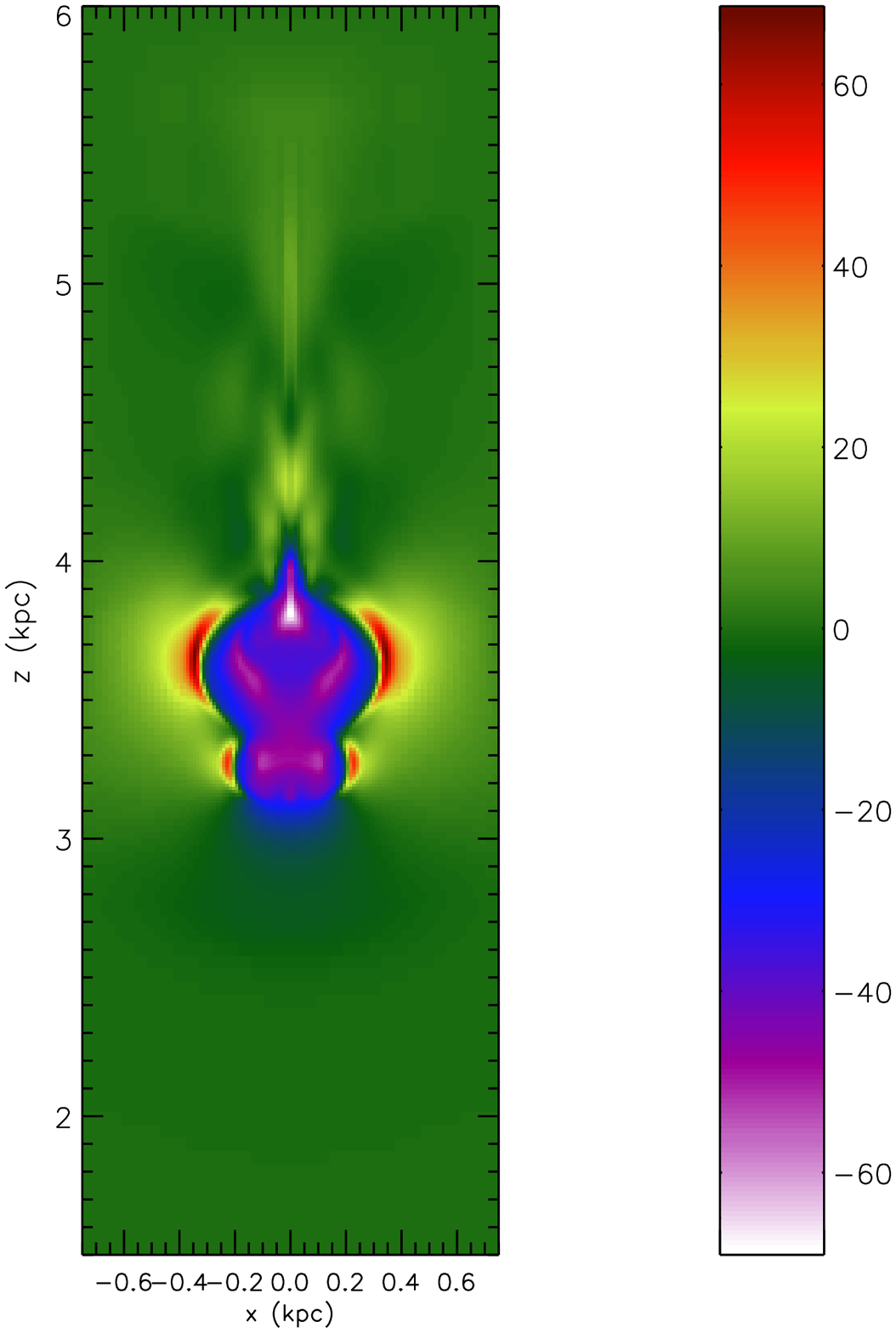}
\caption{Same as Figure \ref{modelA1_P}, but for Model B1 
($n_{cloud}=0.01 ~\mbox{H atoms} ~ \mbox{cm}^{-3}$, 
$B_y = 1.3 ~ \mu \mbox{G}$). 
\label{modelB1_P}}
\end{figure}

\begin{figure}
\centering
\includegraphics[scale=0.25]{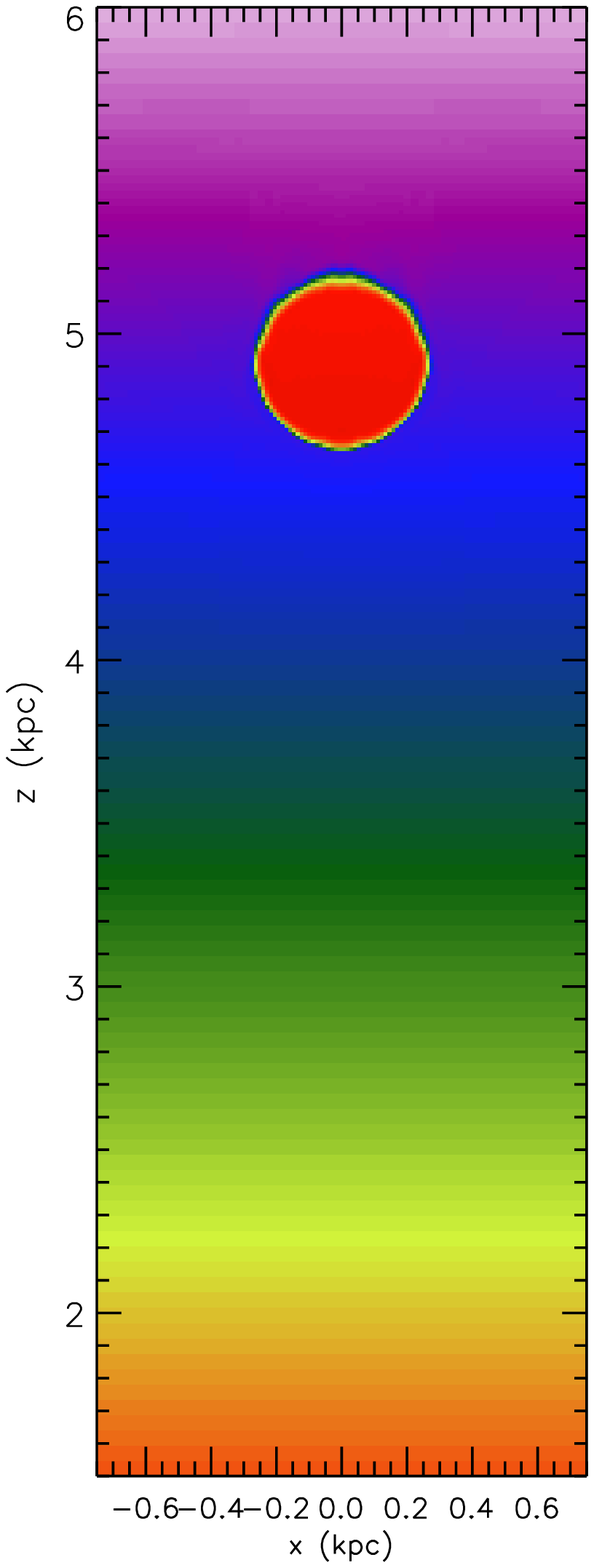}
\includegraphics[scale=0.25]{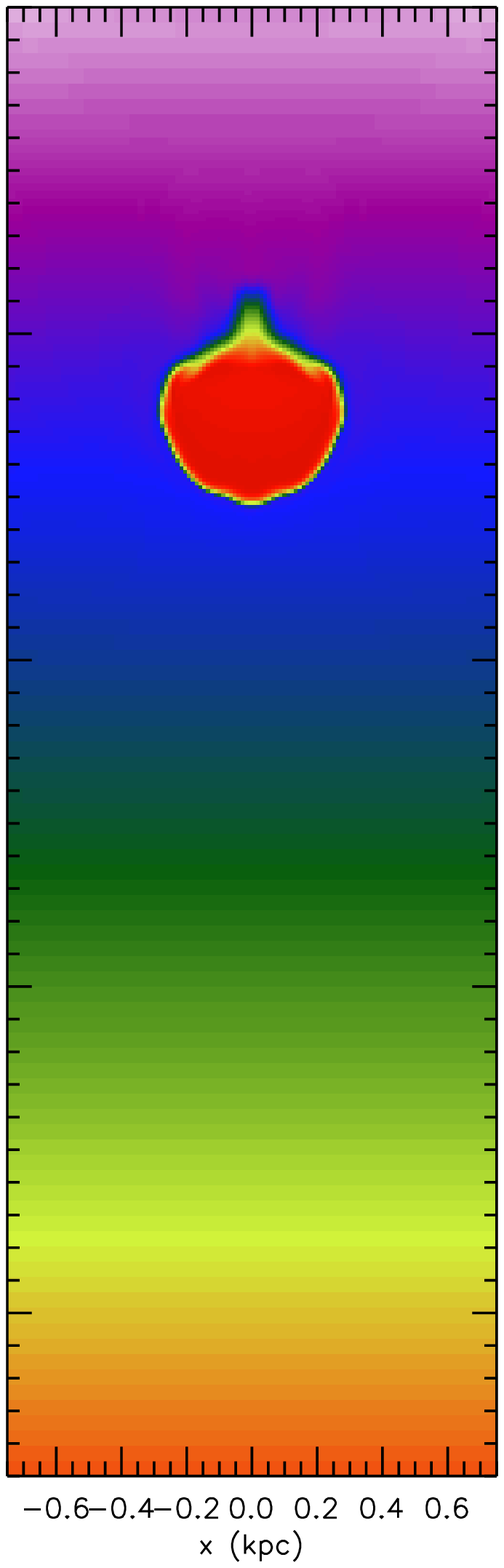}
\includegraphics[scale=0.25]{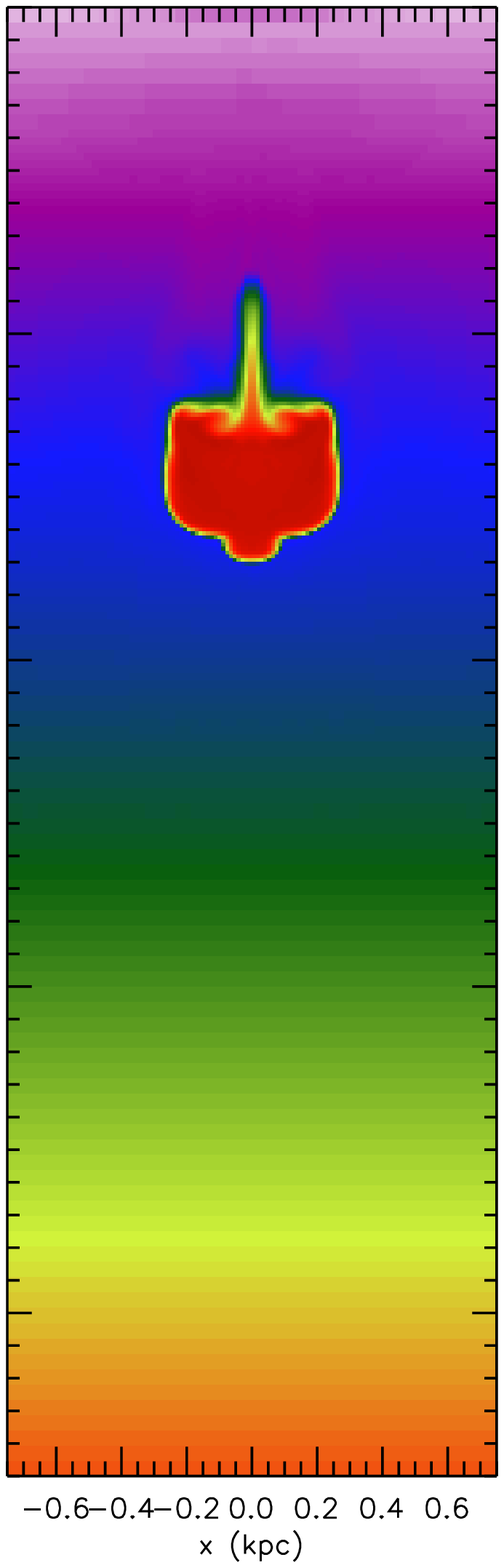}
\includegraphics[scale=0.25]{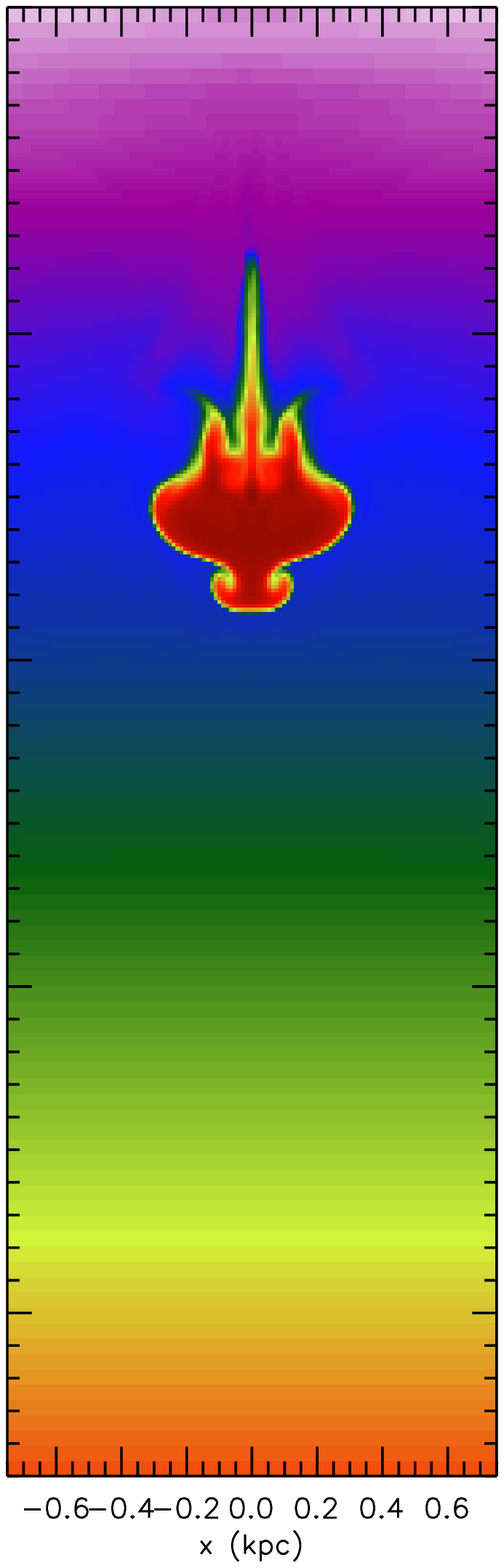}
\includegraphics[scale=0.25]{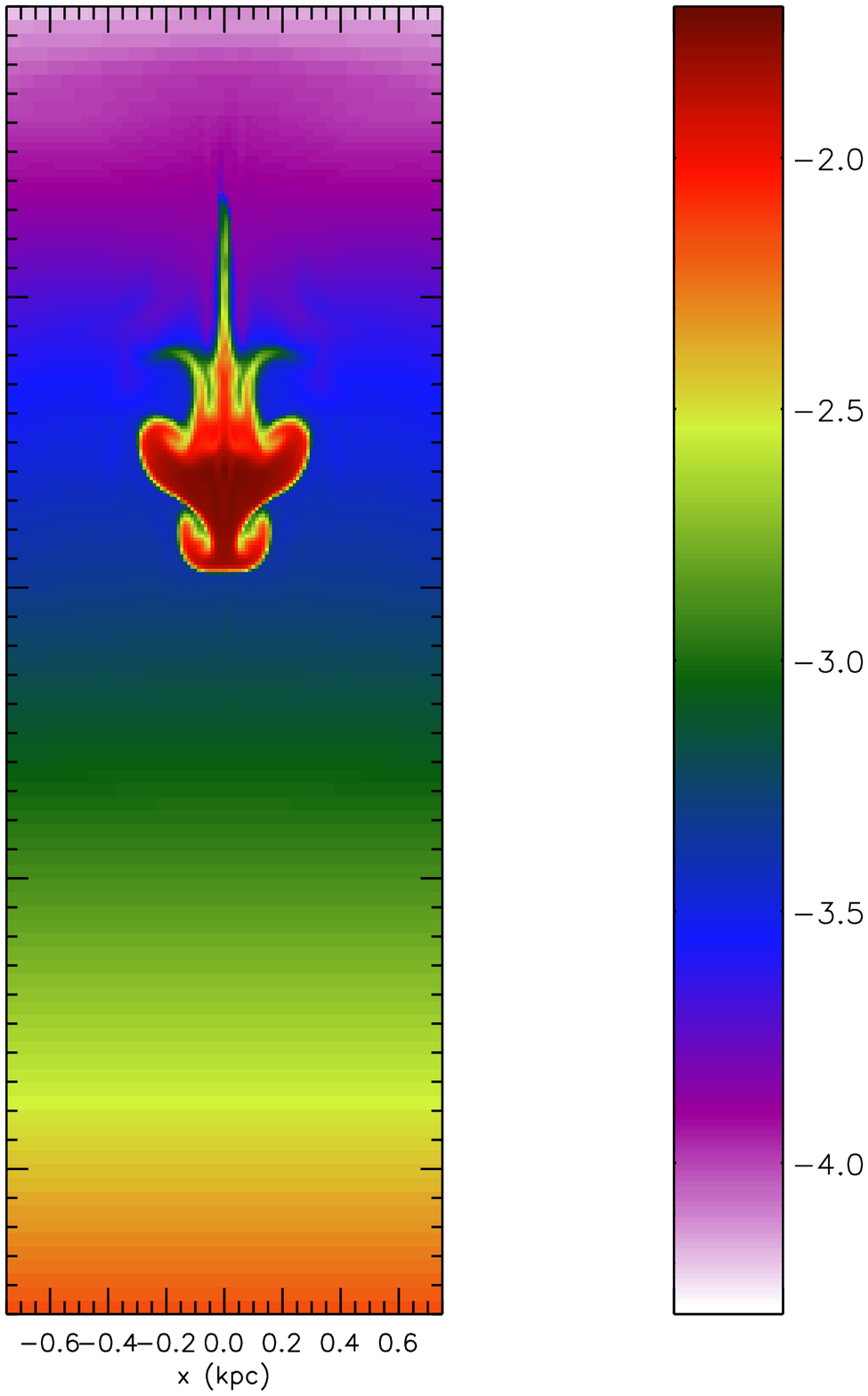}
\includegraphics[scale=0.25]{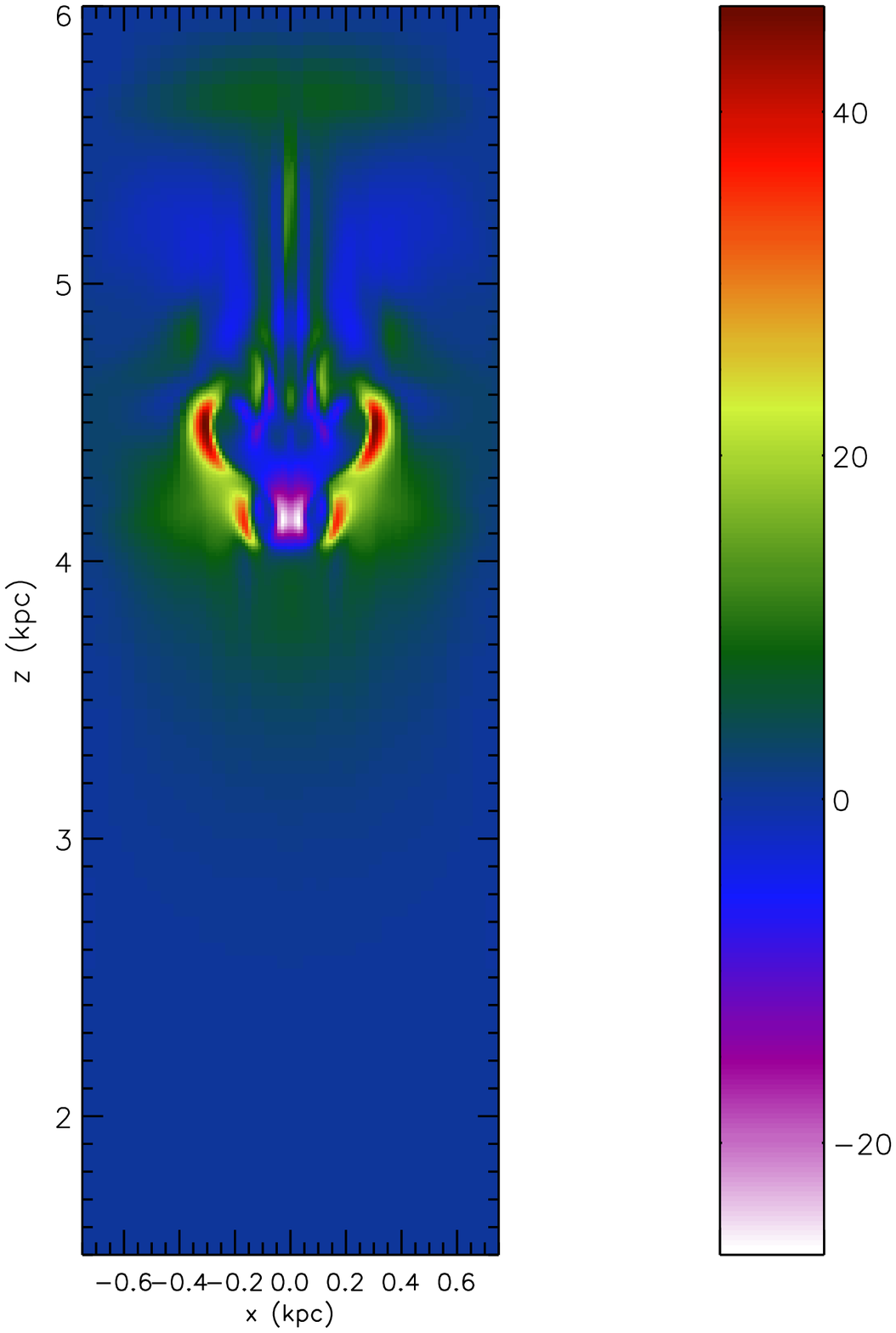}
\caption{Same as Figure \ref{modelA1_P}, but for Model B3 
($n_{cloud}=0.01 ~\mbox{H atoms} ~ \mbox{cm}^{-3}$, 
$B_y = 4.2 ~ \mu \mbox{G}$). 
\label{modelB3_P}}
\end{figure}

\begin{figure}
\centering
\includegraphics[scale=0.25]{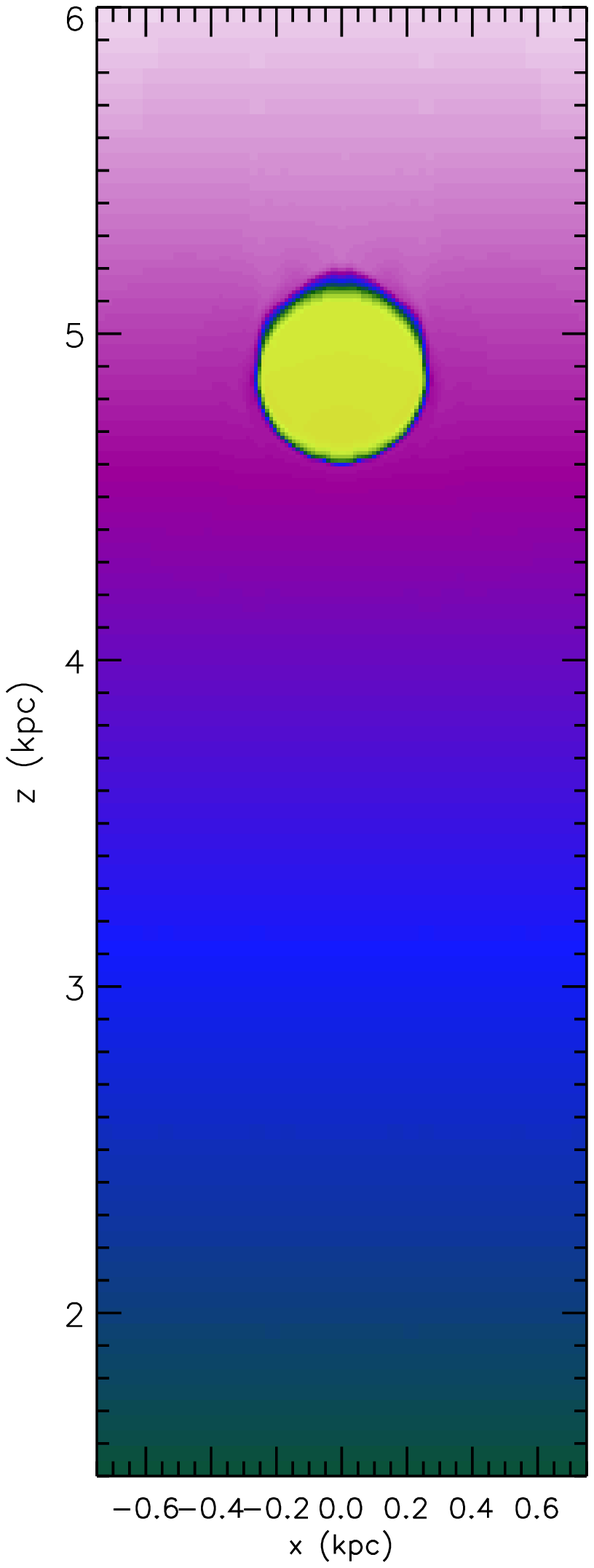}
\includegraphics[scale=0.25]{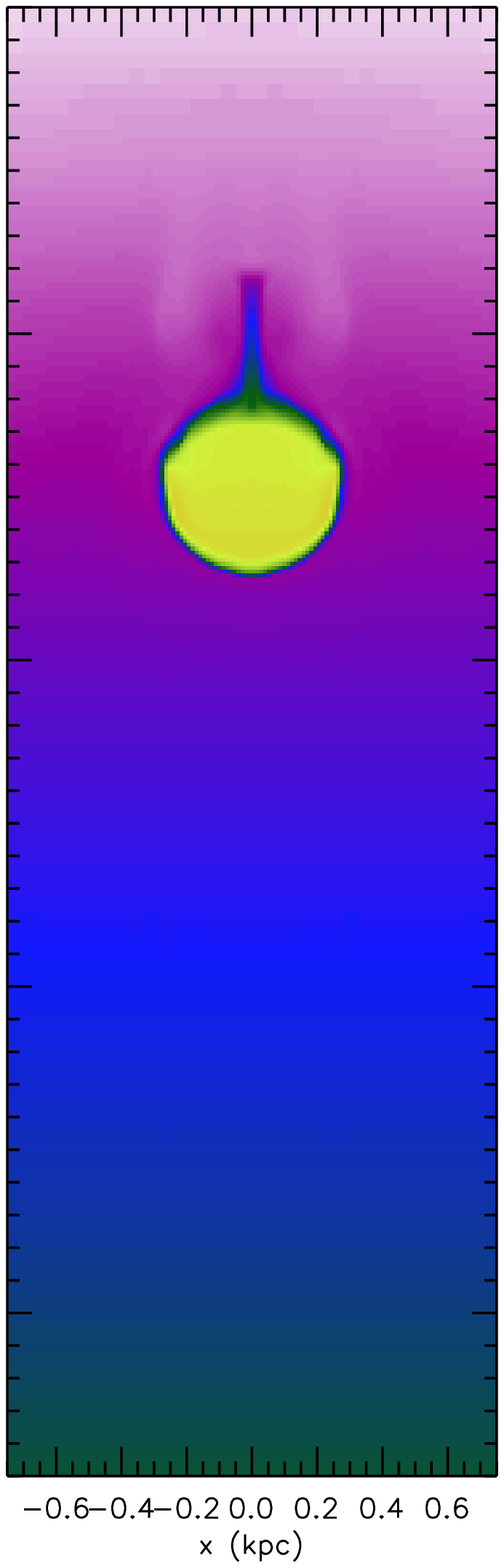}
\includegraphics[scale=0.25]{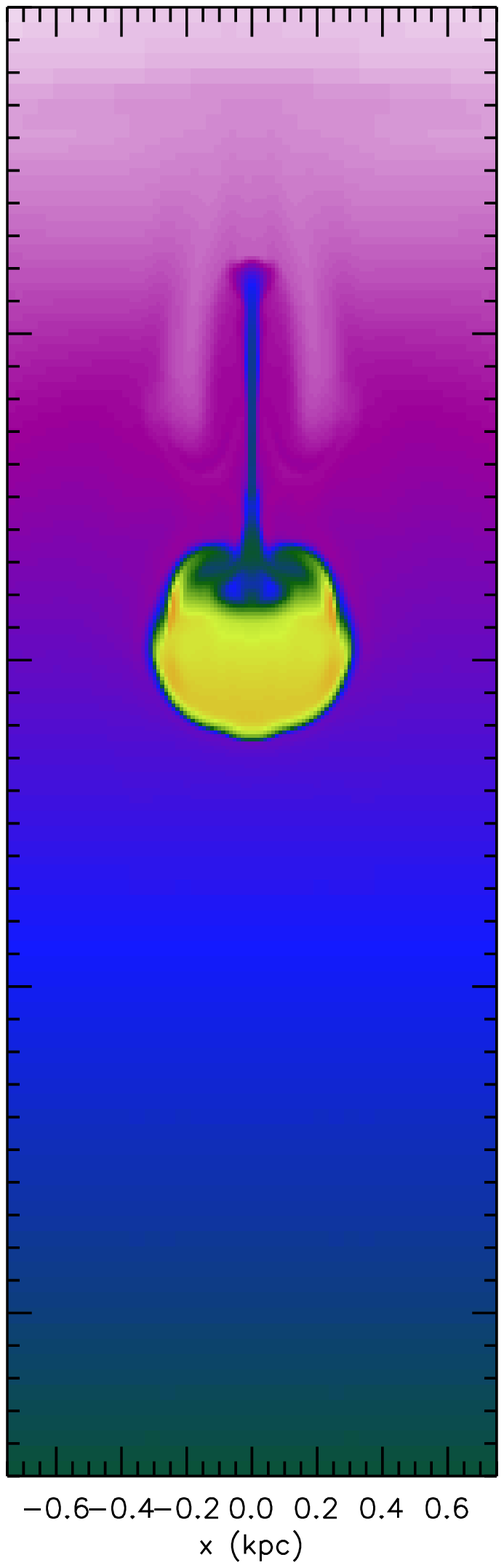}
\includegraphics[scale=0.25]{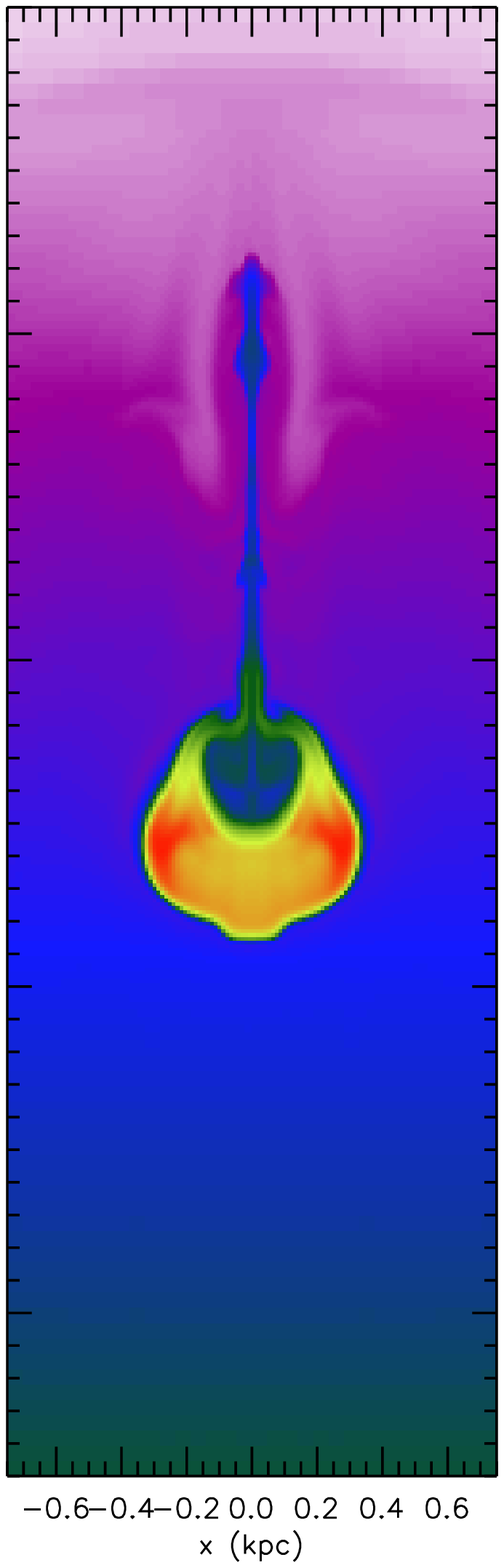}
\includegraphics[scale=0.25]{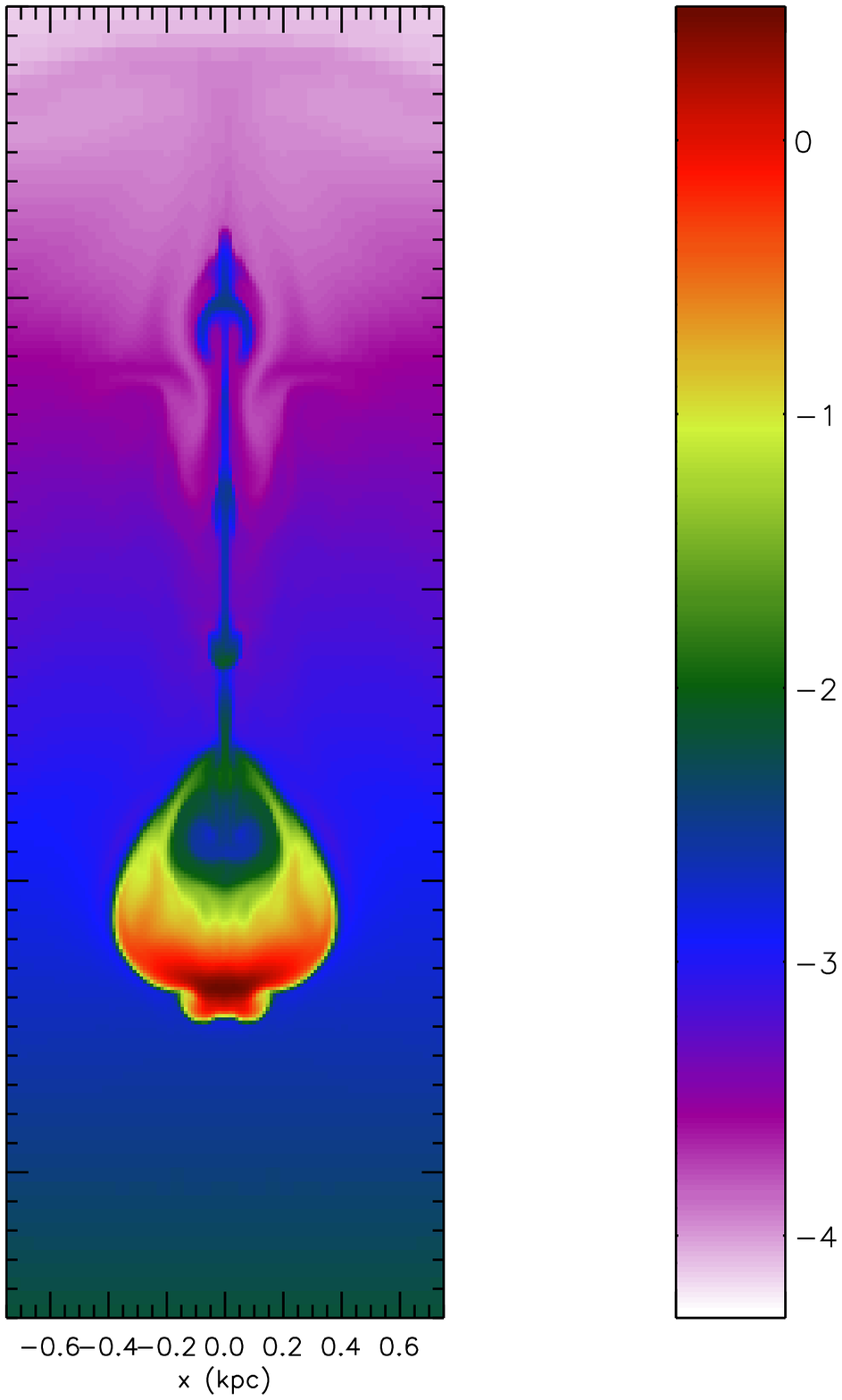}
\includegraphics[scale=0.25]{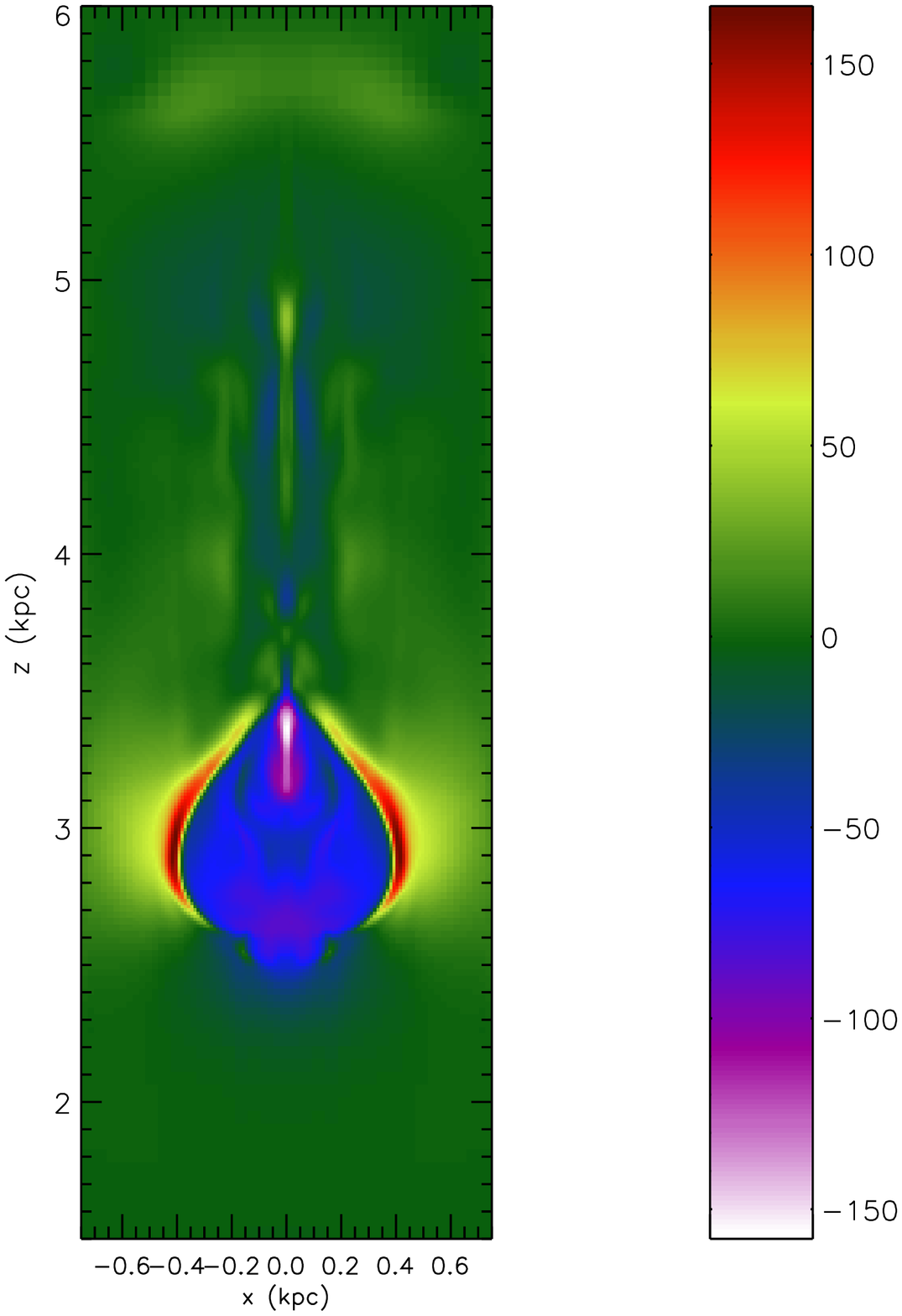}
\caption{Same as Figure \ref{modelA1_P}, but for Model B4 
($n_{cloud}=0.1 ~\mbox{H atoms} ~ \mbox{cm}^{-3}$, 
$B_y = 4.2 ~ \mu \mbox{G}$). 
\label{modelB4_P}}
\end{figure}

\begin{figure}
\centering
\includegraphics[scale=0.25]{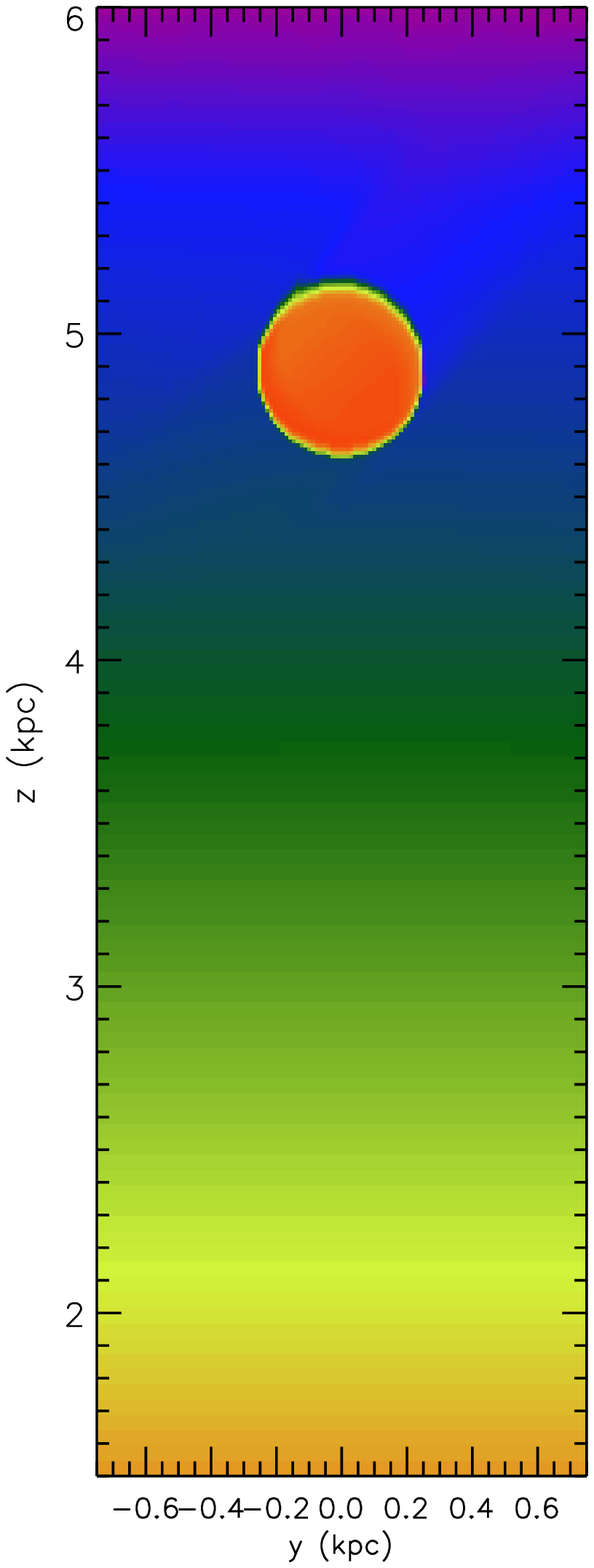}
\includegraphics[scale=0.25]{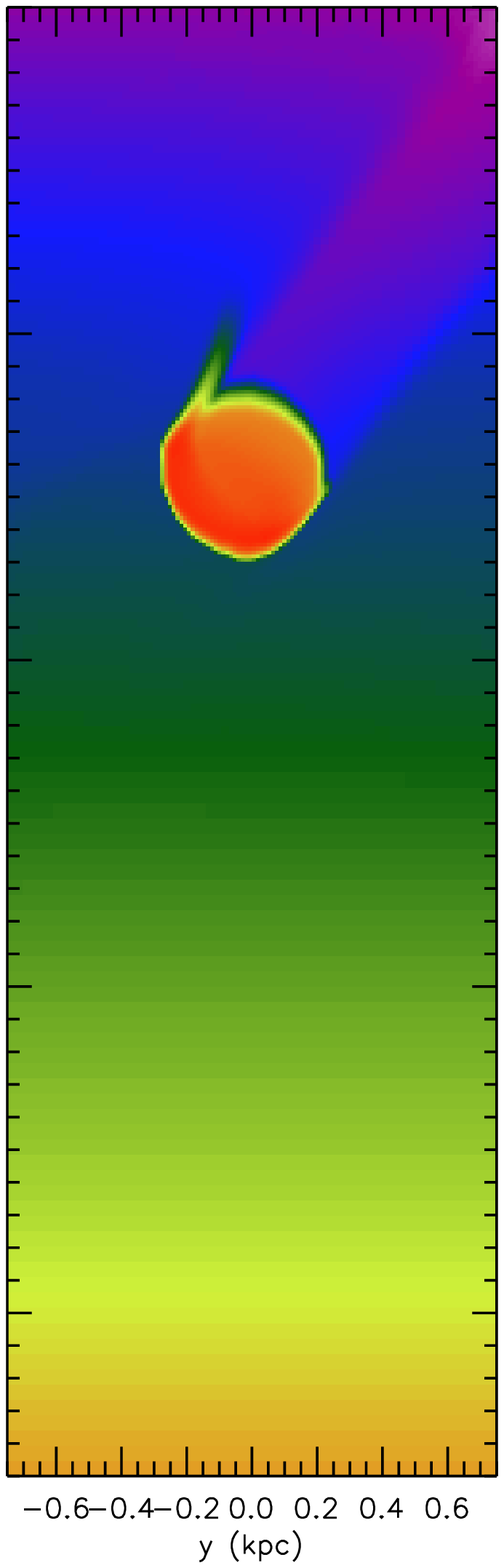}
\includegraphics[scale=0.25]{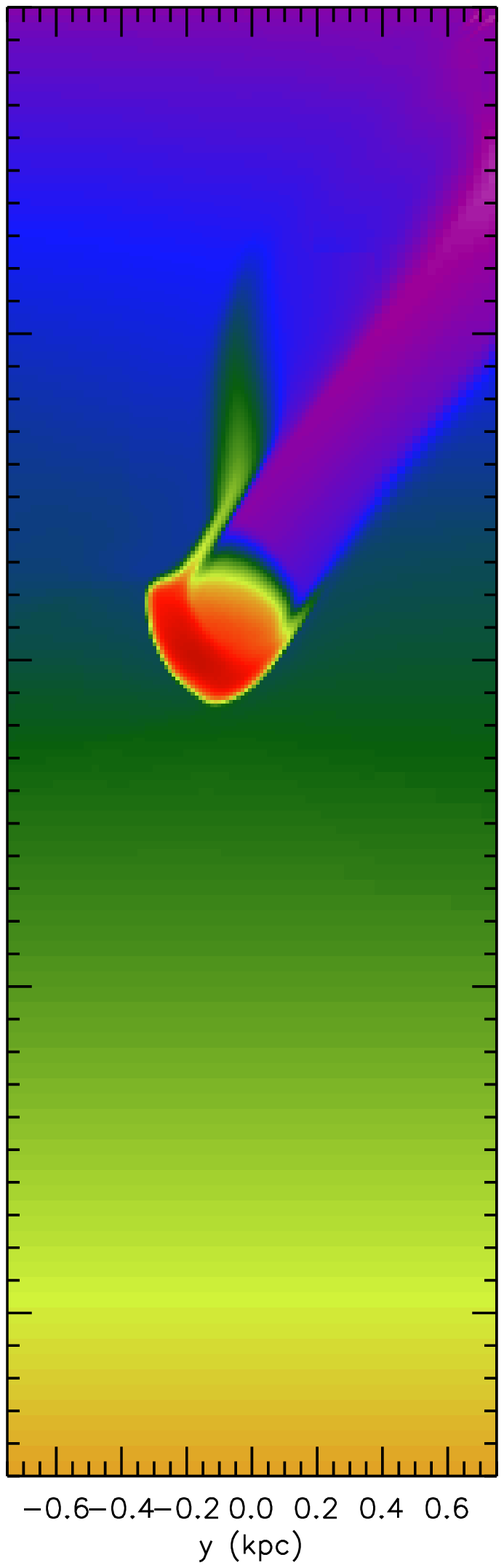}
\includegraphics[scale=0.25]{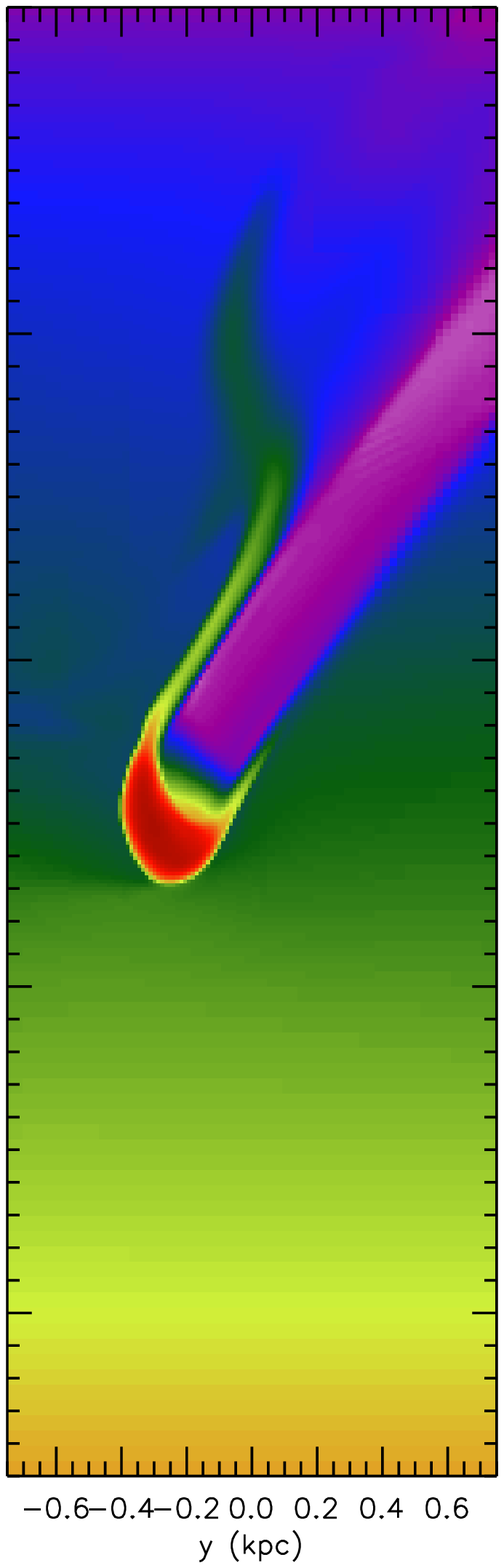}
\includegraphics[scale=0.25]{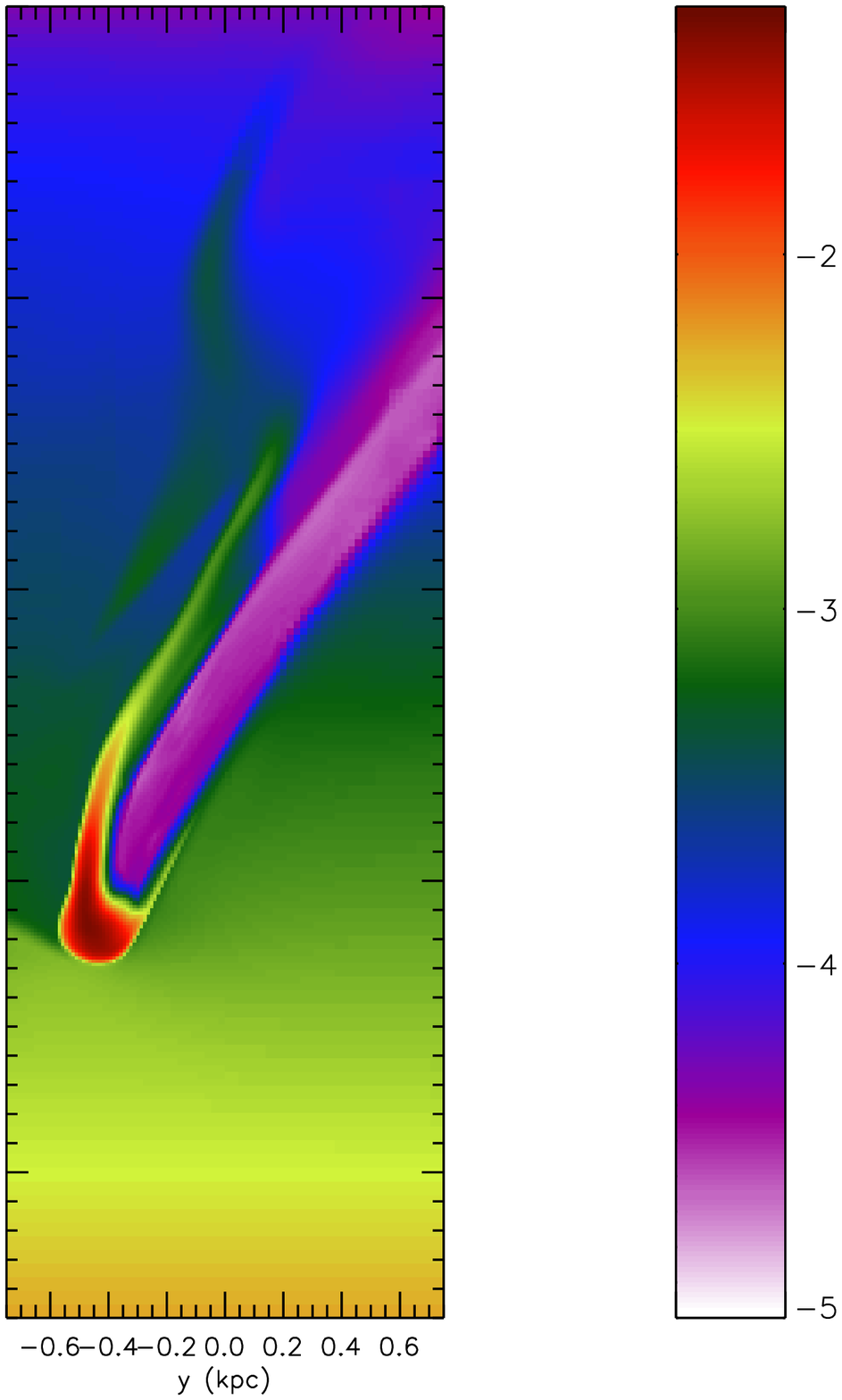}
\includegraphics[scale=0.25]{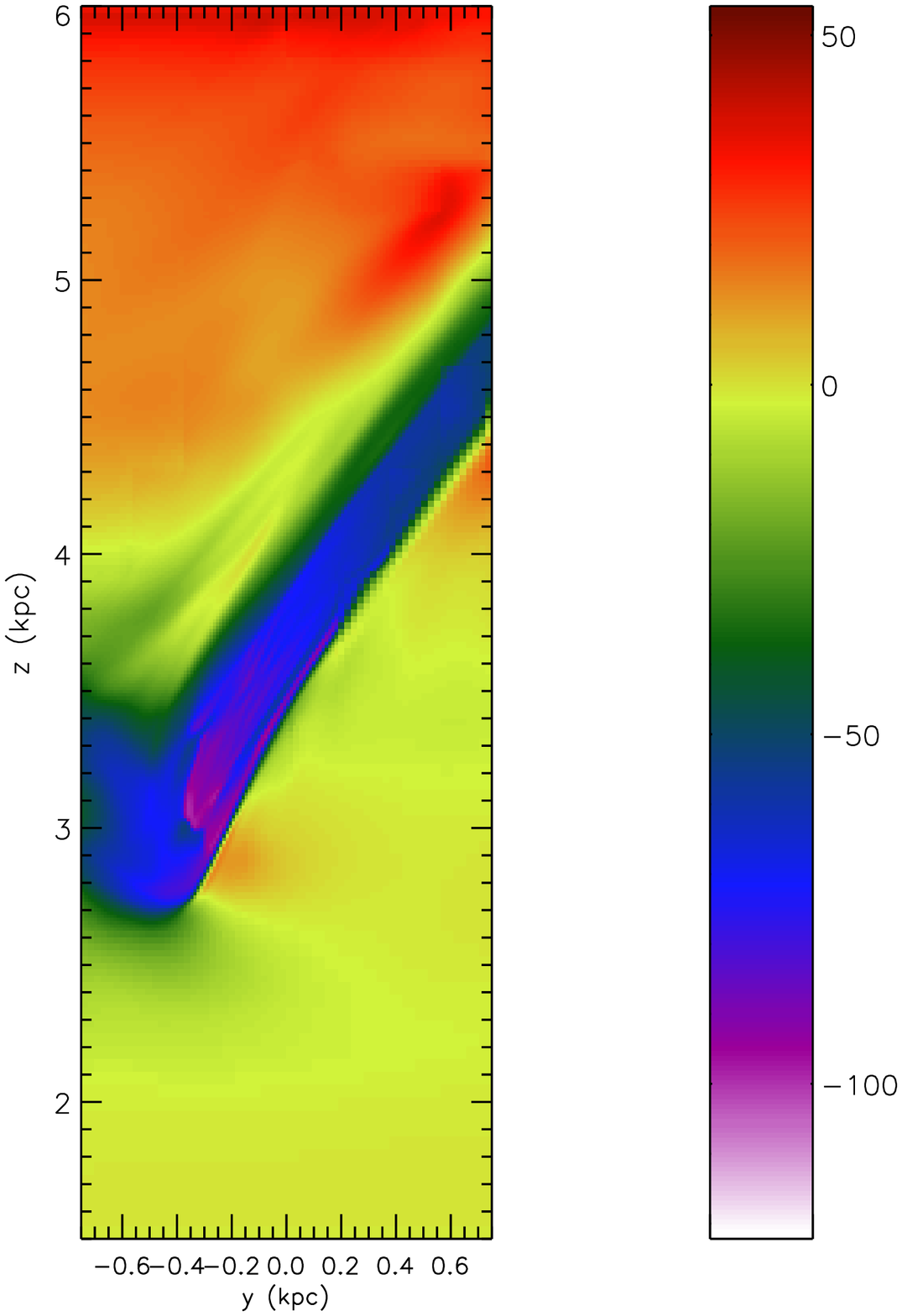}
\caption{Same as Figure \ref{modelA1_P}, but for Model D 
($n_{cloud}=0.01 ~\mbox{H atoms} ~ \mbox{cm}^{-3}$, 
$B_y = B_z = 0.94 ~ \mu \mbox{G}$) 
and the cuts are on the $x=0$ plane. 
\label{modelD_P}}
\end{figure}

In the upper cloud--ISM interface, a rarefaction wave propagates into 
the cloud and into the ISM, respectively. This 
is also similar to another case of 
the one-dimensional shock tube problem, 
in which a high-density region moves away from a neighboring 
low-density region of the same thermal plus magnetic pressure
(our cloud initially has the same thermal plus magnetic pressure 
as the neighboring gas). 
The rarefaction waves result in rarefied density regions 
behind the clouds in our simulations. 
The left five panels of Figures \ref{modelA1_P}-\ref{modelD_P}, 
which display the hydrogen number densities 
for the clouds and ambient media at times of 8, 16, 24, 32, and 
40 Myr, show that these low-density regions develop during the 
cloud's descent. Note that the rarefied regions are most clearly shown 
in Models A1 and A2 which have no ambient magnetic field 
and in Models C1, C2, and D in which the magnetic field 
is oriented parallel to the fluid's velocity vector, 
while they are less clear 
in Models B1-B4 in which the magnetic field 
is oriented perpendicular to the fluid's velocity vector. 
This is because clouds deform differently 
when the magnetic field is oriented perpendicular 
to the cloud's motion. The effect of the magnetic field 
on the cloud's morphology will be discussed in more detail 
in Section \ref{geometry_S}.

Note that the cloud--ISM interfaces in our model simulations 
do not exactly match the one-dimensional shock tube problem. 
Our clouds have spherical shapes, both our interstellar density 
and thermal pressure have gradients, and both the cloud 
and the ISM are affected by the gravity. 
Therefore, the jump in the thermal and magnetic pressure across 
the lower cloud--ISM interface is not constant in our simulations, 
in contrast with the shock tube profiles shown in 
Figure \ref{shock_tube}. In our simulations, 
there are gradients in the thermal and magnetic pressures 
across the lower cloud-ISM interface and these gradients provide 
resistance to the cloud's motion. The shock tube profiles imply 
that this resistance increases with the magnetic field strength 
when the magnetic field lines are perpendicular to the cloud's 
motion.  

\subsection{Results of Model Simulations} \label{model_sim_S}

\subsubsection{Dynamics of the Cloud}\label{dynamics_S}

The cloud's dynamics are affected by two factors, the cloud's initial 
density and the orientation and strength of the ambient magnetic field.  
Denser clouds fall faster than less dense clouds in the same interstellar 
environment. Resistance 
to the cloud's motion increases with the strength of the magnetic field 
when the magnetic field is oriented perpendicular 
to the cloud's motion. 
The magnetic field effects are more important when the cloud's initial 
density is small. 

Physically, the acceleration of the cloud is determined from the competition 
between the gravitational force of the Galaxy and the resisting force 
developed via the cloud-ISM interaction. 
The net acceleration of the cloud ($\vec{a}$) is 
\begin{equation}\label{accel_eq}
\vec{a} =  \vec{g} - \frac{\hat{z}}{\rho_{cl}} (\frac{d p_{tot}}{dz}) , 
\end{equation}
where $\rho_{cl}$ is the cloud's density, $p_{tot}$ is the total 
pressure, $p_{tot}=p_{th}+p_{mag}$, $p_{th}$ is the thermal pressure, and 
$p_{mag}$ is the magnetic pressure. The values of $(a/g)$, calculated 
from $(\frac{d p_{tot}}{dz})$, $\rho_{cl}$, and $g(z)$ 
which are garnered from 
the simulations at $t=$8, 16, 24, and 32 Myr, are shown in 
Table \ref{ag_T}. The distances 
that the clouds have fallen by $t=$16 and 24 Myr 
are also estimated for comparison. 

The estimated values of ($a/g$) decrease over time because resistance 
due to the cloud-ISM interaction increases over time. Comparison between 
Models A1 and A2, Models B1 and B2, and Models B3 and B4, in which 
the cloud's initial densities vary while the ambient magnetic 
field has the same orientation and strength, confirms that 
the heavier clouds fall faster. 
The negative correlation between the magnetic field strength and the resistance 
to motion is confirmed by comparing Model A1 in which there is no 
magnetic field with Models B1 and B3 which have small and large 
magnetic fields, respectively. These three models have small initial cloud 
densities. The comparison between Models A2, B2, and B4, in which 
the clouds have large initial densities, generally exhibits the 
same trend, but the fields are too weak to significantly decelerate 
the heavy cloud at early times. 

\subsubsection{Geometrical Consequences for the Cloud and the ISM}\label{geometry_S}

Our simulations show that the magnetic field affects not only the cloud's 
dynamics but also the cloud's morphology. In this subsection, we discuss 
four geometrical consequences for the cloud and the ISM; 
creation of tail structures, flattening of the cloud, 
asymmetric shape change, and the effect of a tilted magnetic field.  

\begin{figure}
\centering
\includegraphics[scale=0.25]{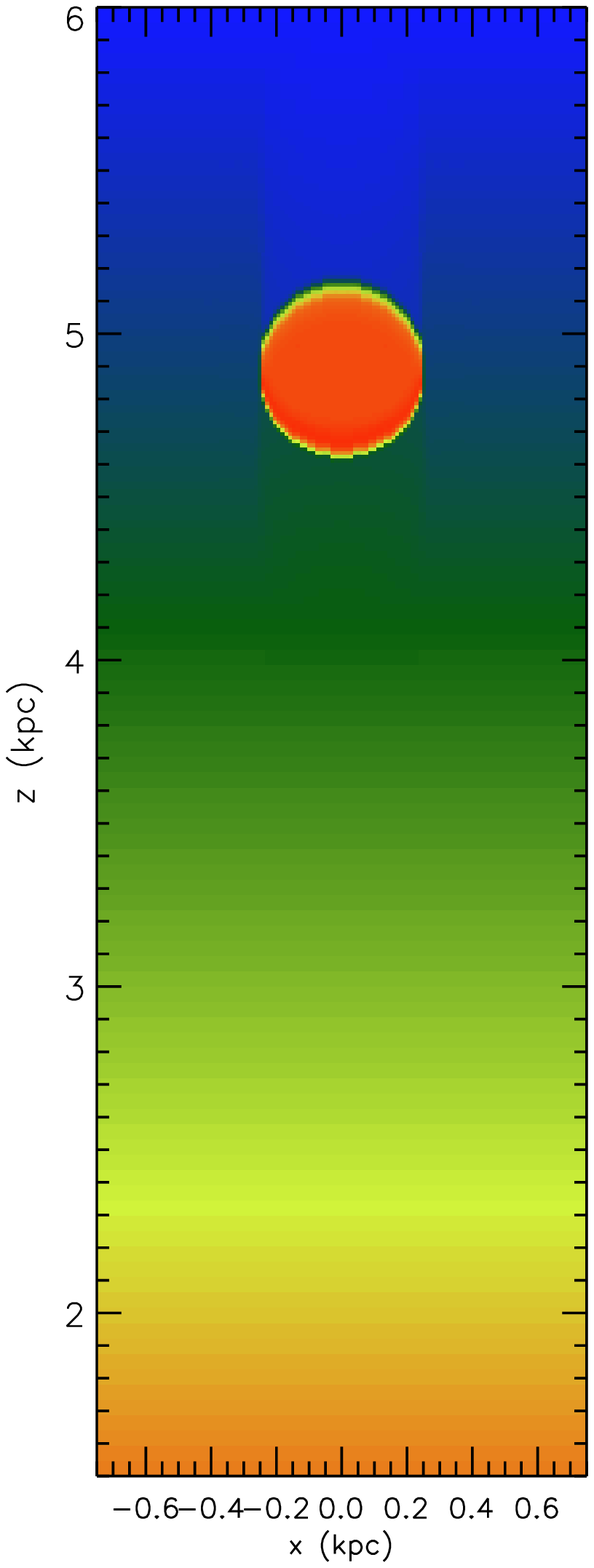}
\includegraphics[scale=0.25]{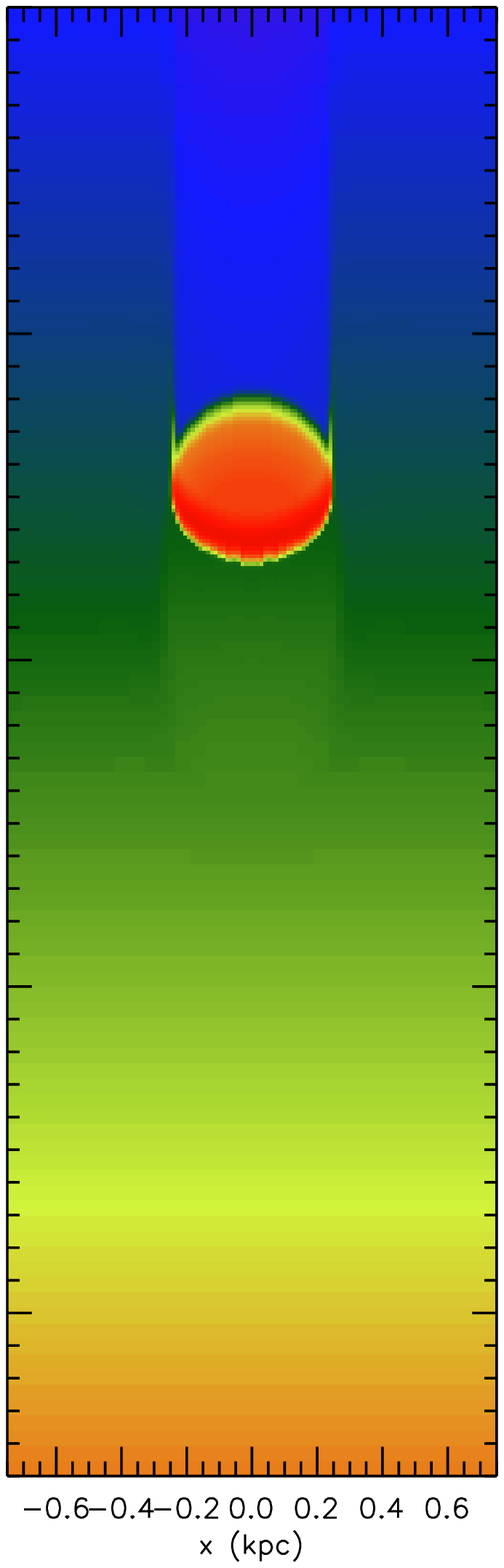}
\includegraphics[scale=0.25]{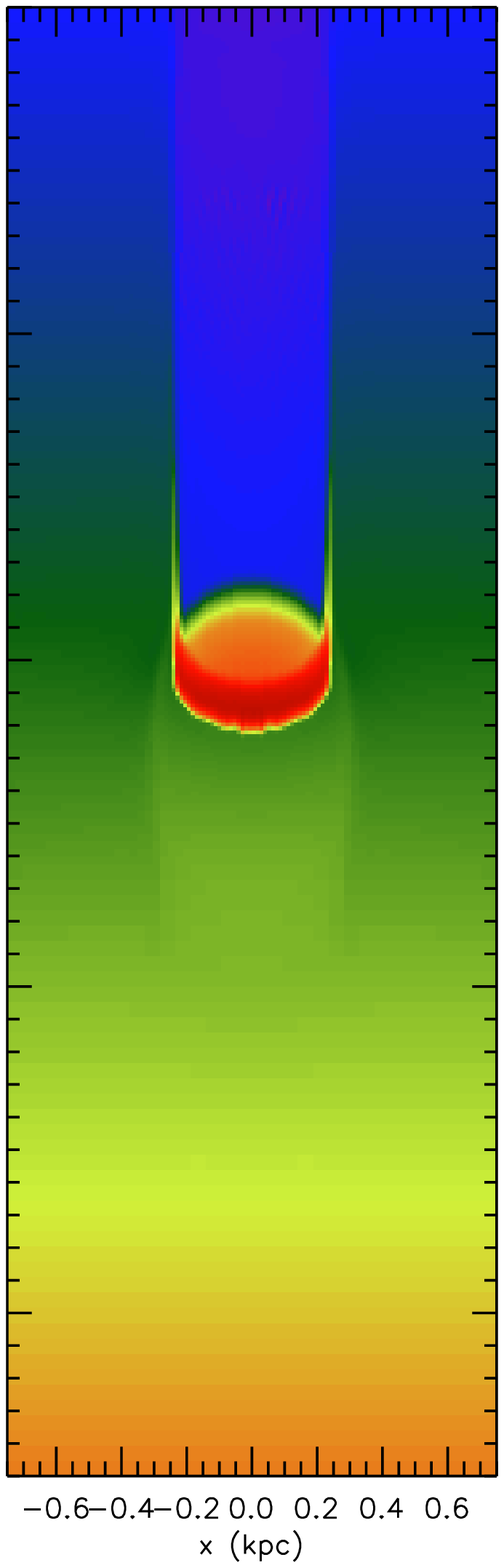}
\includegraphics[scale=0.25]{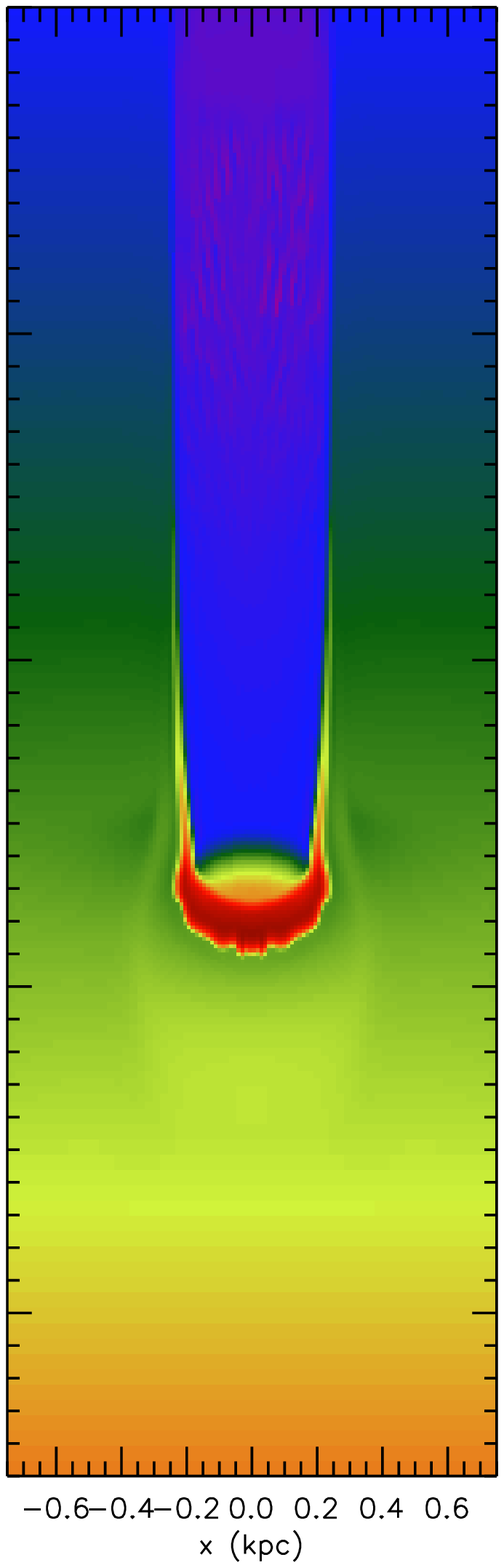}
\includegraphics[scale=0.25]{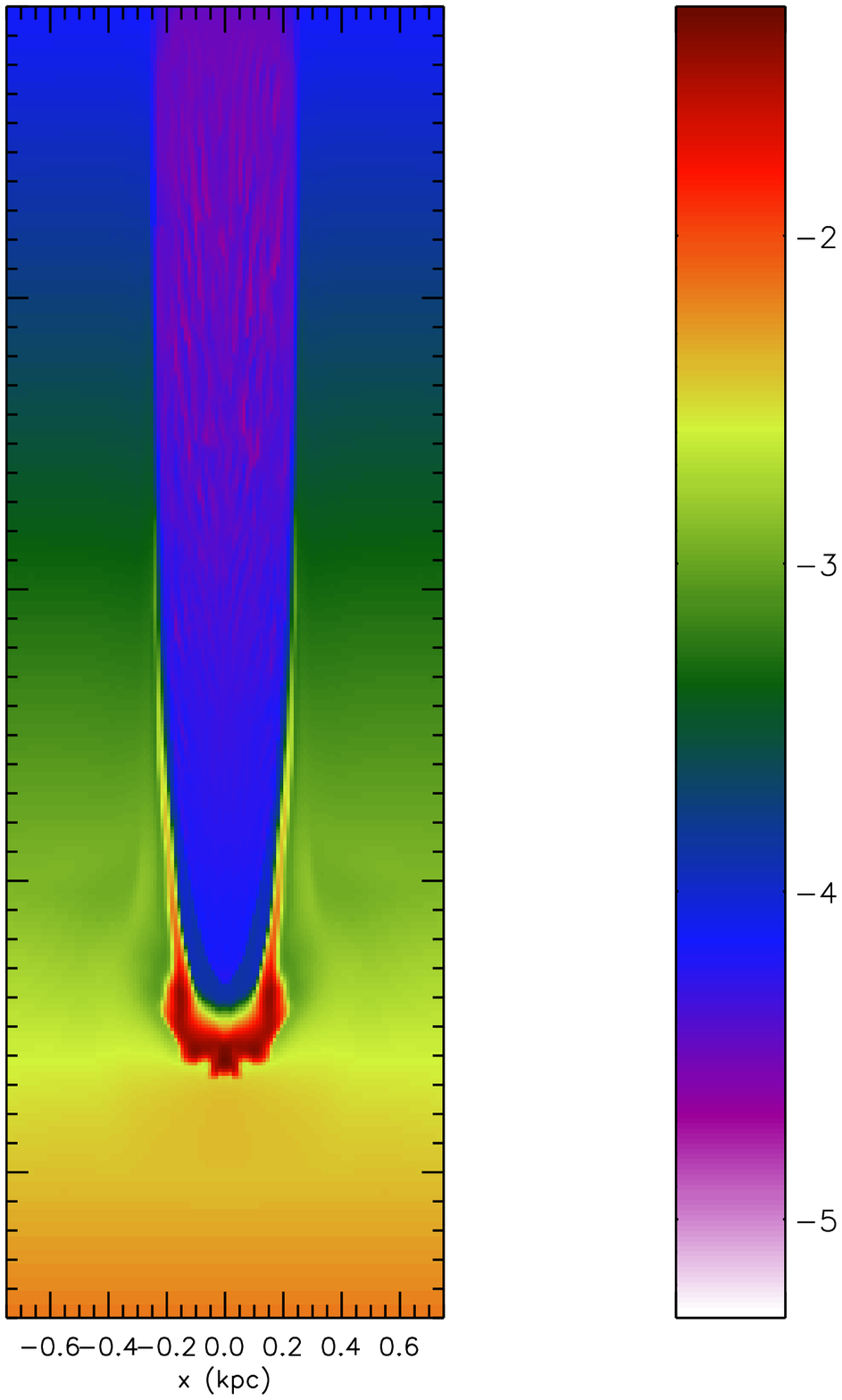}
\includegraphics[scale=0.25]{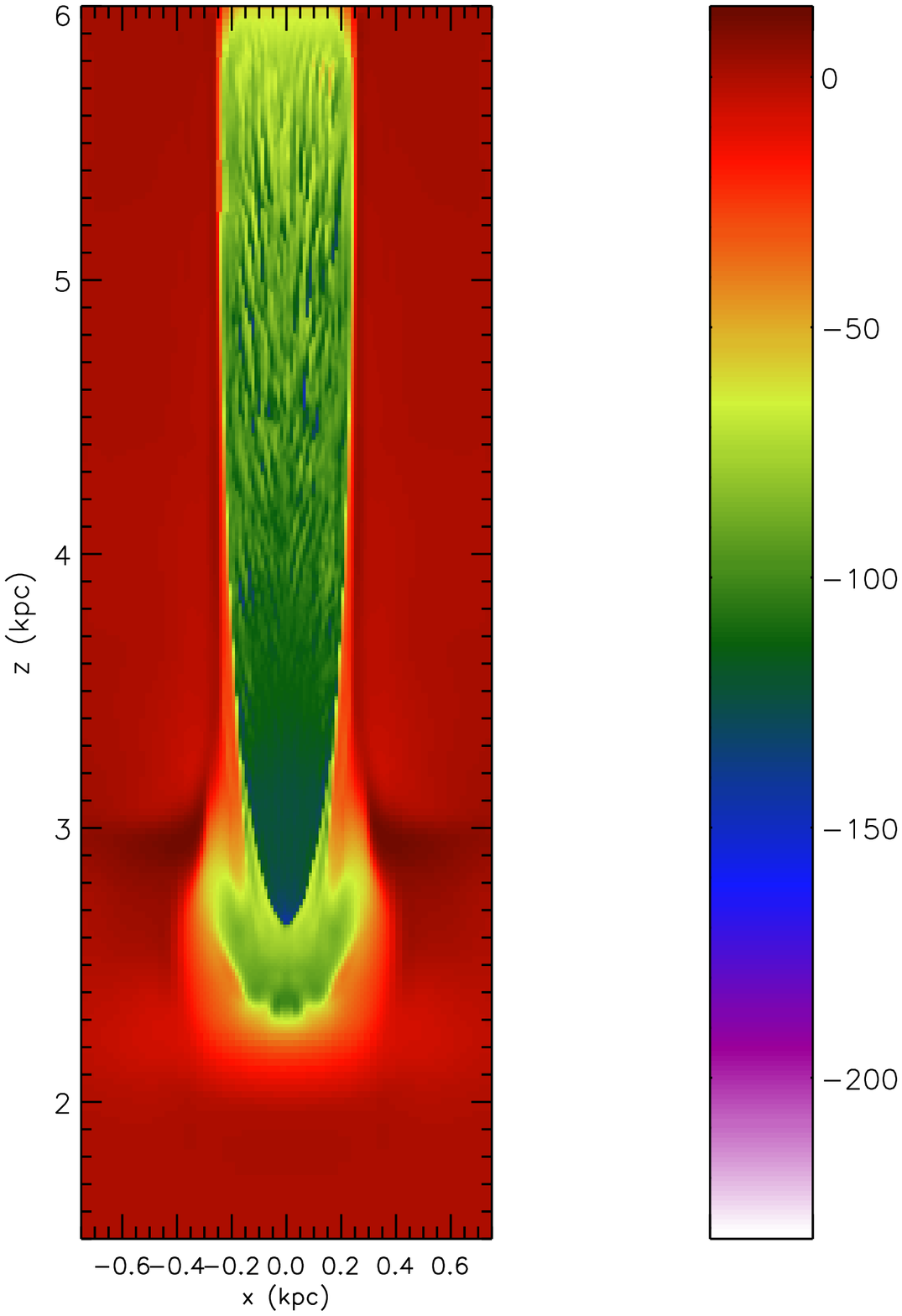}
\caption{Same as Figure \ref{modelA1_P}, but for Model C1 
($n_{cloud}=0.01 ~\mbox{H atoms} ~ \mbox{cm}^{-3}$, 
$B_z = 1.3 ~ \mu \mbox{G}$). 
\label{modelC1_P}}
\end{figure}

\begin{figure}
\centering
\includegraphics[scale=0.25]{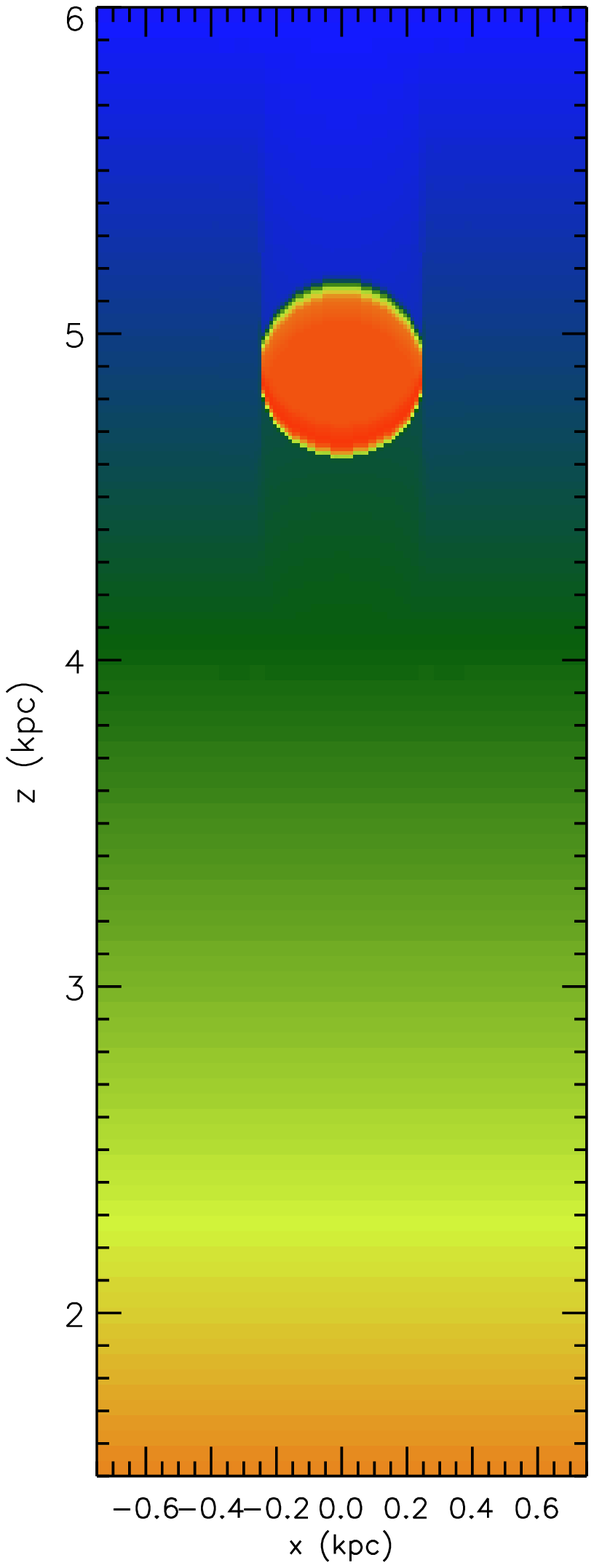}
\includegraphics[scale=0.25]{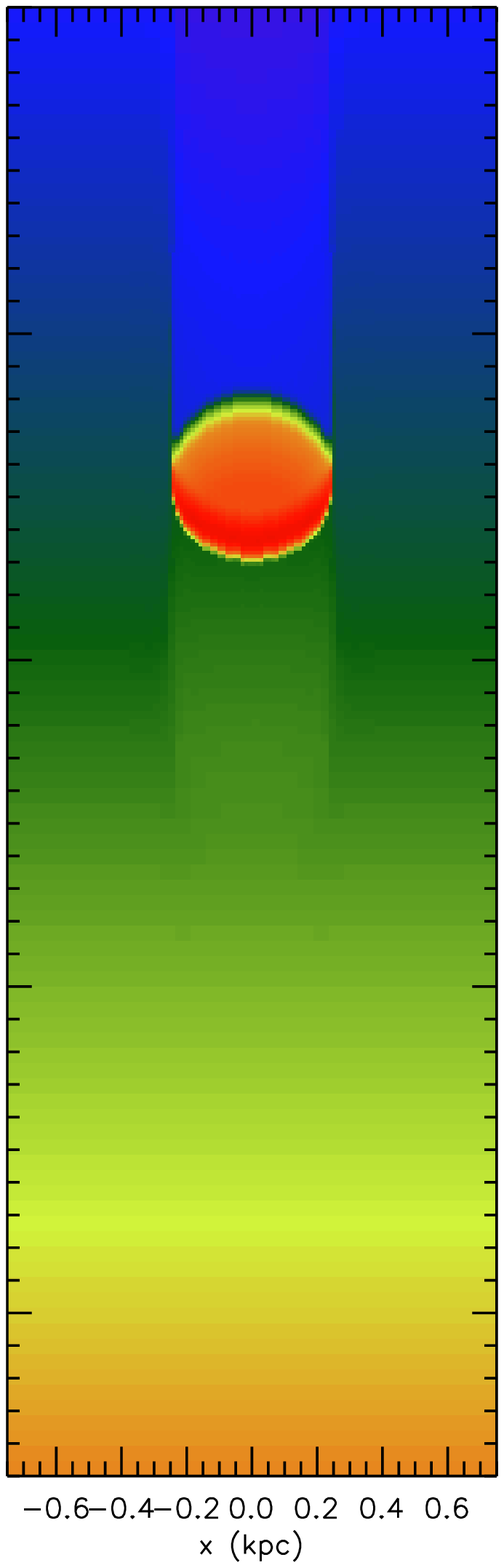}
\includegraphics[scale=0.25]{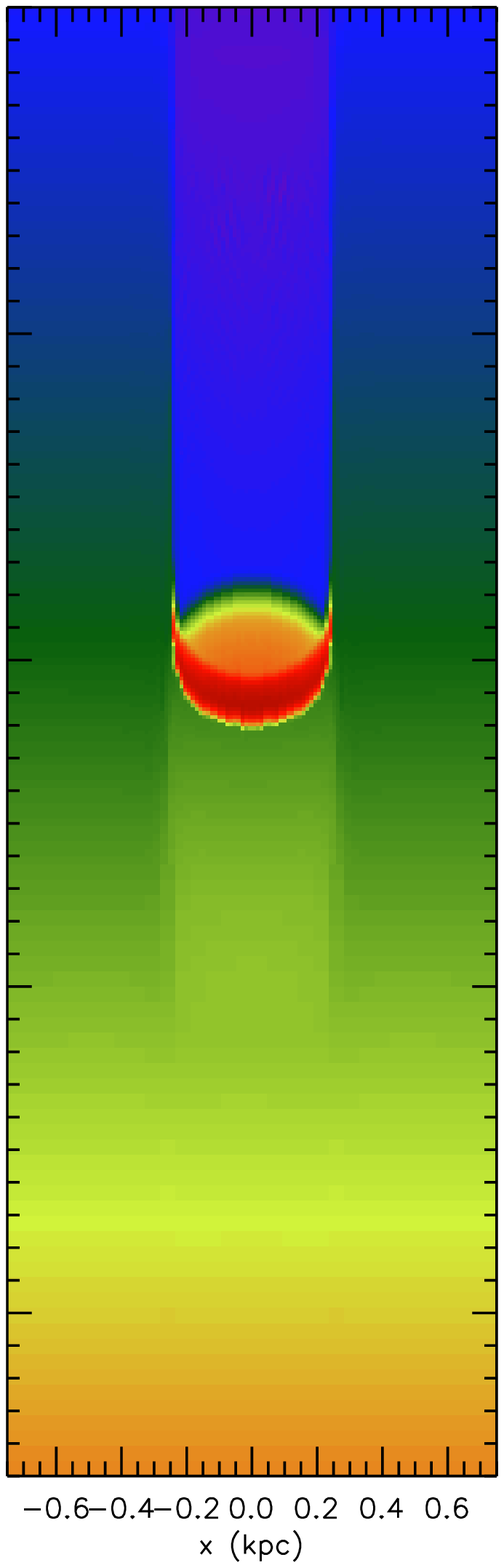}
\includegraphics[scale=0.25]{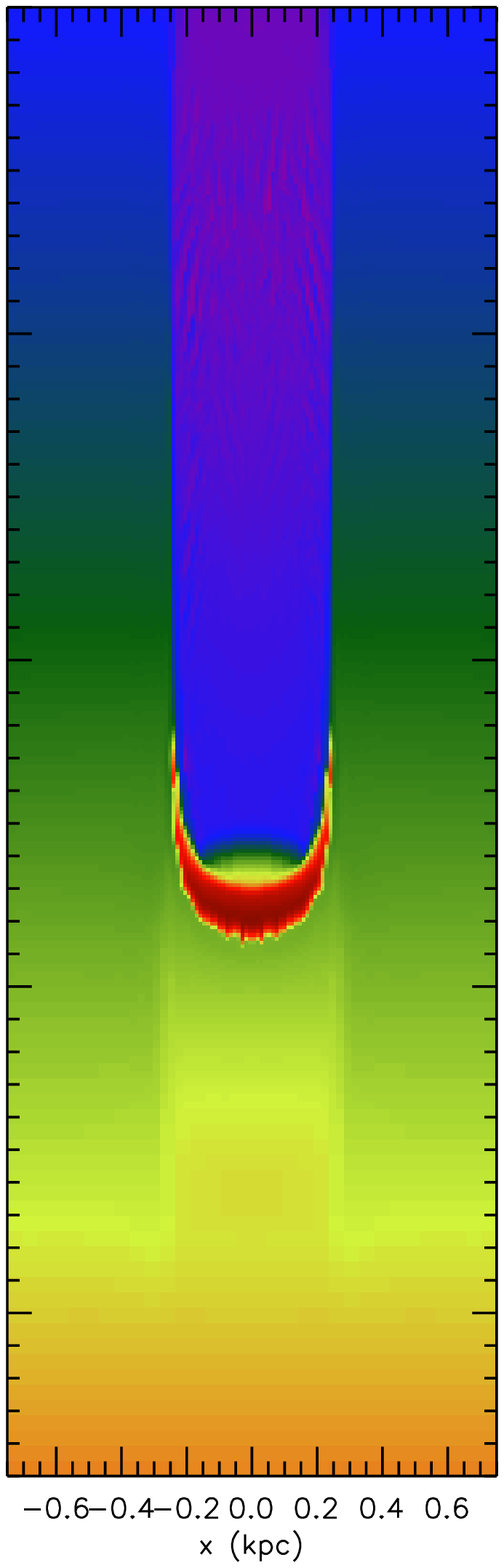}
\includegraphics[scale=0.25]{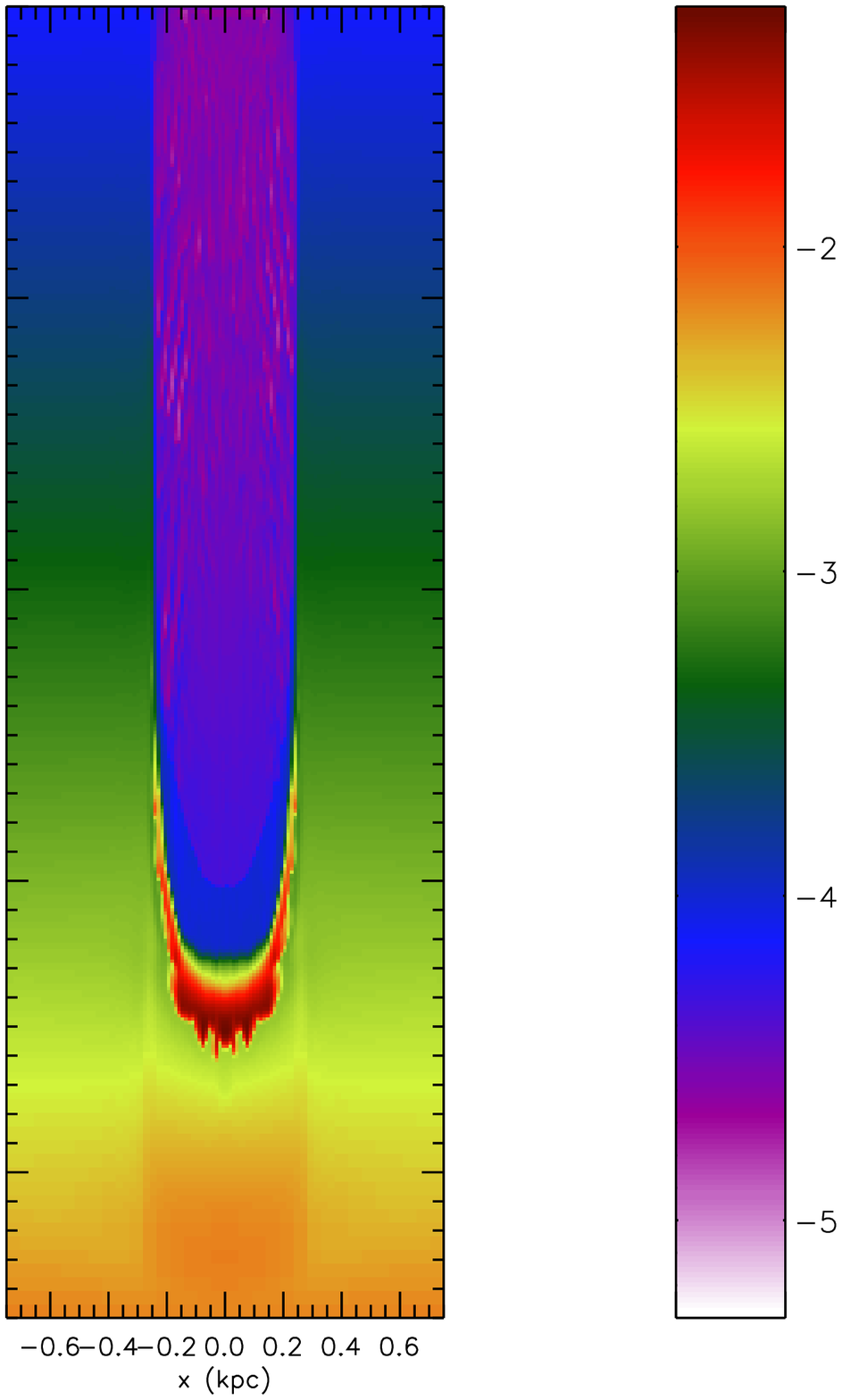}
\includegraphics[scale=0.25]{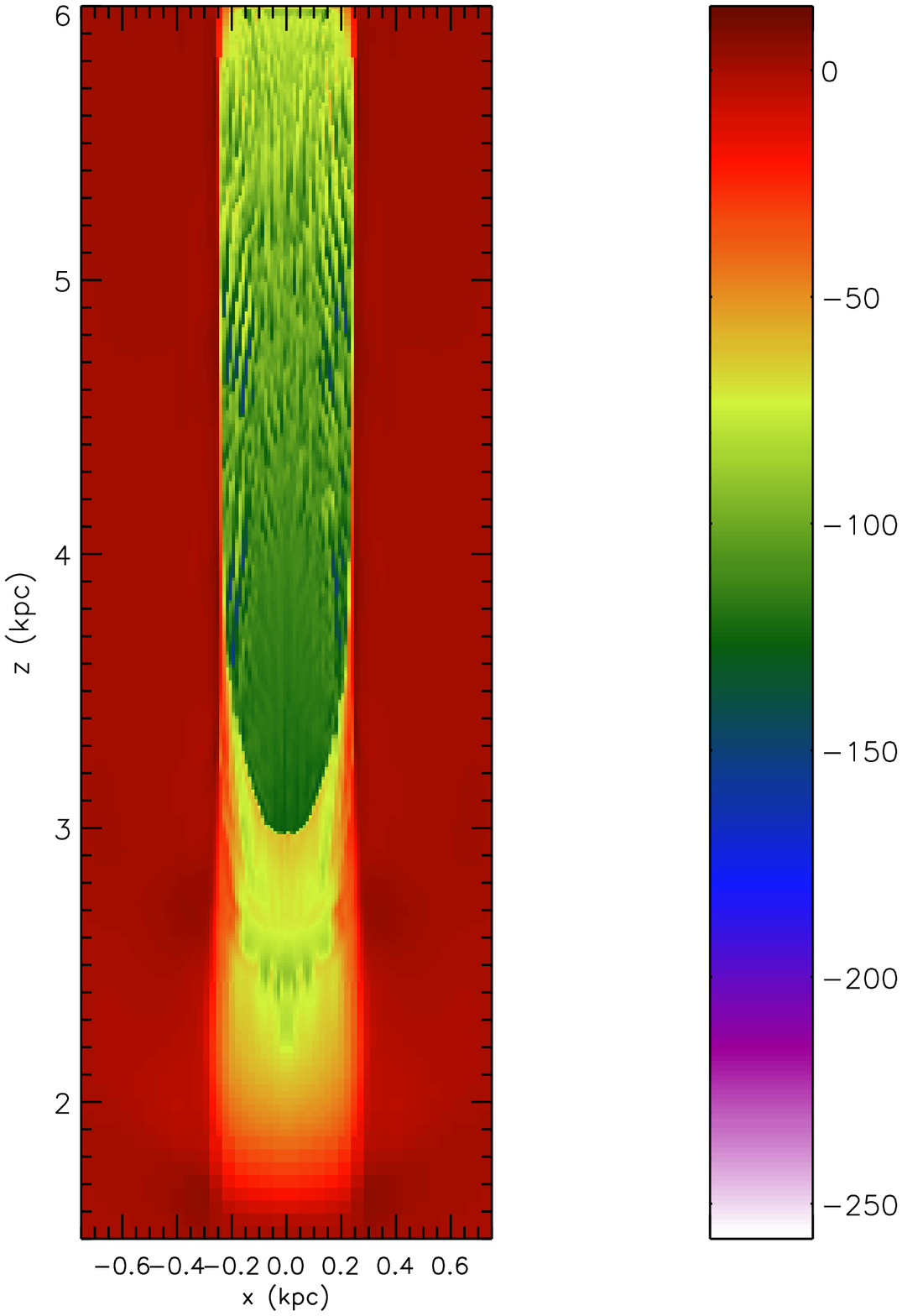}
\caption{Same as Figure \ref{modelA1_P}, but for Model C2 
($n_{cloud}=0.01 ~\mbox{H atoms} ~ \mbox{cm}^{-3}$, 
$B_z = 4.2 ~ \mu \mbox{G}$). 
\label{modelC2_P}}
\end{figure}

In each model, the cloud 
develops a tail due to the shear between the cloud and the ISM, but 
the shape and length of the tail is affected by the magnetic field 
orientation and strength. Especially in Models C1 and C2, where  
the magnetic field is oriented parallel to the cloud's motion, 
a very thin tail develops along the sides of the columnar-shaped 
region behind the cloud. 
This region has a higher magnetic field strength than 
the surrounding gas and can be identified as a low-density region 
behind the cloud in the density profiles of Figures \ref{modelC1_P} 
and \ref{modelC2_P}. 
Because its strong magnetic field protects the region behind the cloud 
from substantial inflows, the material sheared off of the cloud 
cannot invade this region and, instead, is confined along 
the sides of the high magnetic field wall. 

The clouds deform as they fall. Due to both the ISM's force on the cloud 
and the gradient in the Galaxy's gravitational acceleration, the clouds 
flatten into disk- or cigar-shaped objects depending on the orientation 
of the interstellar magnetic field. For example, by 40 Myr 
the cloud in Model A2 deforms into a disk shape 
which has a radius of $\sim0.23$ kpc and 
a depth of $\sim0.1$ kpc. By the same time, the cloud in Model B2 
has deformed into a cigar shape which has lengths 
of $\sim0.65$ kpc along $x$-axis and 
$\sim0.26$ kpc along $y$-axis and a height of $\sim0.13$ kpc. 
We will discuss the observational effect of these flattened clouds in 
Section \ref{observation_S}. In addition to flattening, instabilities result in 
downward protruding nodules. 

\begin{figure}
\centering
\includegraphics[scale=0.25]{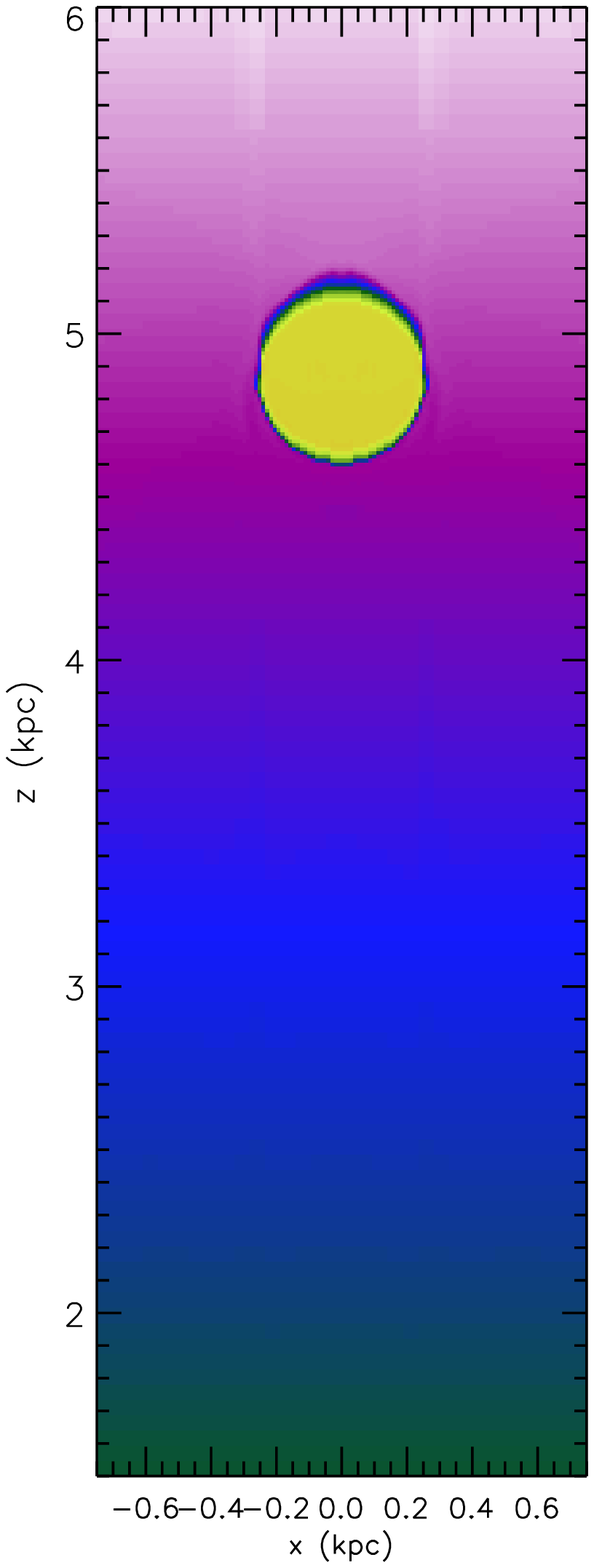}
\includegraphics[scale=0.25]{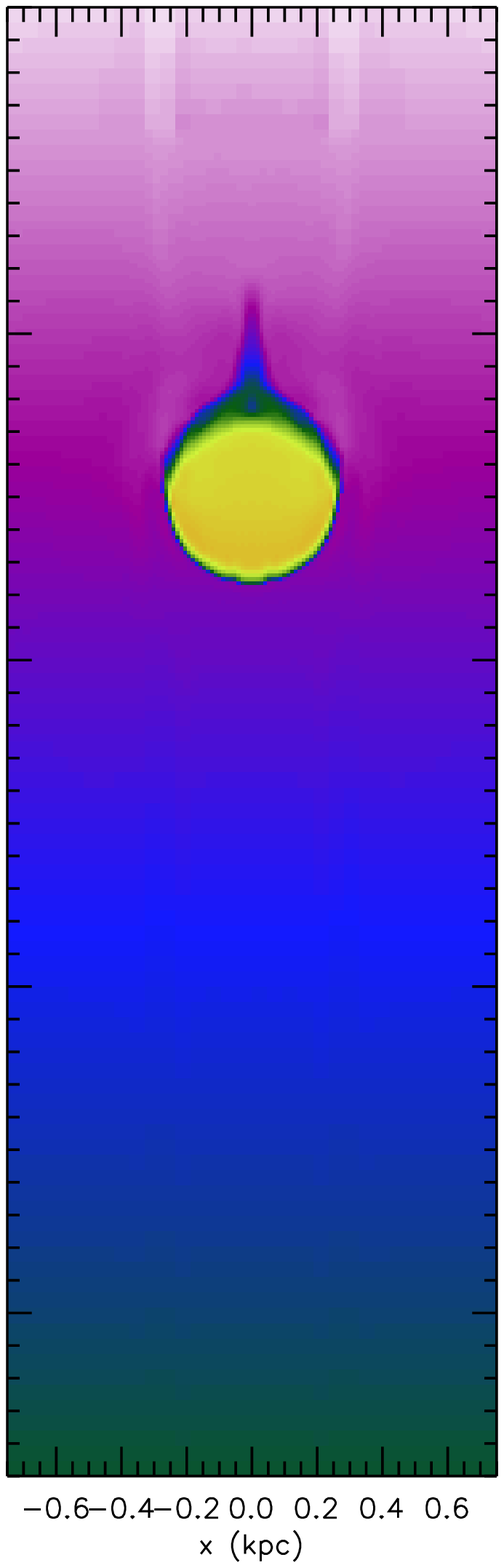}
\includegraphics[scale=0.25]{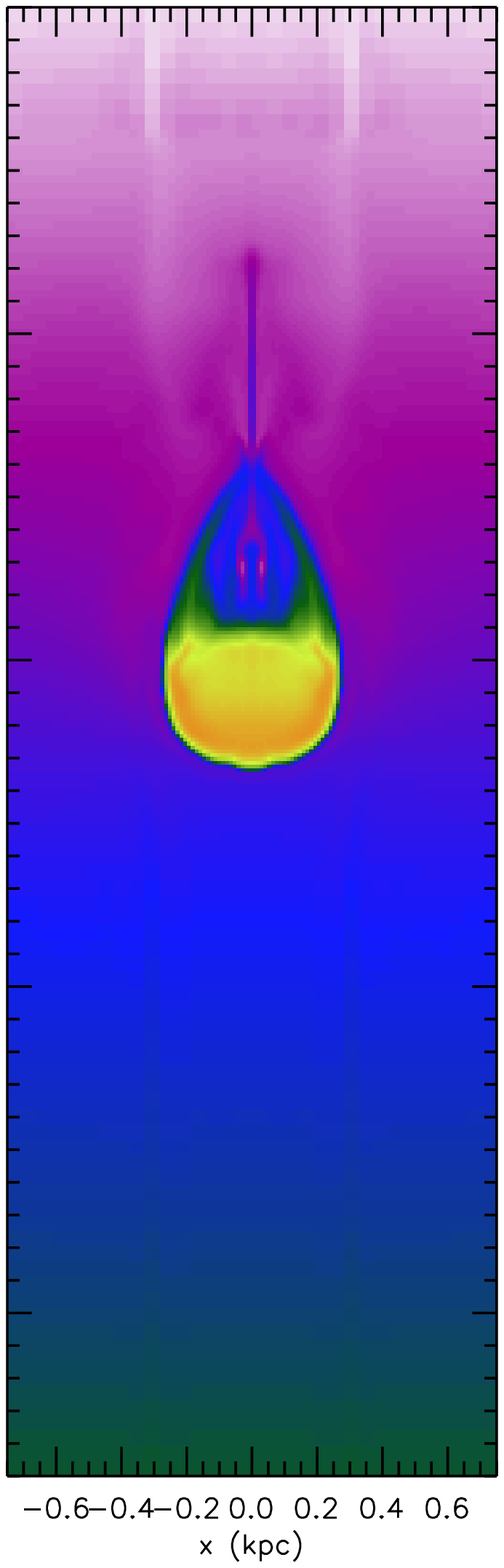}
\includegraphics[scale=0.25]{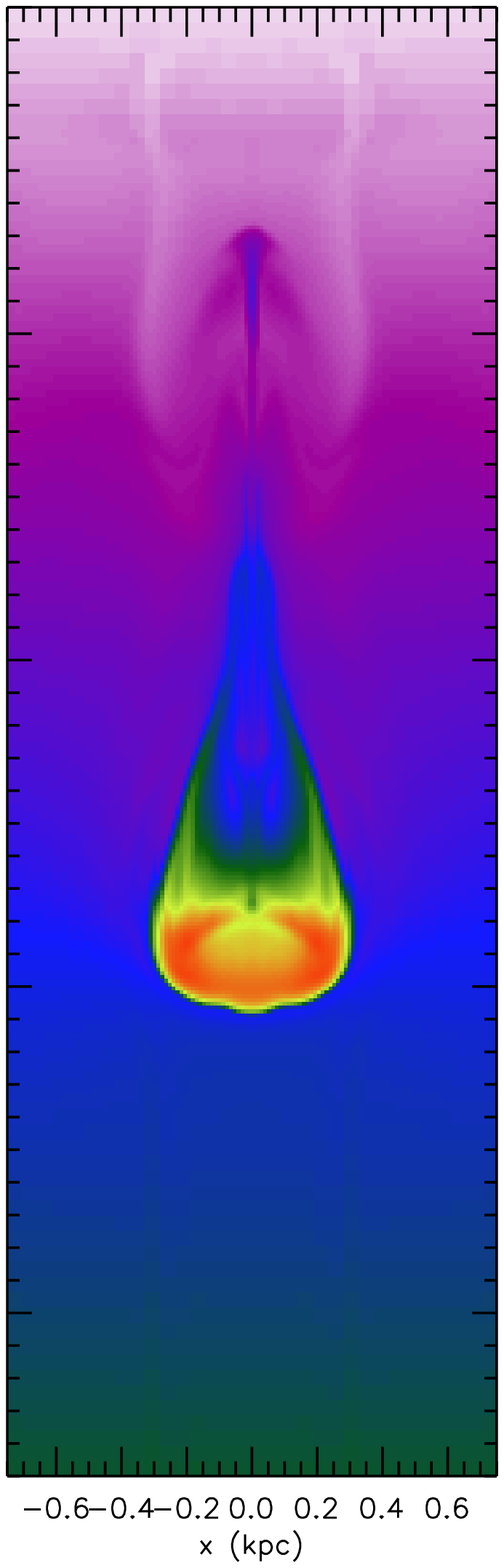}
\includegraphics[scale=0.25]{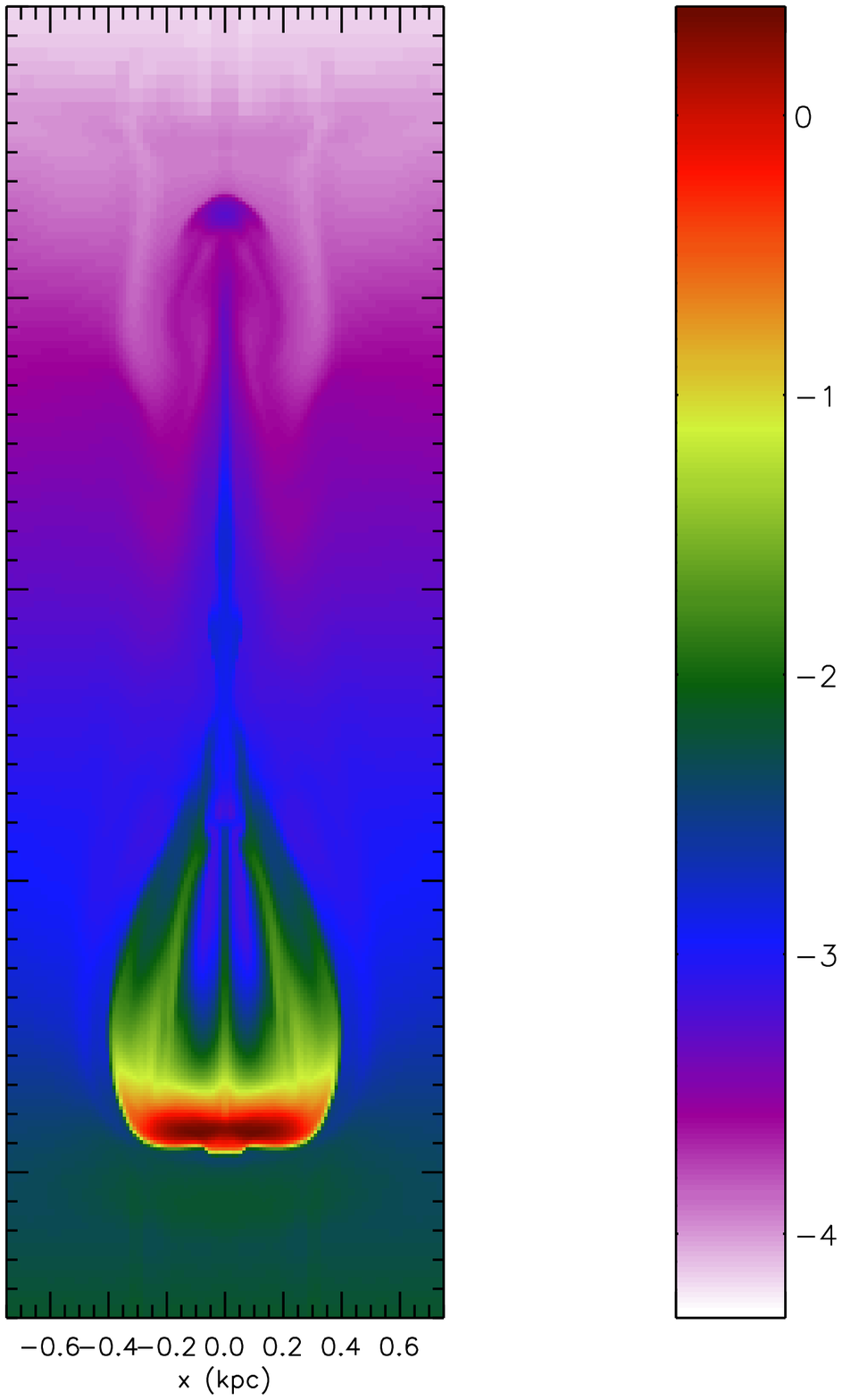}
\includegraphics[scale=0.25]{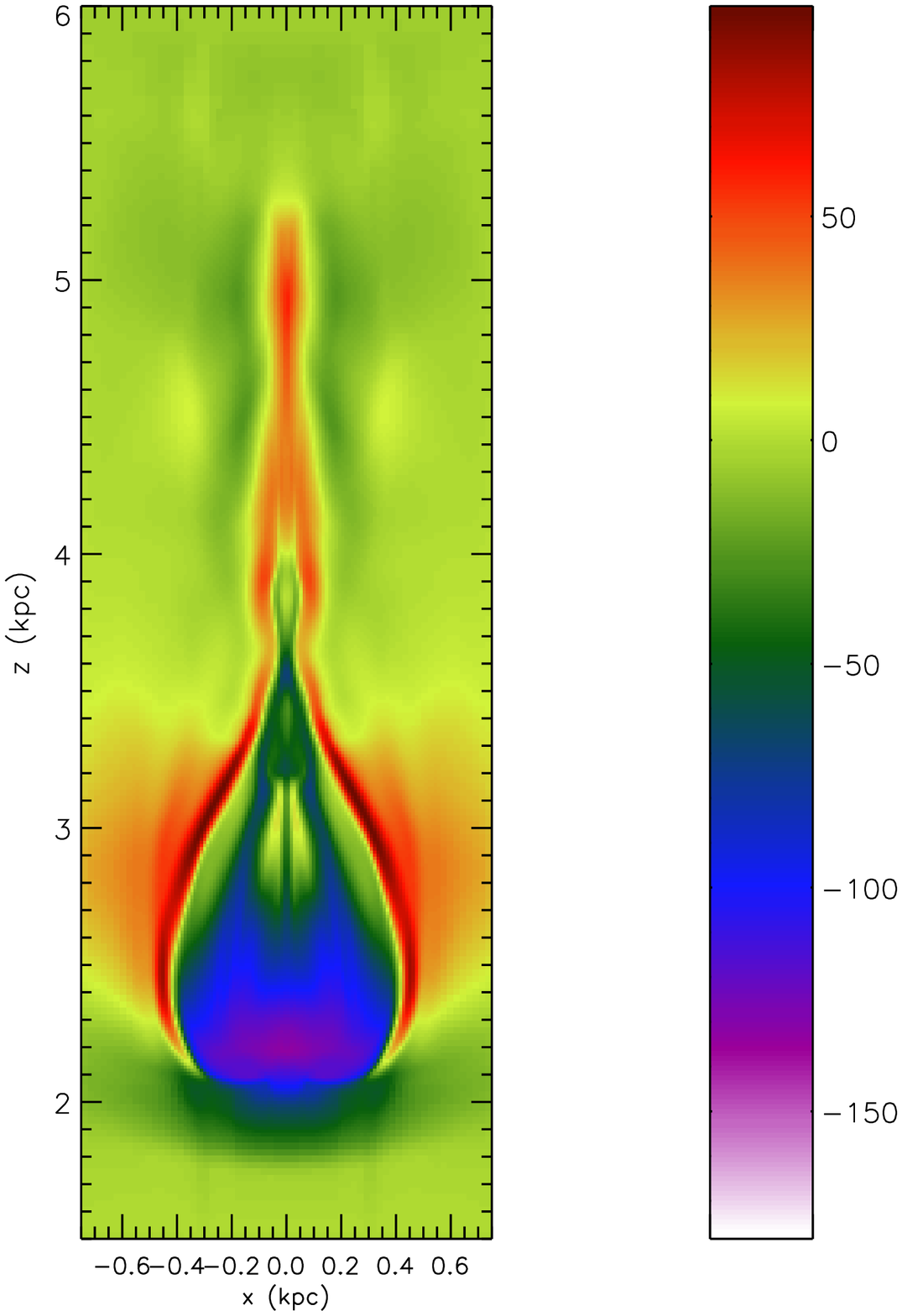}
\caption{Same as Figure \ref{modelA1_P}, but for Model B2 
($n_{cloud}=0.1 ~\mbox{H atoms} ~ \mbox{cm}^{-3}$, 
$B_y = 1.3 ~ \mu \mbox{G}$). 
\label{modelB2_P}}
\end{figure}

\begin{figure}
\centering
\includegraphics[scale=0.25]{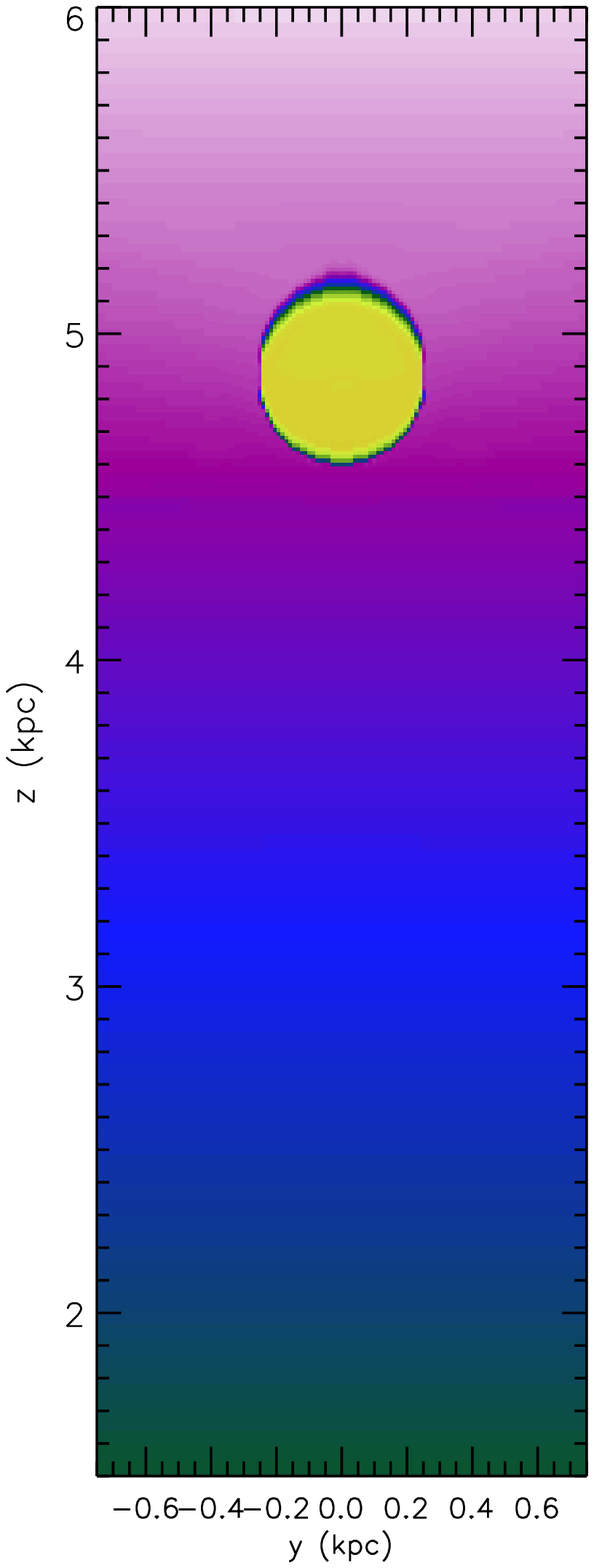}
\includegraphics[scale=0.25]{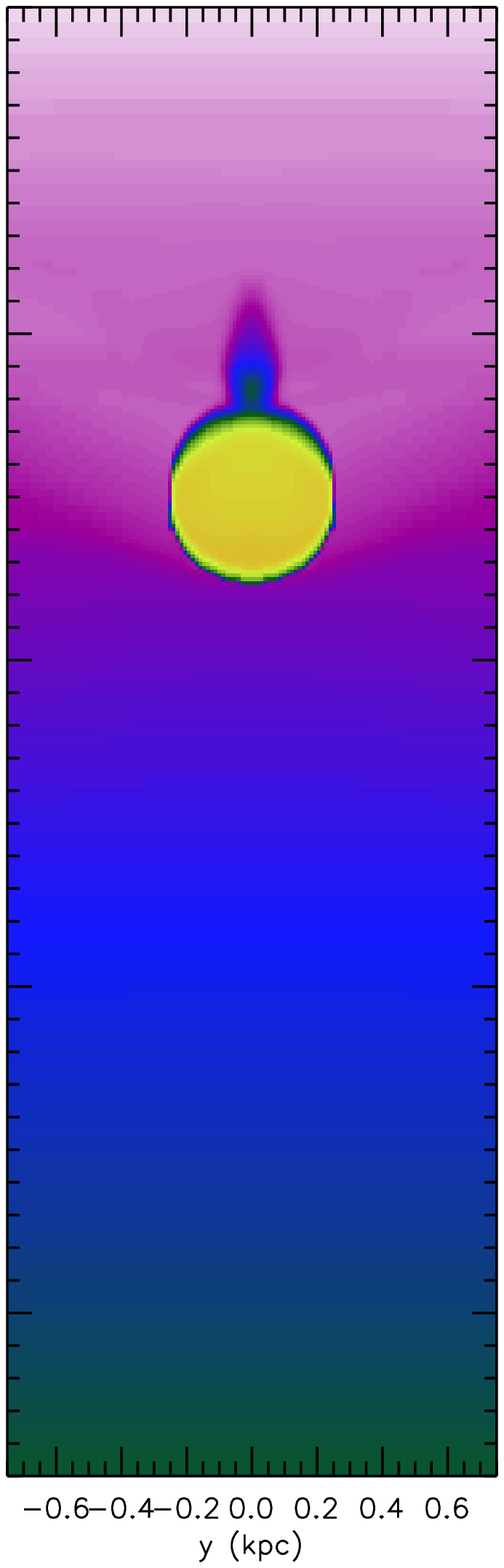}
\includegraphics[scale=0.25]{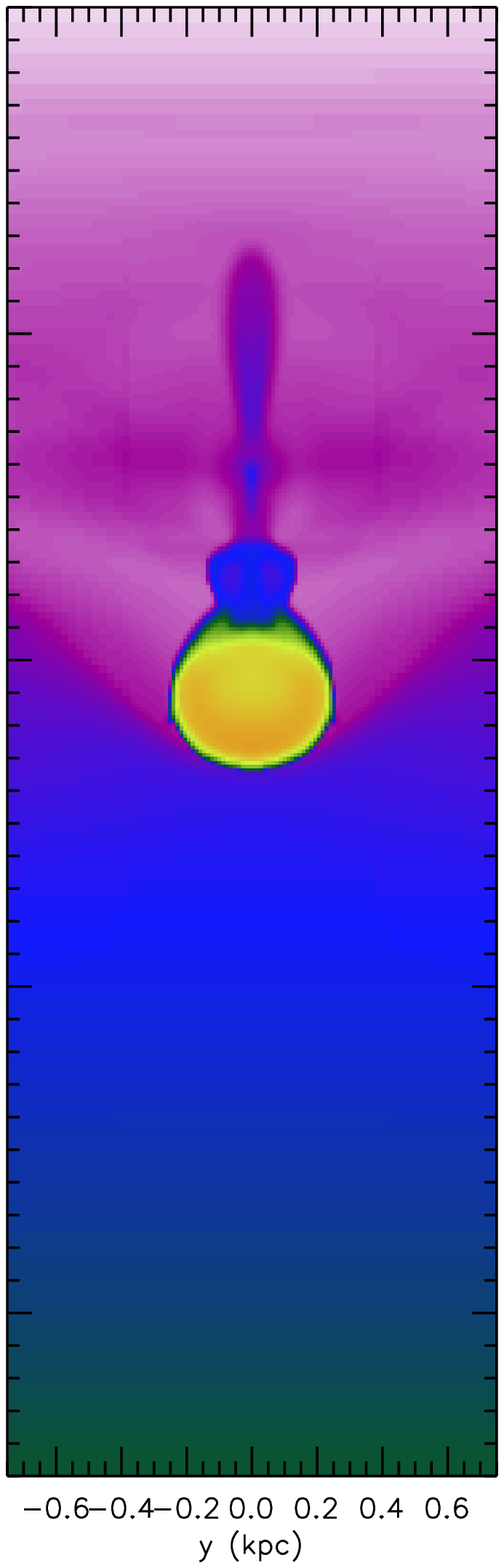}
\includegraphics[scale=0.25]{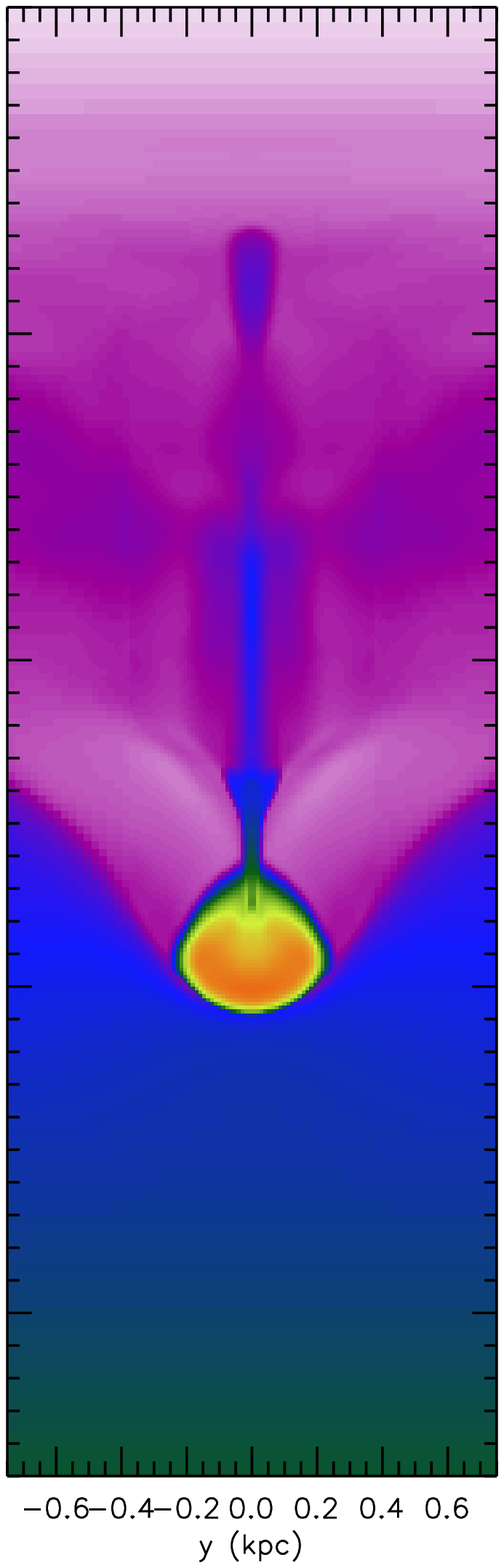}
\includegraphics[scale=0.25]{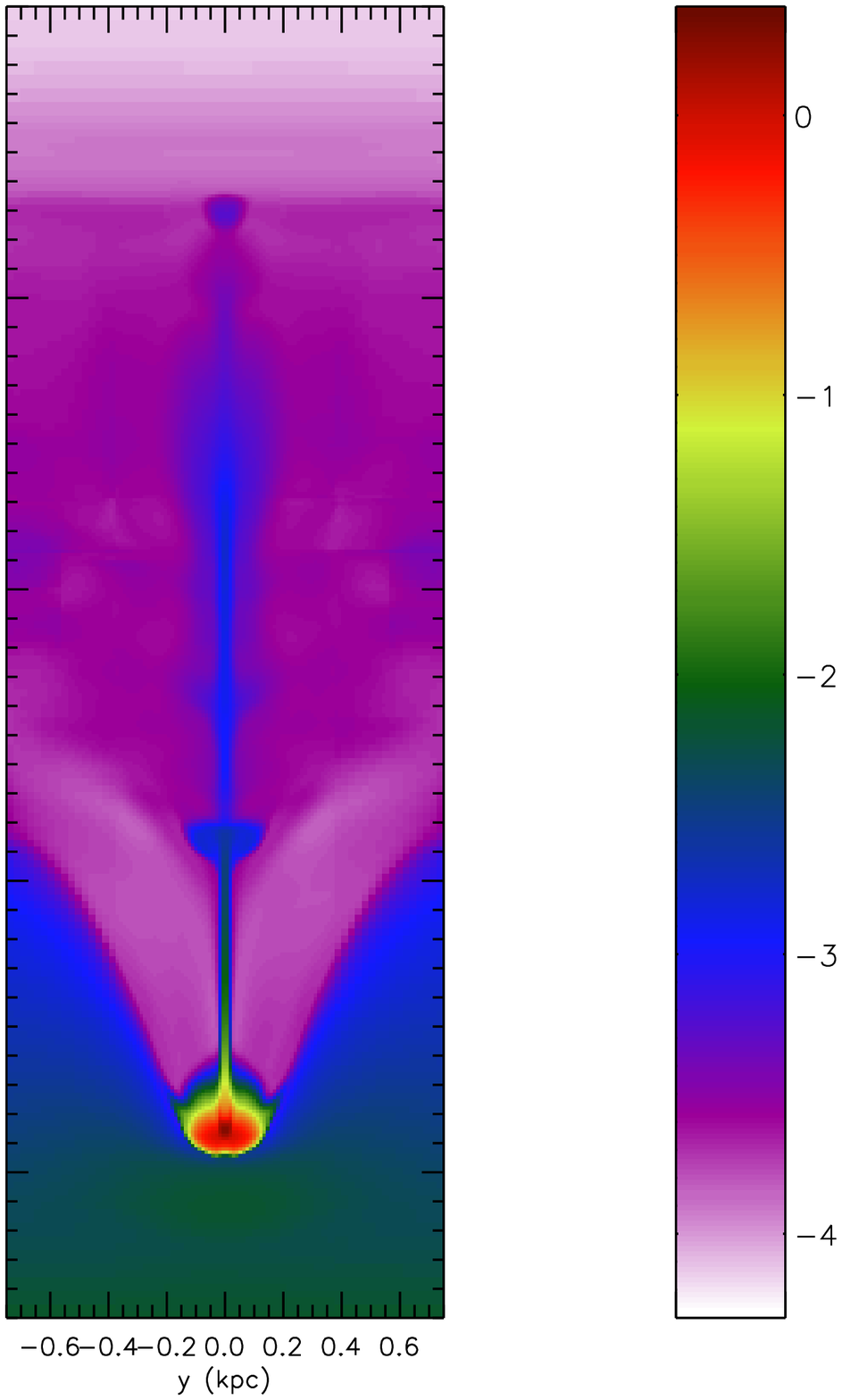}
\includegraphics[scale=0.25]{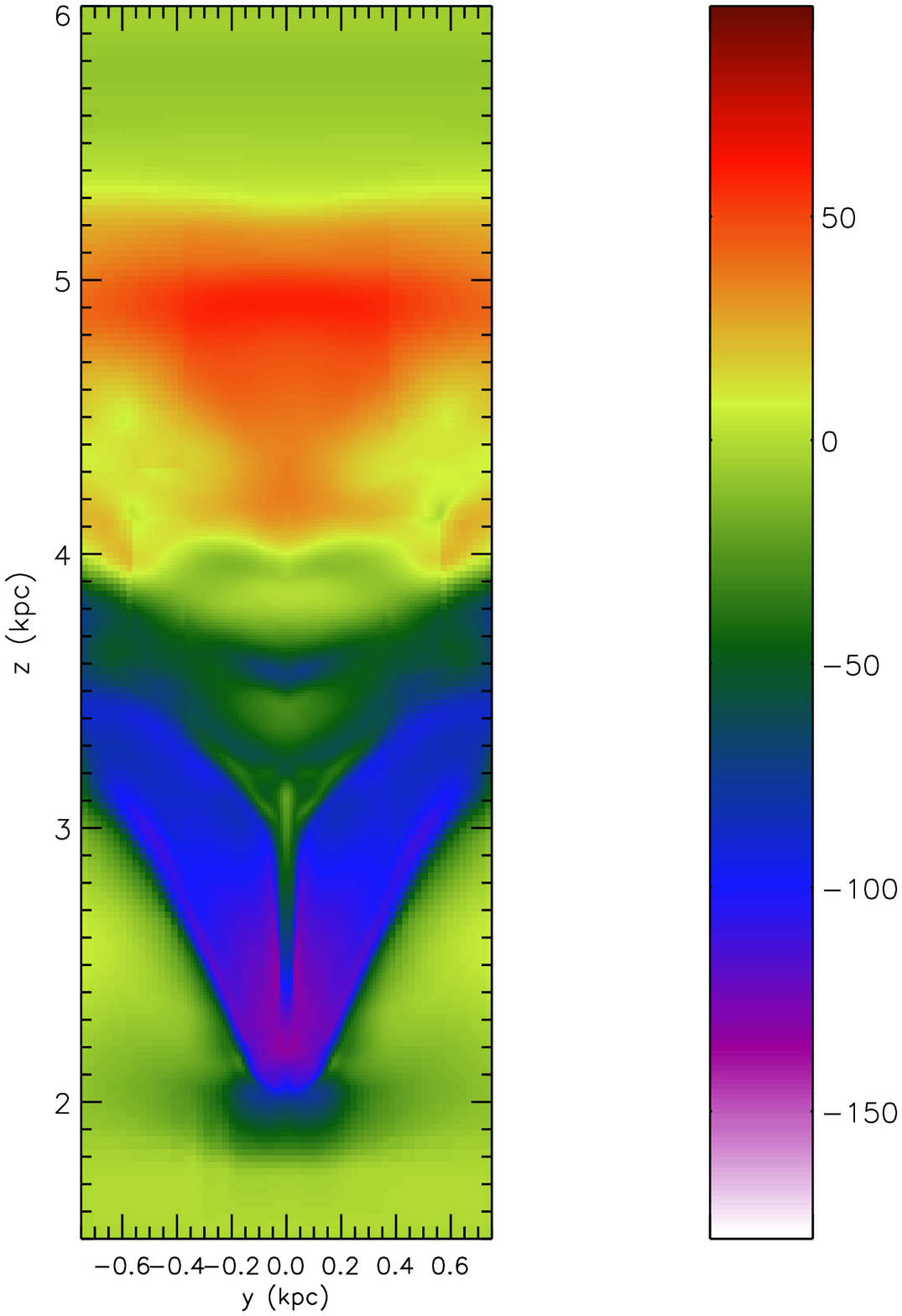}
\caption{Same as Figure \ref{modelB2_P} for Model B2, but the cuts are 
along the $x=0$ plane.  
\label{modelyzB2_P}}
\end{figure}

In the cases where the magnetic field is oriented perpendicular to 
the cloud's motion (Models B1-B4 and Model D which also has 
a magnetic field component that is perpendicular to $\hat{z}$), 
the cloud loses its symmetry about the $z$-axis. 
This is shown, for example, 
by comparing Figures \ref{modelB2_P} and 
\ref{modelyzB2_P}. Both figures show the three-dimensional simulation results 
of Model B2, but Figure \ref{modelB2_P} displays the cut through 
the $y$-$z$ plane ($x=0$ plane), while Figure \ref{modelyzB2_P} shows 
the cut through the $x$-$z$ plane ($y=0$ plane). 
The magnetic field is oriented along the $y$-axis 
(i.e., $B_y \neq 0$) in this simulation, 
thus the magnetic field lines are coming into the plane 
in Figure \ref{modelB2_P}, 
while they are parallel to the horizontal axis 
in Figure \ref{modelyzB2_P}. 
The information in Figures \ref{modelB2_P} and \ref{modelyzB2_P} 
helps demonstrate the cloud's three-dimensional morphology and how it evolves. 
As time elapses, the initially spherical cloud is squeezed 
along the direction of the initial magnetic field ($y$-axis), 
but is allowed to expand along the direction 
perpendicular to the initial magnetic field ($x$-axis). 
Note that this trend is the opposite of that experienced by 
bubbles in the ISM \citep{Gaensler1998ApJ,raley2007}, in which 
the elongation is parallel to the magnetic field direction. 
In our simulations, the cloud's deformation grows with time 
so that by $t=40~\mbox{Myr}$, the cloud's 
footprint looks like a bar which is elongated along the $x$-axis. 
The ambient magnetic field folds around the deformed cloud and 
makes a deep ``V'' shape in the $y$-$z$ plane. In contrast, 
in the cases where there is no magnetic field component 
parallel to the Galactic 
plane, the cloud retains cylindrical symmetry about the $z$-axis. 
Thus, the simulation results of Models A1, A2, C1, and C2 
for cuts through the $y$-$z$ plane ($x=0$ plane) are identical 
to the results for cuts through the $x$-$z$ plane ($y=0$ plane) shown in 
Figures \ref{modelA1_P}, \ref{modelA2_P}, \ref{modelC1_P}, and 
\ref{modelC2_P}. 

In Model D, the magnetic 
field lines are oriented at a $45\degr$ angle with respect to 
the Galactic plane. The important consequence of this magnetic field 
configuration is that the cloud's path is guided by the magnetic field 
lines. However, the cloud's downward acceleration is 
still affected only by the magnetic 
field component that is perpendicular to $\hat{z}$
(i.e., $B_y$). The strength 
of this $B_y$ component in Model D is about $0.94~\mu\mbox{G}$ 
which is $1/\sqrt{2}$ times as strong as the magnetic strength in Model B1. 
As a result, the cloud's vertical acceleration in Model D is between 
those of Model A1 and Model B1. The effect of the magnetic field component 
parallel to $\hat{z}$ ($B_z$) is to create a high magnetic 
pressure region behind the cloud, along whose sides tails develop. 
However, unlike in Models C1 and C2, the tail is asymmetric. 
The tail near the bottom of the cloud is suppressed, 
while the tail near the top of the cloud grows and spreads widely.

\subsubsection{Vertical Velocities}\label{mass_ratio_S}

\begin{figure}
\epsscale{1.0}
\plotone{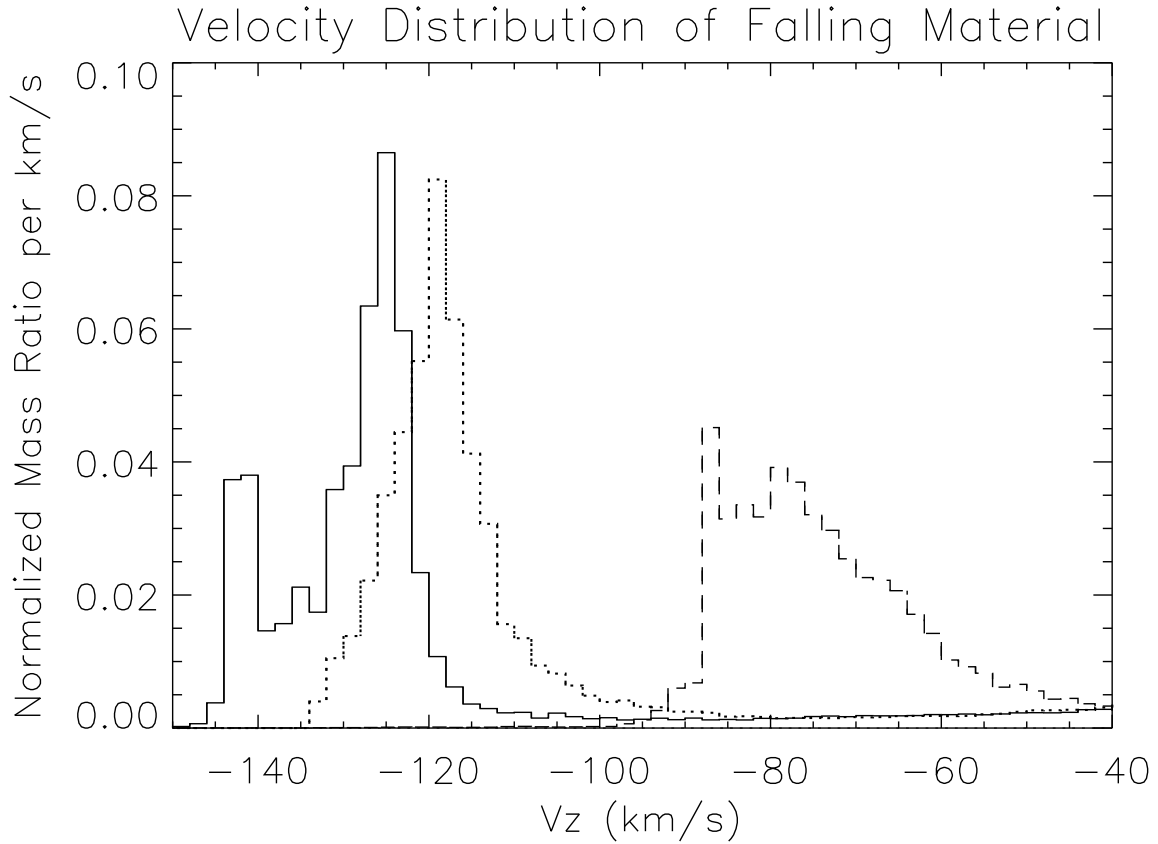}
\caption{Mass of material moving vertically with 
a velocity of $v_z$ at 40 Myr for models with a high initial 
cloud density: A2 (solid line), B2 (dotted line) 
and B4 (dashed line). 
The normalized mass ratios on the $y$-axis are calculated from  
the mass of gas material within a velocity range by dividing by 
the initial mass of the cloud and 
by the width of the velocity bin 
($\Delta v_z=2.0~\mbox{km}~\mbox{s}^{-1}$ in this plot). 
The total mass of material within the 
$-40 \le v_z \le -150 ~\mbox{km}~\mbox{s}^{-1}$ velocity 
range exceeds the cloud's mass in cases where the ambient 
gas has been accelerated. 
\label{ratio_A2B2B4_fig}}
\end{figure}

\begin{figure}
\epsscale{1.0}
\plotone{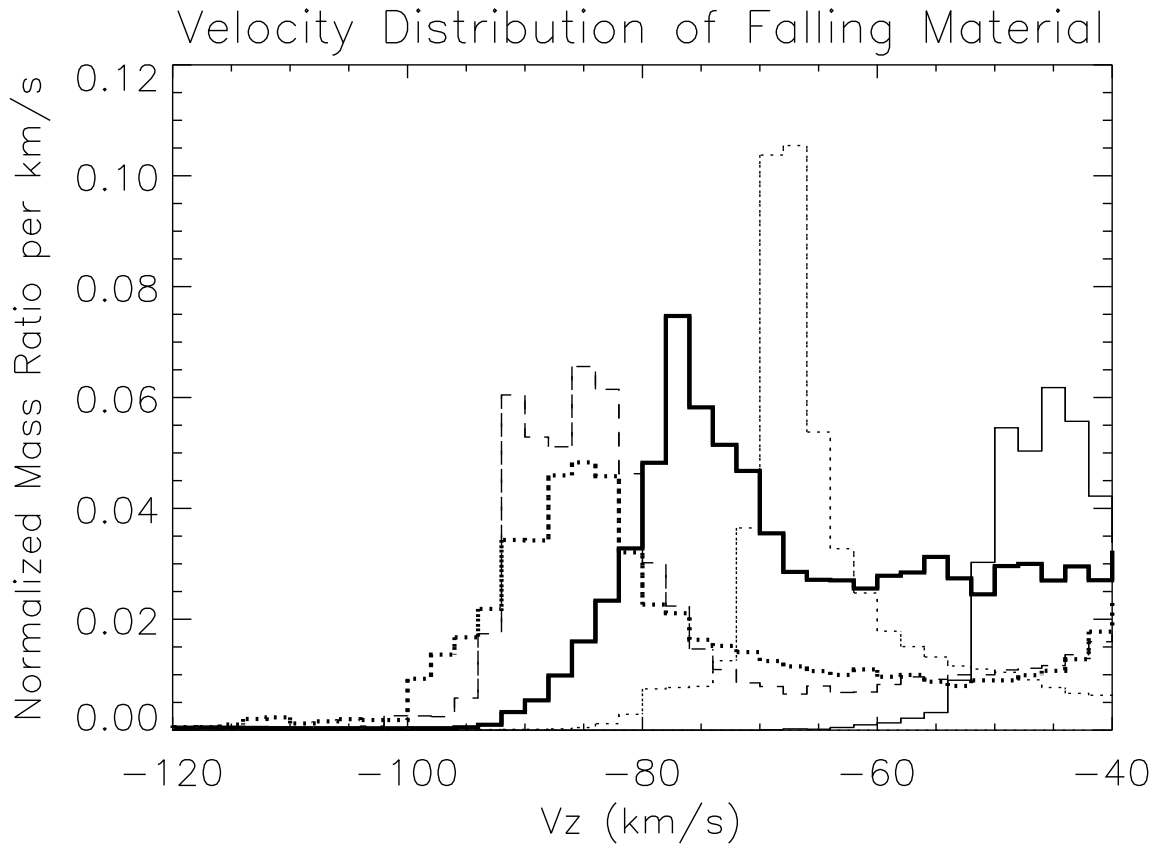}
\caption{Mass of material moving with a velocity of $v_z$ at 40 Myr for 
models with a low initial cloud density: 
A1 (thick dotted line), B1 (thin solid line), C1 (dashed line), 
C2 (thick solid line), and D (thin dotted line). 
Note that there is no material moving within this velocity range 
in Model B3. 
The normalized mass ratios are calculated in the same way as those in 
Figure \ref{ratio_A2B2B4_fig}. 
\label{ratio_A1C1C2_fig}}
\end{figure}

In order to investigate which model clouds achieve 
fast enough speeds to be identified as HVCs, 
we examine their downward velocities. We examine the clouds 
late in their descents ($t=40~\mbox{Myr}$) because they move 
faster at this time than at any previous time. At later times, some of 
the clouds will cross the Galactic plane or slow dramatically. 
Figures \ref{ratio_A2B2B4_fig} and \ref{ratio_A1C1C2_fig} 
show the distribution of material across 
the velocity range $-150 \leq v_z \leq -40$ $\mbox{km}~\mbox{s}^{-1}$.
Table \ref{mass_ratio_T} presents the ratio of the gas mass 
to the initial cloud's mass in five $v_z$ ranges for each model. 

Among all of the simulated clouds, those in Models A2 and B2 (which
have initial densities of $0.1~\mbox{H atoms}~\mbox{cm}^{-3}$) 
are most likely to be identified as HVCs after falling from rest for 40 Myr. 
In these simulations, the amount of 
material with $v_z \leq -90$ $\mbox{km}~\mbox{s}^{-1}$ is 
comparable to the cloud's initial mass. 
The cloud in Model B4 (which has the same initial density as those in 
Models A2 and B2) would be observed as an IVC 
rather than as an HVC because most of the material has vertical velocities 
between $-90$ and $-40$ $\mbox{km}~\mbox{s}^{-1}$. 
In comparison, the low density analogs to the clouds in Models A2 and B2 
(i.e., Models A1 and B1, respectively) would appear to the observers as 
IVCs and the low density analog to the Model B4 cloud (i.e., the Model B3 
cloud) would appear as an even slower class of cloud. 

The results for Models B1-B4, and D listed in Table 
\ref{mass_ratio_T} confirm that the ability of the ISM to retard the 
cloud's motion increases with the strength of the component of the 
magnetic field that is perpendicular to the cloud's velocity vector. 
While the Model B2 cloud moves at HVC velocities, its high magnetic 
field analog (the Model B4 cloud) moves at IVC velocities. 
Similarly, the Model B1 cloud mostly moves at IVC velocities while 
its high magnetic field analog (the Model B3 cloud) moves much slower. 
In Model D, whose $B_y$ is smaller than that of B1, the mass fraction 
in the range of IVC velocities is larger than that in Model B1. The cloud 
in Model D would also be identified as an IVC. 

The results for Models C1 and C2 listed in Table \ref{mass_ratio_T} 
confirm that magnetic fields that run parallel to the cloud's motion 
do not slow the clouds as effectively as fields that cross the cloud's 
motion vector. 
However, the more detailed velocity distributions for these models 
shown in Figure \ref{ratio_A1C1C2_fig} reveal that the effects of 
magnetic fields that are oriented parallel to the cloud's motion 
increase with magnetic field strength. To wit, the velocity distribution for 
Model C2 is skewed to lower velocities, implying that the cloud 
in this model experienced more resistance than those in Models C1 and A1, 
which, respectively, have $1/\sqrt{10}$ and 0 times 
the magnetic field strength of Model C2. 
Note that a significant amount of material in Model C2 travels 
at $-60 \leq v_z \leq -40$ $\mbox{km}~\mbox{s}^{-1}$. Most of this material is ambient ISM 
located ahead of the cloud. Models C1 and A1 also have accelerated ambient 
gas, though less of it. 

The stronger magnetic field in Model C2 more tightly 
confines the lateral spread of both the cloud and the compressed 
ISM below the cloud. 
The pressure of the small confined region 
below the Model C2 cloud provides slightly larger resistance than 
the pressure of the larger region below the Model C1 cloud. 
In Model A1, the zone of high pressure is spread out even further 
than in Model C1 and provides even less resistance to the cloud's descent. 
Thus, the cloud in Model C2 experiences more resistance than those in 
Models A1 and C1 (compare the $v_z$ profiles in Figures \ref{modelA1_P}, 
\ref{modelC1_P}, and \ref{modelC2_P}). 
In addition, the shock wave propagates faster in the more pressurized, 
more confined region below the cloud in Model C2. But, 
in Models A1 and C1, the shock waves stall in the region 
where the pressure is weakened. The shock wave is important 
because the ambient material in the velocity 
range of $-60 \leq v_z \leq -40$ $\mbox{km}~\mbox{s}^{-1}$ 
gained its velocity when the shock wave passed it. 

In addition to downward-moving gas, we also find 
upward-moving gas by $40$ Myr (Table \ref{mass_ratio_T}, 
sixth column). The upwardly moving material is ambient gas 
that was pushed aside by the falling clouds and now moves 
upward along the sides of the clouds. 
Gravity eventually stops this upwardly moving gas material and 
returns it to HSE. 
The behavior of the ambient gas through which a cloud falls, 
especially in Models A1 and A2 in
which there is no ambient magnetic field, 
is similar to water through which a rock falls. 
At any instant, there is a portion of water that moves downward 
because it was pushed down by the rock; 
but the water is not permanently dragged down 
with the rock. Instead, it returns to its original hydrostatic balance
after the rock passes through it. 
The fourth and sixth column in Table 3 show that
roughly similar amounts of gas material move in the ranges of  
$-40\le v_z \le -10$ and $v_z \ge 10$ $\mbox{km}~\mbox{s}^{-1}$, 
resulting from the above behavior of the ambient
gas in Models A1-B4. However,
the upward motion of the ambient gas in Models C1 and C2 is
significantly inhibited due to the magnetic field oriented along the
$z$-direction. In these cases, the magnetic field acts like a cage that
prevents the compressed ambient gas from moving laterally. 
Because it cannot easily flow around the cloud, the ambient material 
is dragged down with the cloud in Models C1 and C2. Although it is 
very unlikely that many infalling clouds are directed along 
the ambient magnetic field of the lower Galactic halo, this 
additional gas dragged with the cloud theoretically increases 
the mass return rate of the fountain process. 
It is interesting to compare the mass of the ambient gas 
that is dragged down in Models C1 and C2 with the results of
\citet{FraternaliBinney2006MNRAS,FraternaliBinney2008MNRAS}. They
modeled the extraplanar neutral gas for two nearby galaxies, NGC 891
and NGC 2403 and they found that the observed dynamics can be
explained if the extraplanar gas is composed of 
$80 \%$-$90 \%$ of fountain gas that accretes $10 \%$-$20 \%$ 
extra gas material from the ambient medium during its descent. Note that
their model assumes point particles for the fountain gas and 
includes the Galaxy's gravity when 
calculating the trajectories of the particles. 
In their model, the estimated accretion rate was a model parameter and
was determined by comparison with observations. Our simulations show
that dragging of ambient gas can be significant depending on the
configuration of the ambient magnetic field. 

Table \ref{mass_ratio_T} also lists significant mass fractions of 
slow moving gas 
(i.e., $-10 \leq v_z \leq 10 ~\mbox{km}~\mbox{s}^{-1}$). 
This is not astronomically significant. It is due to the 
disruption of the ambient medium and the imperfect 
numerical implementation of HSE. 
However, in Model B3, $80\%$ of the cloud mass is 
in this velocity range; the ISM motions merely mask the 
motion of this very slowly descending cloud. 

In only one of our cases, that with the lowest cloud density 
and the highest magnetic field strength (Model B3), 
does the cloud reach terminal velocity. 
On the surface, this may seem to contradict the analytical 
calculation of \citet{BenjaminDanly1997} who assumed and validated 
the assumption that downward falling fountain gas reaches 
terminal velocity. However, their comparisons between theory 
and observations concentrated on IVCs, 
most of which were observed within 200 pc of the Galactic plane 
and were thought to have fallen from heights of approximately 
$z=1$ kpc. 
In our simulations, the clouds fall from a height of $z=5$ kpc 
where the gravitational acceleration is stronger and 
the interstellar density is weaker than at $z=1$ kpc 
(see Figure \ref{ferriere_profile}) 
and we do not track the clouds below a height 
of $z \approx 2$ kpc. Another difference between our work and 
that of \citet{BenjaminDanly1997} is 
that they assume that the drag coefficient 
remains constant throughout time, but we find that the drag 
changes as the magnetic field and gas are compressed beneath 
the cloud causing $d(p_{mag}+p_{th})/dz$ across the cloud-ISM 
interface to increase. In our simulations, the drag is too small 
to decelerate the clouds to the terminal velocity in all but 
Model B3. 

\section{Discussion}\label{discuss_S}

\subsection{How Column and Volume Densities Relate to Observations}
\label{observation_S}

The calculated column densities of fast-moving gas ($v_z \le -90
~\mbox{km}~\mbox{s}^{-1}$) along the $z$-axis at $x=y=0$ 
in Models A2 and B2 are approximately 
$2.73 \times 10^{20}$ and $5.0 \times 10^{20}$ 
$\mbox{cm}^{-2}$, respectively. They are larger than the 
analytically calculated column density for the equivalent line 
of sight through a ballistically falling cloud having an initial 
density of 0.1 H atoms $\mbox{cm}^{-3}$ and a radius of 0.25 kpc 
($1.5 \times 10^{20} ~\mbox{cm}^{-2}$, see Section \ref{freefall_S}) 
because the simulated clouds have accelerated ambient material. 
The Model B2 cloud has a larger column density 
than the Model A2 cloud because more material accumulates along 
the central sight line when the magnetic field lines are perpendicular to 
the cloud's motion (Section \ref{geometry_S}). Note that the HVC column 
densities in both Models A2 and B2 exceed the detection threshold 
for most HVC observations (a few times $10^{18}~\mbox{cm}^{-2}$). 

An interesting comment needs to be made about calculations of the 
volume density that draw upon observational data. 
In such cases, the volume density, $n$, 
is usually estimated from the column density, $N$, by using 
the equation $n = N / d$, where $d$ is the depth of the cloud. 
Since $d$ is not observable, it is often assumed to 
be equal to the cloud's width. 
However, as in our simulations, if a cloud 
is compressed vertically during its descent, or is deformed either due to 
compression or due to the magnetic field, 
then the above assumption is not applicable. 
In the cases where the cloud is more compressed vertically than 
horizontally, its depth is smaller 
than its width and so the volume density calculated with the above 
technique underestimates the real volume density of the cloud. 
For example, this occurs in Models A2 and B2. 
We measured the cloud's dimensions for Models A2 and B2 
at 40 Myr (Section \ref{geometry_S}). If the width of the cloud is used as 
$d$ for the volume density estimation, then the estimated volume density 
is approximately $0.19~\mbox{H atoms}~\mbox{cm}^{-3}$ and 
$0.25~\mbox{H atoms}~\mbox{cm}^{-3}$ for the cloud of Models A2 and 
B2, respectively. However, the mean volume density 
measured from the simulation is approximately 
$0.56~\mbox{H atoms}~\mbox{cm}^{-3}$ and 
$2.13~\mbox{H atoms}~\mbox{cm}^{-3}$ for Models A2 and B2, respectively. 

\subsection{Magnetic Field Geometries in the Galactic Fountain}

\begin{figure}
\epsscale{1.0}
\plottwo{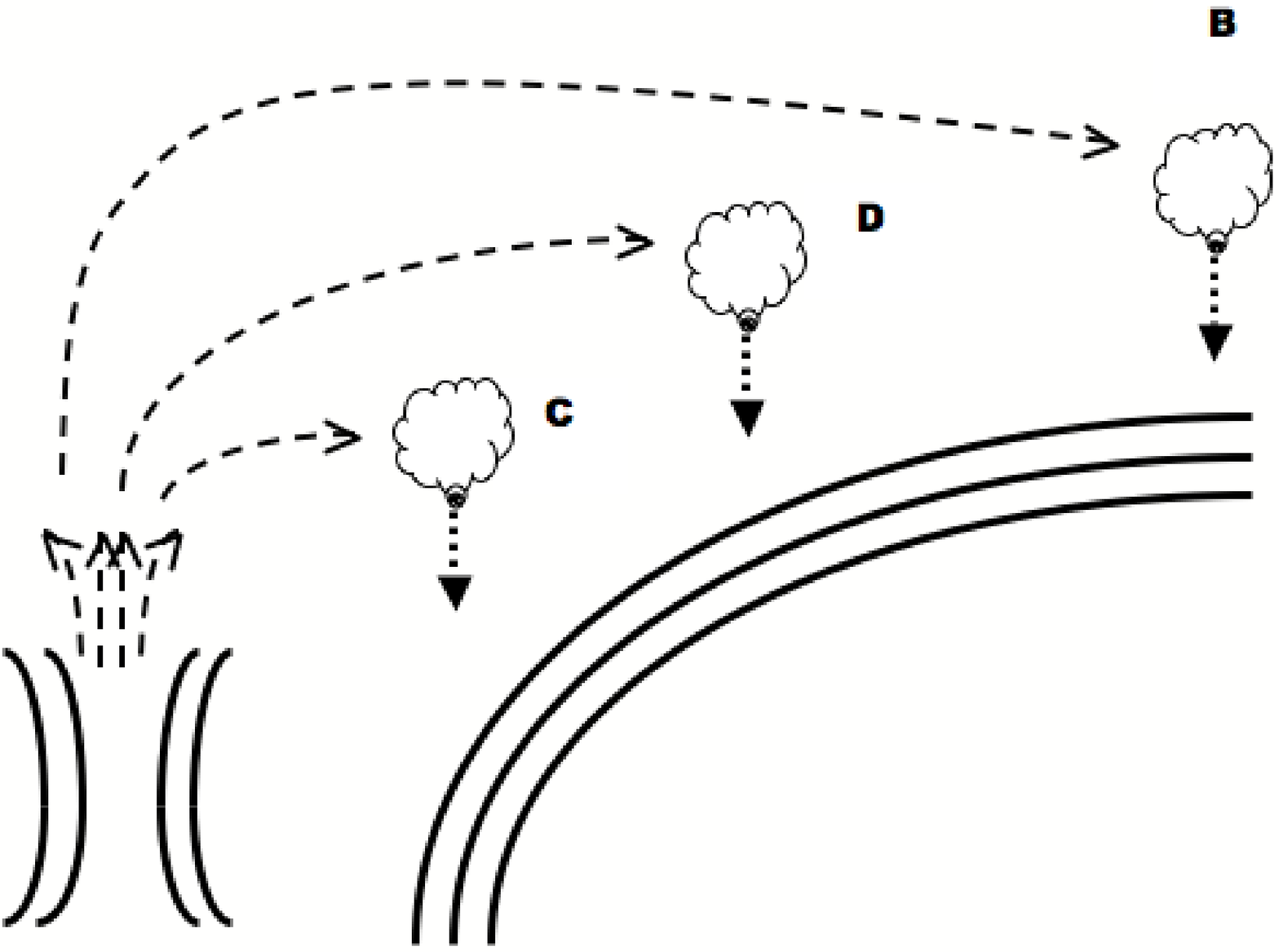}{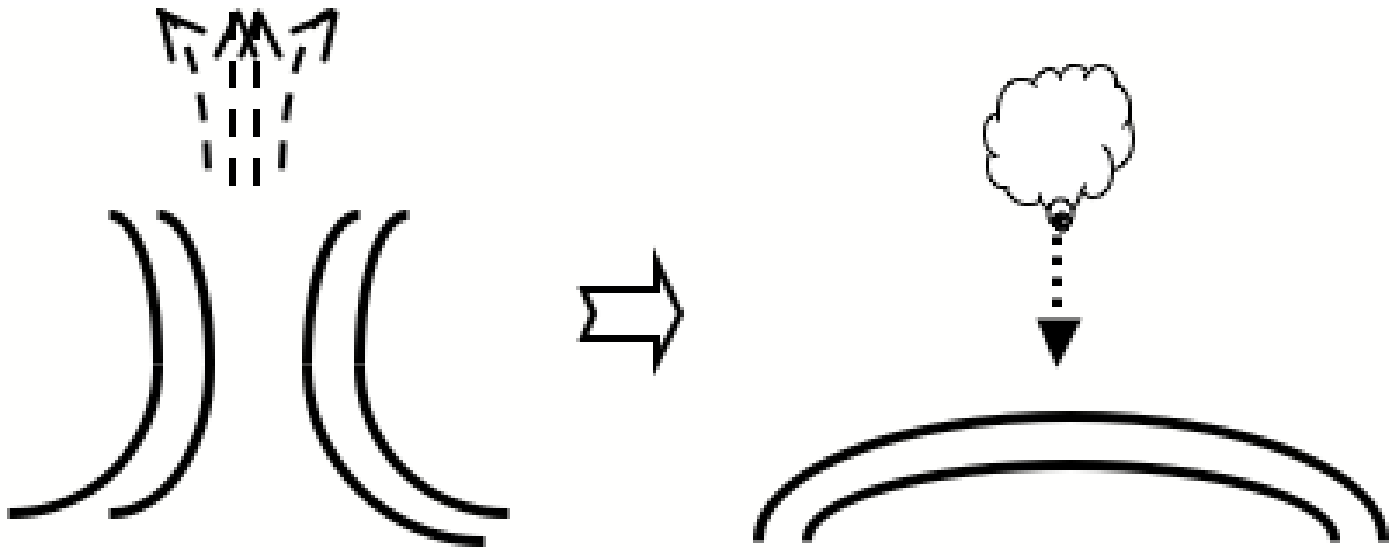}
\caption{Possible magnetic geometries in the context of a Galactic fountain. 
The solid lines are magnetic field lines and the dashed and dotted lines 
show possible paths of the fountain gas. See the text for a detailed 
explanation. 
\label{model_cartoon}}
\end{figure}

Figure \ref{model_cartoon} shows two possible scenarios 
describing the geometry of the magnetic field in the context 
of a Galactic fountain model. 
In both figures, magnetic field lines surrounding interstellar bubbles 
are indicated by the solid lines 
and possible paths of the fountain gas are 
represented by the dashed lines. 
In the left figure, hot gas escapes from a hot bubble, 
moves upward, and possibly to the side, cools down, and 
condenses into clouds at various possible locations
with respect to a neighboring bubble. 
The magnetic field lines of the neighboring bubble 
are assumed to be aligned along its periphery. 
As a result, clouds B-D (corresponding to the 
models in Table \ref{models_T} whose names 
begin with the letters B-D) 
encounter different magnetic field geometries during their descents. 
The magnetic field geometries are parallel, perpendicular and 
tilted at $45\degr$ with respect to the Galactic plane 
in Models B-D, respectively. 

The right figure shows a second possible scenario. 
In it, hot gas escapes from a hot bubble. 
While this gas cools down and condenses, the bubble's magnetic field lines, 
that had been breached to let the hot gas out, reform. In this case, the cloud falls back down near the region where the hot gas escaped. If the condensed gas cloud falls onto the top of the bubble, then it encounters a magnetic field that is parallel to the Galactic plane, like in Model B. If the escape region is on the sides of the bubble or somewhere between the top and the side, then the cloud will encounter a magnetic field that is perpendicular to or at a $45\degr$ angle to the plane, as in Model C or D, respectively. 

\subsection{Recombination of Ionized Gas}

The magnetic field only acts directly upon the ionized gas. 
It does not act directly on neutral particles, although it does act on 
ionized particles that are mixed in with neutral particles 
and the ionized particles will influence the neutral particles. 
Technically, those of our simulations that include magnetic fields assume that both the cloud and the ambient ISM are ionized. If the ionized cloud recombines later, it could be observed as a neutral HVC. Even if the ionized cloud does not fully recombine, the MHD simulations are still relevant because partially and fully ionized HVCs have been observed 
\citep{sembach2003,Foxetal2006ApJS}. 

\subsection{Hot Ambient Medium}\label{hot_ISM_effect_S}

Within 5 kpc of the Milky Way's midplane, the ISM contains 
hot, warm, and cool gas. Recent observations revealed that 
even cold gas (\ion{H}{1}) contributes 
a significant portion of the halo 
ISM in the Milky Way \citep{KalberlaDedes2008AA} and an 
external galaxy, NGC 891 \citep{Oosterlooetal2007AJ}.  
However, there is evidence that 
hot ($>10^6$ K) gas plays a more dominant role in 
some galaxies. The halos of starburst galaxies, for example, 
contain significant quantities of hot gas 
(e.g., \citet{Stricklandetal2004ApJS}). 
In addition, \citet{FukugitaPeebles2006ApJ} suggested that 
the very extended halos of normal spiral galaxies are hot. 
If a cloud was to fall through a hotter more rarefied gas than 
we assumed in our simulations, then it would experience less 
resistance due to the thermal pressure gradient. This effect 
would be most relevant if the magnetic field was absent or 
oriented parallel to the cloud's motion and would allow 
the cloud to fall slightly faster than the clouds in Models A1, 
A2, C1, and C2. If, in contrast, a non-negligible magnetic 
field oriented perpendicular to the cloud's motion was to 
permeate the halo, then the thermal pressure gradient 
would be unimportant. The buildup of magnetic pressure 
beneath the cloud would dominate the cloud's dynamics. 
This magnetic pressure would slow the cloud, regardless of 
the reduced thermal pressure. 


\section{Conclusion}\label{conclusion_S}

Our simulations show that some clouds that form in the Galactic fountain process 
can be observed as HVCs during their descent toward the Galactic plane. 
According to our simulations, a cloud is more likely to accelerate to 
HVC-class velocities if its initial density is large enough 
to overcome the interaction with the ambient ISM. 
The interaction with the ambient ISM depends upon the configuration of the magnetic 
field in the Galactic halo. 
Magnetic fields that are parallel to the Galactic plane decelerate the falling cloud most effectively, while magnetic fields that are perpendicular to the plane and thus 
parallel to the motion of the cloud merely confine the material 
and guide its trajectory, but do not significantly slow the cloud. 
The strength of the magnetic field is important to the motion of the cloud 
only when the magnetic field is perpendicular to the cloud's motion. In this case, 
the cloud's downward movement compresses and thus strengthens 
the magnetic field. 
Unless the density of the cloud is sufficiently large, the resulting magnetic 
pressure controls the cloud's dynamics. 

We find that an initial cloud density of $0.1~\mbox{H atoms}~\mbox{cm}^{-3}$ 
is large enough for the cloud to overcome the interaction with the ambient ISM
if the ambient magnetic field strength is $\le 1.3 ~ \mu$G, regardless of 
the magnetic field's orientation. 
Since an ambient magnetic field with this strength or less is very likely 
to exist at the cloud's maximum height above the Galactic plane, clouds with this density or greater that fall from the height of $z=5$ kpc should accelerate to $\vert v \vert >
90$ $\mbox{km}~\mbox{s}^{-1}$ and therefore be easily identified as HVCs after they fall for about 40 Myr. 
However, if the ambient magnetic field is 
perpendicular to the motion of the cloud and has a relatively large strength
($\ge 4.2 ~ \mu$G), then the cloud's interaction with the ISM is stronger and 
significantly hinders the cloud's motion. 
A cloud falling in such an environment for 40 Myr 
is more likely to be identified as an IVC 
than as an HVC. 
If the initial cloud density is 1/10 of the above density, then the cloud 
reaches HVC velocities during its fall only if the magnetic field is absent 
or parallel to the cloud's motion. 



\acknowledgments

The FLASH code used in this work was in part developed 
by the DOE-supported ASC/Alliance Center for Astrophysical Thermonuclear Flashes at the University of Chicago. 
The simulations were performed at the Research Computing Center (RCC) of the University of Georgia. We thank Dr. Shan-Ho Tsai for her assistance in obtaining RCC computing resources, Dr. David Henley for his comments on the draft, 
and the anonymous referee for his or her valuable comments.

\bibliography{apj-jour,ref_hvc}

\begin{thebibliography}{47}
\expandafter\ifx\csname natexlab\endcsname\relax\def\natexlab#1{#1}\fi

\bibitem[{{Agertz} {et~al.}(2007){Agertz}, {Moore}, {Stadel}, {Potter},
  {Miniati}, {Read}, {Mayer}, {Gawryszczak}, {Kravtsov}, {Nordlund}, {Pearce},
  {Quilis}, {Rudd}, {Springel}, {Stone}, {Tasker}, {Teyssier}, {Wadsley}, \&
  {Walder}}]{Agertzetal2007MNRAS}
{Agertz},  O., et al. 2007, \mnras,   380, 963

\bibitem[{{Belokurov} {et~al.}(2007){Belokurov}, {Zucker}, {Evans}, {Kleyna},
  {Koposov}, {Hodgkin}, {Irwin}, {Gilmore}, {Wilkinson}, {Fellhauer},
  {Bramich}, {Hewett}, {Vidrih}, {De Jong}, {Smith}, {Rix}, {Bell}, {Wyse},
  {Newberg}, {Mayeur}, {Yanny}, {Rockosi}, {Gnedin}, {Schneider}, {Beers},
  {Barentine}, {Brewington}, {Brinkmann}, {Harvanek}, {Kleinman}, {Krzesinski},
  {Long}, {Nitta}, \& {Snedden}}]{belokurov2007}
{Belokurov},  V., et al. 2007, \apj, 654, 897

\bibitem[{{Benjamin} \& {Danly}(1997)}]{BenjaminDanly1997}
{Benjamin}, R.~A., \& {Danly}, L. 1997, \apj, 481, 764

\bibitem[{{Blitz} {et~al.}(1999){Blitz}, {Spergel}, {Teuben}, {Hartmann}, \&
  {Burton}}]{Blitz1999}
{Blitz}, L., {Spergel}, D.~N., {Teuben}, P.~J., {Hartmann}, D., \& {Burton},
  W.~B. 1999, \apj, 514, 818

\bibitem[{{Collins} {et~al.}(2007){Collins}, {Shull}, \&
  {Giroux}}]{collins2007}
{Collins}, J.~A., {Shull}, J.~M., \& {Giroux}, M.~L. 2007, \apj, 657, 271

\bibitem[{{Cowie} \& {McKee}(1977)}]{CowieMcKee1977ApJ}
{Cowie}, L.~L., \& {McKee}, C.~F. 1977, \apj, 211, 135

\bibitem[{{Danly} {et~al.}(1993){Danly}, {Albert}, \& {Kuntz}}]{danly1993}
{Danly}, L., {Albert}, C.~E., \& {Kuntz}, K.~D. 1993, \apjl, 416, L29

\bibitem[{{Davis} {et~al.}(1996){Davis}, {Mulchaey}, {Mushotzky}, \&
  {Burstein}}]{Davisetal1996ApJ}
{Davis}, D.~S., {Mulchaey}, J.~S., {Mushotzky}, R.~F., \& {Burstein}, D. 1996,
  \apj, 460, 601

\bibitem[{{de Avillez}(2000)}]{avillez2000HVCs}
{de Avillez}, M.~A. 2000, \apss, 272, 23

\bibitem[{{de Avillez} \&
  {Breitschwerdt}(2007)}]{deAvillezBreitschwerdt2007ApJL}
{de Avillez}, M.~A., \& {Breitschwerdt}, D. 2007, \apjl, 665, L35

\bibitem[{{Ferriere}(1998)}]{ferriere1998}
{Ferriere}, K. 1998, \apj, 497, 759

\bibitem[{{Fox} {et~al.}(2006){Fox}, {Savage}, \& {Wakker}}]{Foxetal2006ApJS}
{Fox}, A.~J., {Savage}, B.~D., \& {Wakker}, B.~P. 2006, \apjs, 165, 229

\bibitem[{{Fraternali} \& {Binney}(2006)}]{FraternaliBinney2006MNRAS}
{Fraternali}, F., \& {Binney}, J.~J. 2006, \mnras, 366, 449

\bibitem[{{Fraternali} \& {Binney}(2008)}]{FraternaliBinney2008MNRAS}
---. 2008, \mnras, 386, 935

\bibitem[{{Fryxell} {et~al.}(2000){Fryxell}, {Olson}, {Ricker}, {Timmes},
  {Zingale}, {Lamb}, {MacNeice}, {Rosner}, {Truran}, \& {Tufo}}]{fryxell2000}
{Fryxell},  B., et al. 2000, \apjs, 131, 273

\bibitem[{{Fukugita} \& {Peebles}(2006)}]{FukugitaPeebles2006ApJ}
{Fukugita}, M., \& {Peebles}, P.~J.~E. 2006, \apj, 639, 590

\bibitem[{{Gaensler}(1998)}]{Gaensler1998ApJ}
{Gaensler}, B.~M. 1998, \apj, 493, 781

\bibitem[{{Gibson} {et~al.}(2000){Gibson}, {Giroux}, {Penton}, {Putman},
  {Stocke}, \& {Shull}}]{Gibson2000}
{Gibson}, B.~K., {Giroux}, M.~L., {Penton}, S.~V., {Putman}, M.~E., {Stocke},
  J.~T., \& {Shull}, J.~M. 2000, \aj, 120, 1830

\bibitem[{{Jones} {et~al.}(1996){Jones}, {Ryu}, \&
  {Tregillis}}]{Jonesetal1996ApJ}
{Jones}, T.~W., {Ryu}, D., \& {Tregillis}, I.~L. 1996, \apj, 473, 365

\bibitem[{{Kalberla} \& {Dedes}(2008)}]{KalberlaDedes2008AA}
{Kalberla}, P.~M.~W., \& {Dedes}, L. 2008, \aap, 487, 951

\bibitem[{{Keenan} {et~al.}(1995){Keenan}, {Shaw}, {Bates}, {Dufton}, \&
  {Kemp}}]{keenan1995}
{Keenan}, F.~P., {Shaw}, C.~R., {Bates}, B., {Dufton}, P.~L., \& {Kemp}, S.~N.
  1995, \mnras, 272, 599

\bibitem[{{Kudoh} \& {Basu}(2004)}]{kudoh2004}
{Kudoh}, T., \& {Basu}, S. 2004, \aap, 423, 183

\bibitem[{{Kuntz} \& {Danly}(1996)}]{KuntzDanly1996}
{Kuntz}, K.~D., \& {Danly}, L. 1996, \apj, 457, 703

\bibitem[{LeVeque(2002)}]{leveque2002}
LeVeque, R.~J. 2002, Finite Volume Methods for Hyperbolic Problems (Cambridge
  University Press)

\bibitem[{{Lu} {et~al.}(1998){Lu}, {Sargent}, {Savage}, {Wakker}, {Sembach}, \&
  {Oosterloo}}]{Lu1998}
{Lu}, L., {Sargent}, W.~L.~W., {Savage}, B.~D., {Wakker}, B.~P., {Sembach},
  K.~R., \& {Oosterloo}, T.~A. 1998, \aj, 115, 162

\bibitem[{{Mac Low} {et~al.}(1994){Mac Low}, {McKee}, {Klein}, {Stone}, \&
  {Norman}}]{MacLowetal1994ApJ}
{Mac Low}, M.-M., {McKee}, C.~F., {Klein}, R.~I., {Stone}, J.~M., \& {Norman},
  M.~L. 1994, \apj, 433, 757

\bibitem[{MacNeice {et~al.}(2000)MacNeice, Olson, Mobarry, deFainchtein, \&
  Packer}]{mac00}
MacNeice, P., Olson, K.~M., Mobarry, C., deFainchtein, R., \& Packer, C. 2000,
  Comput. Phys. Commun., 126, 330

\bibitem[{{McKee} \& {Begelman}(1990)}]{McKeeBegelman1990ApJ}
{McKee}, C.~F., \& {Begelman}, M.~C. 1990, \apj, 358, 392

\bibitem[{{Muller} {et~al.}(1963){Muller}, {Oort}, \& {Raimond}}]{muller1963}
{Muller}, C.~A., {Oort}, J.~H., \& {Raimond}, E. 1963, C.R. Acad. Sci. Paris,
  257, 1661

\bibitem[{{Oort}(1966)}]{oort1966}
{Oort}, J.~H. 1966, \bain, 18, 421

\bibitem[{{Oosterloo} {et~al.}(2007){Oosterloo}, {Fraternali}, \&
  {Sancisi}}]{Oosterlooetal2007AJ}
{Oosterloo}, T., {Fraternali}, F., \& {Sancisi}, R. 2007, \aj, 134, 1019

\bibitem[{{Putman} {et~al.}(2004){Putman}, {Thom}, {Gibson}, \&
  {Staveley-Smith}}]{putman2004}
{Putman}, M.~E., {Thom}, C., {Gibson}, B.~K., \& {Staveley-Smith}, L. 2004,
  \apjl, 603, L77

\bibitem[{{Raley} {et~al.}(2007){Raley}, {Shelton}, \& {Plewa}}]{raley2007}
{Raley}, E.~A., {Shelton}, R.~L., \& {Plewa}, T. 2007, \apj, 661, 222

\bibitem[{{Santill{\'a}n} {et~al.}(1999){Santill{\'a}n}, {Franco}, {Martos}, \&
  {Kim}}]{santillan1999}
{Santill{\'a}n}, A., {Franco}, J., {Martos}, M., \& {Kim}, J. 1999, \apj, 515,
  657

\bibitem[{{Schwarz} {et~al.}(1995){Schwarz}, {Wakker}, \& {van
  Woerden}}]{Schwarzetal1995AA}
{Schwarz}, U.~J., {Wakker}, B.~P., \& {van Woerden}, H. 1995, \aap, 302, 364

\bibitem[{{Sembach} {et~al.}(2003){Sembach}, {Wakker}, {Savage}, {Richter},
  {Meade}, {Shull}, {Jenkins}, {Sonneborn}, \& {Moos}}]{sembach2003}
{Sembach},  K.~R., et al. 2003,   \apjs, 146, 165

\bibitem[{{Shapiro} \& {Field}(1976)}]{shapiro1976}
{Shapiro}, P.~R., \& {Field}, G.~B. 1976, \apj, 205, 762

\bibitem[{{Slavin} {et~al.}(1993){Slavin}, {Shull}, \&
  {Begelman}}]{Slavinetal1993ApJ}
{Slavin}, J.~D., {Shull}, J.~M., \& {Begelman}, M.~C. 1993, \apj, 407, 83

\bibitem[{{Strickland} {et~al.}(2004){Strickland}, {Heckman}, {Colbert},
  {Hoopes}, \& {Weaver}}]{Stricklandetal2004ApJS}
{Strickland}, D.~K., {Heckman}, T.~M., {Colbert}, E.~J.~M., {Hoopes}, C.~G., \&
  {Weaver}, K.~A. 2004, \apjs, 151, 193

\bibitem[{{Tenorio-Tagle} {et~al.}(1986){Tenorio-Tagle}, {Bodenheimer},
  {Rozyczka}, \& {Franco}}]{tenorio1986}
{Tenorio-Tagle}, G., {Bodenheimer}, P., {Rozyczka}, M., \& {Franco}, J. 1986,
  \aap, 170, 107

\bibitem[{{Tenorio-Tagle} {et~al.}(1987){Tenorio-Tagle}, {Franco},
  {Bodenheimer}, \& {Rozyczka}}]{tenorio1987}
{Tenorio-Tagle}, G., {Franco}, J., {Bodenheimer}, P., \& {Rozyczka}, M. 1987,
  \aap, 179, 219

\bibitem[{{Thom} {et~al.}(2008){Thom}, {Peek}, {Putman}, {Heiles}, {Peek}, \&
  {Wilhelm}}]{Thometal2008ApJ}
{Thom}, C., {Peek}, J.~E.~G., {Putman}, M.~E., {Heiles}, C., {Peek}, K.~M.~G.,
  \& {Wilhelm}, R. 2008, \apj, 684, 364

\bibitem[{{Tufte} {et~al.}(1998){Tufte}, {Reynolds}, \& {Haffner}}]{Tufte1998}
{Tufte}, S.~L., {Reynolds}, R.~J., \& {Haffner}, L.~M. 1998, \apj, 504, 773

\bibitem[{{Vieser} \& {Hensler}(2007{\natexlab{a}})}]{VieserHensler2007static}
{Vieser}, W., \& {Hensler}, G. 2007{\natexlab{a}}, \aap, 475, 251

\bibitem[{{Vieser} \& {Hensler}(2007{\natexlab{b}})}]{VieserHensler2007dyn}
---. 2007{\natexlab{b}}, \aap, 472, 141

\bibitem[{{Wakker}(2001)}]{Wakker2001}
{Wakker}, B.~P. 2001, \apjs, 136, 463

\bibitem[{{Wakker} {et~al.}(2007){Wakker}, {York}, {Howk}, {Barentine},
  {Wilhelm}, {Peletier}, {van Woerden}, {Beers}, {Ivezi{\'c}}, {Richter}, \&
  {Schwarz}}]{wakker2007_complexC}
{Wakker},  B.~P., et al. 2007, \apjl, 670, L113

\end{thebibliography}

\clearpage

\clearpage

\clearpage

\clearpage

\clearpage

\clearpage

\clearpage

\clearpage

\clearpage

\clearpage

\clearpage

\begin{deluxetable}{ccccccc}

\tabletypesize{\footnotesize}
\tablecaption{Models \label{models_T}}
\tablewidth{0pt}
\tablecolumns{4}

\tablehead{
\colhead{Models} & \colhead{Cloud Density} & \colhead{Magnetic Pressure}  &
\colhead{Magnetic Orientation \tablenotemark{a}} \\
\colhead{} & \colhead{(H atoms $\mbox{cm}^{-3}$)} & 
\colhead{($10^{-14}$ erg $\mbox{cm}^{-3}$)}  &
\colhead{}
}

\startdata
A1 & 0.01 & 0.0 & \nodata \\
A2 & 0.1 & 0.0 & \nodata \\ 
B1 & 0.01 & 7.0 & Perpendicular \\
B2 & 0.1 & 7.0 &  Perpendicular \\ 
B3 & 0.01 & 70.0 & Perpendicular \\
B4 & 0.1 & 70.0 & Perpendicular \\
C1 & 0.01 & 7.0 &  Parallel \\
C2 & 0.01 & 70.0 & Parallel \\ 
D & 0.01 & 7.0 & $45 \degr$ \tablenotemark{b}
\enddata

\tablenotetext{a}{With respect to $\hat{z}$}
\tablenotetext{b}{In our choice of coordinates, 
$B_y = B_z = \sqrt{4 \pi P_{\mbox{mag}}} \approx 0.94 \mu \mbox{G}$, 
where $P_{\mbox{mag}}=7.0\times10^{-14}~\mbox{erg}~\mbox{cm}^{-3}$.}

\end{deluxetable}

\clearpage

\begin{deluxetable}{ccccccccc}

\tablewidth{0pt}
\tabletypesize{\footnotesize}
\tablecaption{ $a/g$ and Distance Fallen \label{ag_T} }
\tablecolumns{9}

\tablehead{
\colhead{} &  
\colhead{Cloud} & 
\colhead{Magnetic} & 
\multicolumn{4}{c}{$a/g$} &
\multicolumn{2}{c}{~~~Distance Fallen (kpc)} \\
\colhead{Models} & \colhead{Density} & \colhead{Field} &
\multicolumn{4}{c}{-----------------------------------------------} & 
\multicolumn{2}{c}{~~~------------------------------} \\ 
\colhead{} & \colhead{(H atoms $\mbox{cm}^{-3}$)} & 
\colhead{($B_y$ in $\mu\mbox{G}$)} & 
\colhead{8 Myr} & \colhead{16 Myr} & \colhead{24 Myr} & 
\colhead{32Myr} &
\colhead{~~~16 Myr} & \colhead{24 Myr}
}

\startdata
A1 & 0.01 & 0.0 & 0.91 & 0.80 & 0.70 & 0.56 & ~~~0.71 & 1.24 \\
A2 & 0.1 & 0.0 & 0.97 & 0.92 & 0.83 & 0.94 & ~~~0.74 & 1.34 \\ 
B1 & 0.01 & 1.3 & 0.88 & 0.48 & 0.38 & 0.22 & ~~~0.61 & 0.97 \\
B2 & 0.1 & 1.3 & 0.96 & 0.95 & 0.87 & 0.73 & ~~~0.71 & 1.28 \\
B3 & 0.01 & 4.2 & 0.74 & 0.73 & 0.55 & 0.06 & ~~~0.49 & 0.66 \\
B4 & 0.1 & 4.2 & 0.96 & 0.77 & 0.69 & 0.20 & ~~~0.67 & 1.19 
\enddata

\end{deluxetable}

\clearpage

\begin{deluxetable}{ccccccc}

\tablewidth{0pt}
\tabletypesize{\footnotesize}
\tablecaption{Fraction of Cloud Mass Traveling within a Given Velocity Range 
at $40$ Myr 
\label{mass_ratio_T}}
\tablecolumns{4}

\tablehead{
\colhead{Models} &  
\colhead{$v_z \leq -90$} & 
\colhead{$-90 \leq v_z \leq -40$} & 
\colhead{$-40 \leq v_z \leq -10$} &
\colhead{$-10 \leq v_z \leq 10$} &
\colhead{$v_z \geq 10$} \\
\colhead{} & 
\colhead{( $\mbox{km}~\mbox{s}^{-1}$ )} &
\colhead{( $\mbox{km}~\mbox{s}^{-1}$ )} & 
\colhead{( $\mbox{km}~\mbox{s}^{-1}$ )} &
\colhead{( $\mbox{km}~\mbox{s}^{-1}$ )} &
\colhead{( $\mbox{km}~\mbox{s}^{-1}$ )} 
}

\startdata
A1 & 0.25 & 0.91 & 2.81 & 16.39 & 1.68 \\
A2 & 1.0 & 0.09 & 0.3 & 1.41 & 0.25 \\ 
B1 & 0.0 & 0.62 & 0.55 & 20.08 & 0.54 \\
B2 & 0.95 & 0.1 & 0.41 & 1.25 & 0.28 \\
B3 & 0.0 & 0.0 & 0.21 & 21.64 & 0.24 \\
B4 & 0.03 & 0.94 & 0.11 & 1.68 & 0.26 \\
C1 & 0.19 & 1.02 & 0.93 & 19.48 & 0.13 \\
C2 & 0.02 & 1.59 & 0.99 & 19.28 & 0.0 \\
D & 0.0 & 1.01 & 0.44 & 16.72 & 0.35 \\
\enddata

\tablecomments{The fractions sum to $> 1.0$ because 
ambient material has been counted along with 
the cloud material.}

\end{deluxetable}

\end{document}